\documentclass[manuscript]{aastex631}

\usepackage[version=3]{mhchem} 
\usepackage{multirow}
\shorttitle{CO modeling paper}
\shortauthors{Yang et al.}
\graphicspath{{./}{figures/}}

\begin{document}

\title{High-fidelity reaction kinetic modeling of hot-Jupiter atmospheres incorporating thermal and UV photochemistry enhanced by metastable CO(a\textsuperscript{3}$\Pi$)}

\correspondingauthor{Murthy S. Gudipati; Jeehyun Yang}
\email{murthy.gudipati@jpl.nasa.gov; jeehyun.yang@jpl.nasa.gov}

\author[0000-0002-1551-2610]{Jeehyun Yang}
\affiliation{Science Division, Jet Propulsion Laboratory, California Institute of Technology, 4800 Oak Grove Drive, Pasadena, CA 91109, USA }

\author[0000-0001-5992-373X]{Murthy S. Gudipati}
\affiliation{Science Division, Jet Propulsion Laboratory, California Institute of Technology, 4800 Oak Grove Drive, Pasadena, CA 91109, USA }

\author[0000-0001-9385-3376]{Bryana L. Henderson}
\affiliation{Science Division, Jet Propulsion Laboratory, California Institute of Technology, 4800 Oak Grove Drive, Pasadena, CA 91109, USA }

\author[0000-0003-1845-6690]{Benjamin Fleury}
\affiliation{Univ Paris Est Creteil and Universite Paris Cit\'e, CNRS, LISA, F-94010 Cr\'eteil, France}



\begin{abstract}
A detailed modeling of simultaneous UV-photochemical and thermochemical processes in exoplanet atmosphere-like conditions is essential for the analysis and interpretation of a vast amount of current and future spectral data from exoplanets. However, a detailed reaction kinetic model that incorporates both UV photochemistry and thermal chemistry is challenging due to the massive size of the chemical system as well as to the lack of understanding of photochemistry compared to thermal-only chemistry. Here, we utilize an automatic chemical reaction mechanism generator to build a high-fidelity thermochemical reaction kinetic model later then incorporated with UV-photochemistry enhanced by metastable triplet-state carbon monoxide (a\textsuperscript{3}$\Pi$). Our model results show that two different photochemical reactions driven by Lyman-$\alpha$ photons (i.e. \ce{H2} + CO(a\textsuperscript{3}$\Pi$) $\rightarrow$ H + HCO and CO(X\textsuperscript{1}$\Sigma^+$) + CO(a\textsuperscript{3}$\Pi$) $\rightarrow$ C(\textsuperscript{3}P) + \ce{CO2}) can enhance thermal chemistry resulting in significant increases in the formation of \ce{CH4}, \ce{H2O}, and \ce{CO2} in \ce{H2}-dominated systems with trace amounts of CO, which qualitatively matches with the observations from previous experimental studies. Our model also suggests that at temperatures above 2000 K, thermal chemistry becomes the dominant process. Finally, the chemistry simulated up to 2500 K does not produce any larger species such as \ce{C3} species, benzene or larger (i.e. PAHs). This might indicate that the photochemistry of \ce{C2} species such as \ce{C2H2} might play a key role in the formation of organic aerosols observed in the previous experimental study.

\end{abstract}



\section{Introduction} \label{sec:intro}
With increasing detection of exoplanets and spectroscopic characterization of exoplanet atmospheres, our knowledge of other stellar systems and planets around these stars is expanding significantly. Many of the spectral exoplanet transmission data are from the atmosphere of hot Jupiters and giant planets with short orbital periods and small semi-major axis value, which indicates that both thermal equilibrium chemistry and UV photochemistry are available in these hot-Jupiter atmospheres (\cite{Moses-2014, Madhusudhan-2016}), and hence simultaneous thermal and photochemical reaction pathways need to be considered when it comes to modeling the atmospheric composition of these UV-rich hot-Jupiter like exoplanets. It has been suggested from previous theoretical studies that thermal chemistry mainly dominates hot-Jupiter atmospheres with temperatures above 1500 K, while disequilibrium chemistry such as photochemistry could play an important role in exoplanet atmospheres whose temperatures are lower than 1500 K (\cite{Moses-2011, Venot_2012, Moses-2014}). 

The chemistry of an exoplanetary atmosphere largely depends on its C/O ratio, where C/O $\geq$ 1 indicates a carbon-rich atmosphere and C/O $<$ 1 indicates an oxygen-rich atmosphere (Solar C/O = 0.5). Along with temperature, these C/O ratios can be used to classify exoplanetary atmospheres and identify the major chemistry products that occur in each case. Several groups have explored the connection between C/O ratio, temperature, and chemistry in the hot-Jupiter atmospheres using modeling, observations, and experiments, including \cite{Madhusudhan_2012, Moses-2013, Venot_2015, Drummond-2019} and \cite{Fleury_2019, Fleury_2020}. \cite{Madhusudhan_2012} combined observations with an atmospheric chemical model (photochemistry not included) to classify how the C/O ratio affects the chemical compositions of the atmospheres, and \cite{Venot_2015} have explored with a 1D thermo-photochemical model how various parameters including the C/O ratio affect the chemistry in hot-Jupiter atmospheres. In \cite{Fleury_2019,Fleury_2020}, they have experimentally explored the influence of the C/O ratio on the formation of trace species and photochemical aerosols in hot-Jupiter atmospheres. 

Among these, the experimental studies by \cite{Fleury_2019, Fleury_2020} clearly showed enhanced formation of \ce{CH4}, \ce{H2O}, and \ce{CO2} through simultaneous thermal- and UV photochemistry compared to the product formation with thermal-only chemistry at the temperature conditions below 1500 K. They also observed the formation of non-volatile hydrocarbon aerosols as solid thin films after the UV irradiation of carbon-enriched (C/O = 1) gas mixture at the temperature of 1473 K (\cite{Fleury_2019}).

These two previous experimental studies provided us with new and first experimental insights into potential importance of photochemistry in hot-Jupiter-like exoplanet atmospheres as well as interesting questions with regard to the formation pathways for both observed gaseous products (i.e. \ce{CH4}, \ce{H2O}, and \ce{CO2}) and aerosol products. By tracking these formation pathways for both gaseous and condensed phase products from laboratory experimental data points, we can identify key chemical species including intermediates that are formed during UV irradiation of the laboratory analogs of exoplanet atmospheres. Eventually, these reactions can be implemented into 1D T-P profile supported atmospheric photochemical models (also known as atmospheric photochemical models) that address various physical (e.g. vertical diffusion of molecules, dry and wet deposition, atmospheric escape, condensation and sedimentation of species, etc.) and chemical (e.g. photochemistry, kinetics of the reactions between atmospheric components, etc.) principles (\cite{Hu_2012}). Adding a more complete picture of reaction chemistry into this atmospheric photochemical model will benefit analysis and interpretation of a vast amount of current and future spectral data from exoplanets as well as designing future space missions.

For example, the combination between the recent observational data of the atmosphere of WASP-39b by the James Webb Space Telescope (JWST) and the atmospheric chemical modeling works has suggested the first evidence of photochemistry in an exoplanet atmosphere (\cite{Tsai_2022}). Indeed, the JWST observational data of the atmosphere of WASP-39b shows a peak at 4.05 $\mu$m which has been attributed to \ce{SO2} molecules. Using multiple atmospheric photochemical model results, \cite{Tsai_2022} strongly suggest
that \ce{SO2} is formed by photochemistry in the atmosphere of WASP-39b after tracking its formation pathways using reaction kinetic modeling. Therefore, it is clear that photochemistry plays an important role in the composition and spectra of observed exoplanet atmospheres, which gives a rationale that precise reaction kinetic modeling is essential when it comes to atmospheric photochemical modeling. When this precise reaction kinetic modeling is coupled with experimental work as presented in this study, it can significantly benefit the astronomical community by decreasing uncertainties in chemical reactions that need to be implemented in atmospheric chemical models used to interpret observational data.

Tracking chemical formation pathways of laboratory experiments is very challenging without the aid of reaction kinetic modeling. This is because the timescales of the initial chemical reactions are too short even for the well-controlled pseudo-first order radical chemistry (e.g. $\sim$30 $\mu$s) (\cite{Golan-2013}), so that the products that are observed in gas-phase static cell experiments (e.g. \cite{Fleury_2019, Fleury_2020}) are formed far beyond tertiary or even further reaction chemistry (including surface chemistry on the wall). For this reason, reaction kinetic modeling has been actively utilized in a variety of fields (e.g. astrochemistry, combustion research, etc.) to interpret various observational and experimental data. Particularly, the advancement of computer-aided automatic construction of reaction kinetic modeling in combustion research is impressive. For example, \cite{Liu_2020} used Reaction Mechanism Generator (RMG, \cite{Gao_2016, Liu-2021}) to automatically construct the acetylene pyrolysis model (with a temperature range of 1000--1500 K, a pressure of 0.2 atm, and a reaction time of 0.5 s) that contains 1594 species and 8924 reactions and successfully described up to pyrene formations observed from the previous acetylene (\ce{C2H2}) pyrolysis experiment by \cite{Norinaga_2008}.

RMG is an open-source software that automatically constructs networks with relevant core reactions based on its own algorithm to choose reaction rates (e.g. experimentally measured rates would be prioritized followed by less reliable sources in order) and a rate-based iterative algorithm for model generation (\cite{Gao_2016}). This approach has a few advantages over the traditional way (i.e. manually choosing reaction rates from previous references) of building reaction chemical networks: (i) since reaction rates are chosen based on a rate-based iterative algorithm, compared to manually selecting reaction rates, it is less likely that the reaction mechanism will miss important (i.e. relevant) reactions as long as the reaction libraries are solid, (ii) the approach has multiple self-feed back and refinement steps that are based on a solid algorithm, thus providing the model with more reliability, and (iii) this automatic approach enables us to describe larger and more complex chemical system that cannot be constructed by the traditional way.

As we can see from \cite{Liu_2020}'s \ce{C2H2} pyrolysis model, in order to reasonably describe the chemistry of exoplanet atmosphere-like conditions (i.e. T = 1000--1500 K and P = 0.2 atm) with even relatively simple starting material (i.e. \ce{C2H2}), it requires a model size that is too enormous to construct manually. This enormous model size gives us a rationale to utilize this computer-aided modeling technique to precisely interpret experimental studies of hot-Jupiter atmospheric chemistry.

On top of this, including photochemistry into reaction kinetic modeling is also important. Although the temperature, pressure, and chemical conditions of hot-Jupiter atmospheres are similar to those of flame (i.e. combustion), the existence of UV photons (mainly Lyman-$\alpha$ from their parent stars) in hot-Jupiter atmospheres is the major difference that distinguishes hot-Jupiter atmosphere from combustions (\cite{France-2013, Miguel-2015}). Molecules in the top layers of the hot-Jupiter atmospheres will interact with accessible UV-photons and contribute to whole reaction chemistry of the system. Among a variety of molecules available in hot-Jupiter atmospheres, carbon monoxide (CO) is one of major interest to astrochemists since CO is predicted to be one of the most abundant species in hot-Jupiter-type exoplanet atmospheres whose temperatures are higher than 1000 K (\cite{Moses-2013, Venot_2015, Drummond-2019}).

In addition to this, the recent laboratory experimental results by \cite{Fleury_2019,Fleury_2020} strongly suggest CO as a possible photochemical precursor driven by CO photoexcitation, but lacked a detailed reaction kinetic model to explain reaction pathways to the observed products. Indeed, CO can be electronically excited to stay chemically reactive with a relatively long-lifetime through UV irradiation (\cite{Fournier_1980, Gudipati_1998, Lee_2020}). Although photochemically excited CO has been raised as an important precursor to photochemical pathways (\cite{Fleury_2019, Roudier-2021}), as far as we know, no previous study has ever assessed the reaction kinetic role of this electronically excited CO in these hot-Jupiter-like atmospheres in detail. 

With this background, in this paper, we will utilize an automatic chemical reaction mechanism generator to build a high-fidelity chemical network that can assess the chemical importance of photoexcited carbon monoxide and qualitatively rationalize the increase of the production yields of \ce{CH4}, \ce{H2O}, \ce{CO2}, and aerosols during UV irradiation compared to thermal-only chemistry, which were observed in the previous studies by \cite{Fleury_2019, Fleury_2020}. Our work reported here is the first of its kind to incorporate simultaneous thermally and photochemically excited CO-induced chemical reaction pathways, which will provide not only a better insight into reaction mechanisms in hot-Jupiter like exoplanet atmospheres, but also provide a tool to confidently predict major and minor atmospheric molecular species under different conditions.

\section{Methods} \label{sec:methods}
\subsection{Kinetic and thermodynamic parameter libraries}\label{rmg_library}
In order to consider the conversion of carbon monoxide into methane, a part of kinetic and thermodynamic parameters were taken from a recent experiment and kinetic modeling combined study of methane oxidation by \cite{Hashemi_2016} and chosen as the seed mechanism's kinetic parameters that are included in the reaction kinetic model as main chemistry. All the other kinetic and thermodynamic parameters were taken from the libraries included in the previous acetylene pyrolysis model by \cite{Liu_2020} that successfully described up to pyrene (4-ring polycyclic aromatic hydrocarbons) formation observed from the previous experiment by \cite{Norinaga_2008} starting from acetylene for the residence time of 0.5 s at the temperatures of 1073--1373 K, at the pressure of 80 mbar. In doing so, we can describe any larger molecules that might have been formed but not observed in the previous experiments by \cite{Fleury_2019} and \cite{Fleury_2020}. All these kinetic and thermodynamic parameters can be found in CHEMKIN format in the Supplementary Materials.
With regard to thermodynamic parameters of the triplet carbon monoxide (i.e. CO(a\textsuperscript{3}$\Pi$)) that is not available in the library (since RMG's library usually contains the thermodynamic parameters of molecules in the ground state), we first carried out electronic structure calculations at the CBS-QB3 level of theory using Gaussian 09 (\cite{g09}) to determine geometric conformations, energies, and vibrational frequencies of the triplet carbon monoxide. Then the thermodynamic parameters of this molecule were calculated by Arkane (\cite{Allen_2012}), a package included in the open-source software RMG v3.1.0 (\cite{Gao_2016, Liu-2021}), with atomic energy corrections, bond corrections, and spin orbit corrections, based on the CBS-QB3 level of theory as the model chemistry. These thermodynamic parameters of the triplet carbon monoxide is given in CHEMKIN format in the Supplementary Materials as well as molecular parameter outputs (Gaussian 09 output file) and Arkane input file.

\subsection{Automoatic thermochemical reaction model generation}\label{rmg_generation}
RMG was used to generate the thermal chemistry model that can simulate the experiments of \cite{Fleury_2019, Fleury_2020}. An initial molar composition of 99.7 \% of hydrogen and 0.3 \% of carbon monoxide was used for the experimental condition of \cite{Fleury_2019}, while an initial molar composition of 99.26 \% of hydrogen, 0.26 \% of carbon monoxide, and 0.48 \% of water was used for the experimental condition of \cite{Fleury_2020}. Batch reactor conditions were set with a temperature range of 300--1800 K and a pressure of 15 mbar for both models. These physical conditions are relevant to the atmosphere of hot Jupiters and gas giant exoplanets. For example, according to \cite{Tsai_2022}, the temperature profile of the atmosphere of WASP-39b (which is a hot-Jupiter type exoplanet) at 10 mbar ranges from 700 K to 1300 K, which is very similar to the physical conditions simulated in this study.

The pressure dependence feature of RMG was enabled to automatically construct pressure-dependent networks for species with up to 10 heavy atoms. Species constraints were set to limit the maximum number of carbon atoms in any molecule to 16 and the maximum number of radicals to 3 in order to keep the model generation realistic but conserve computing time as well. After the model generation completed, the final model contained 475 species and 1284 reactions (forward-reverse reaction pairs), which can be found in the Supplementary Materials as well as the RMG input file.

\subsection{Rate coefficients of photochemical reactions and phosphorescence quenching of CO(a\textsuperscript{3}$\Pi$)} \label{photoexcit_rate_calc}
\begin{figure}
    \centering
    \includegraphics[width=1.0\textwidth]{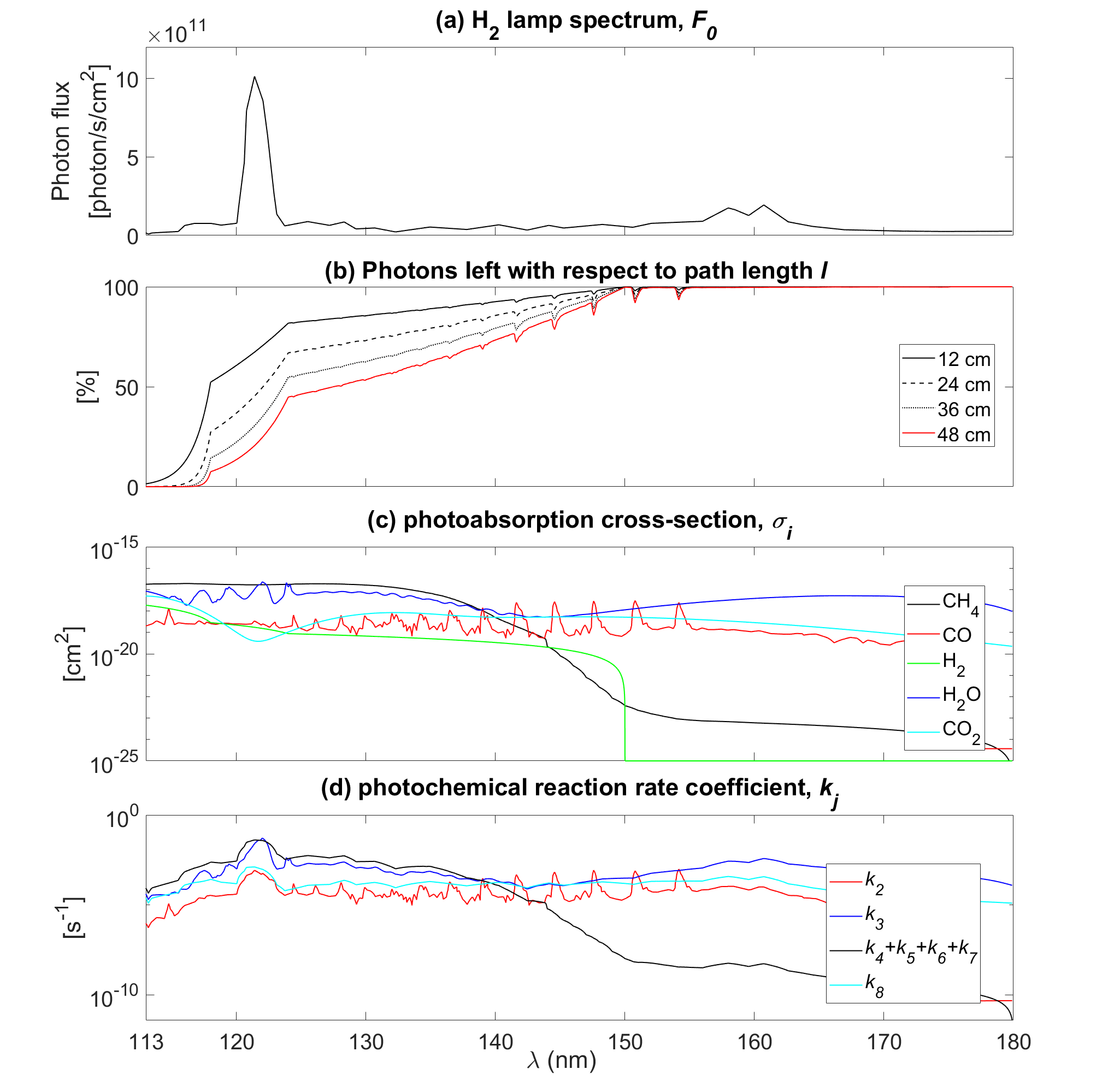}
    \caption{\footnotesize Various parameters necessary to estimate photochemical reaction rate coefficients of major species used in this study. Detail of how we have obtained these parameters in this Figure is described in the section \ref{photoexcit_rate_calc}. (a) VUV spectra in the 112--180 nm range of F-type MDHL after scaling down of the spectral profile taken from \cite{Ligterink_2015} for a \ce{H2} pressure of 1.2 mbar, flowing rate of 0.5 sccm, and an input power of 70 W; (b) the proportion of photons of corresponding $\lambda$ left at 12, 24, 36, and 48 cm of the path length calculated at 1473 K and 15 mbar of total gas pressure of 99.7 \% \ce{H2} and 0.3 \% of CO using the Equation \ref{eq:photodissrate}; (c) photoabsorption cross-sections (\textit{$\sigma$}) profiles of corresponding gas species between 110--180 nm. Each \textit{$\sigma$} of (i) \ce{CH4} was taken from \cite{Laufer_2011, Mount_1977}, (ii) \ce{CO} was taken from \cite{MyerSamson_1970, Thompson_1963}, (iii) \ce{H2} was taken from \cite{Backx_1976}, (iv) \ce{H2O} was taken from \cite{Mota_2005}, and (v) \ce{CO2} was taken from \cite{Venot_2018}; (d) wavelength-dependent rate coefficient \textit{k}$_j(\lambda)$ of corresponding photochemical reactions which are defined in Table \ref{tab:photochem_rxns}}
    \label{fig:photo_rxns}
\end{figure}

Since RMG has been developed to simulate the combustion chemistry, it doesn't include any photochemical reactions in its library. However, it is essential to include photochemical reactions in order to assess the importance of CO photochemistry in the reaction kinetic model. For this reason, we selectively calculated photochemical reaction rate coefficients (\textit{k}\textsubscript{\textit{j}}) of major species (that were observed in the previous experiments) as a function of the path length (\textit{l}) and wavelength ($\lambda$) under a given experimental total gas pressure (i.e. 15 mbar in \cite{Fleury_2019,Fleury_2020}):
\begin{equation}
    \textit{k}\textsubscript{\textit{j}}(\lambda,\textit{l})=\Phi\textsubscript{\textit{j}}(\lambda)\sigma\textsubscript{\textit{i}}(\lambda)\textit{F}\textsubscript{0}(\lambda)\textit{e}\textsuperscript{-$\sum$$\sigma$\textsubscript{\textit{i}}($\lambda$)\textit{n\textsubscript{\textit{i}}l}}
    \label{eq:photodissrate}
\end{equation}
where \textit{i} represents each gas species (i.e. \ce{CO}, \ce{H2}, \ce{H2O}, \ce{CH4}, or \ce{CO2}), $\Phi\textsubscript{\textit{j}}(\lambda)$ is the quantum yield of the photochemical reaction \textit{j}, $\sigma$\textsubscript{\textit{i}} is the photoabsorption cross-sections of the gas species \textit{i} (see Figure \ref{fig:photo_rxns}c), and \textit{F}\textsubscript{0}($\lambda$) is a photon flux at the wavelength $\lambda$ at the photochemical reaction cell window (i.e. \textit{l}=0 or zero optical depth). Unfortunately, the exact VUV photon flux, \textit{F}\textsubscript{0}($\lambda$) profile used in the experiments conducted by \cite{Fleury_2019, Fleury_2020} was not available. So instead, \textit{F}\textsubscript{0}($\lambda$) profile was taken from \cite{Ligterink_2015} which used F-type microwave discharge hydrogen-flow lamps (MDHLs) similar to the system used in the previous experiments and scaled down for input power of 70 W (simulating stellar UV photons) instead of 100 W, and \ce{H2}-pressure of 1.2 mbar instead of 0.41 mbar (refer to Fig. 4 in \cite{Ligterink_2015}). The resulting VUV spectra is shown in Figure \ref{fig:photo_rxns}a. The exponential term is to consider the optical depth caused by the gas species absorbing photons emitted from the UV light source, where \textit{n\textsubscript{\textit{i}}} is the number density of corresponding gas species \textit{i}. As shown in Figure \ref{fig:photo_rxns}b, under the experimental condition of 99.7 \% \ce{H2} and 0.3 \% of \ce{CO}, 15 mbar, and 1473 K, at least more than 40 \% of Ly-$\alpha$ photons are available even at the end of the path length (i.e. \textit{l} = 48 cm). This Equation (\ref{eq:photodissrate}) is then integrated to the photochemical cell length (i.e. \textit{l} = 0 to 48 cm) as shown in Figure \ref{fig:photo_rxns}d, followed by integration to the wavelength range (i.e. $\lambda$ = 112 to 180 nm) to calculate corresponding photochemical reaction rate coefficient (\textit{k}\textsubscript{\textit{j}}) as listed in Table \ref{tab:photochem_rxns}.

A few assumptions have been made with respect to calculating the carbon monoxide photoexcitation (i.e. X\textsuperscript{1}$\Sigma^+$ $\rightarrow$ a\textsuperscript{3}$\Pi$) rate coefficient and phosphorescence (i.e. a\textsuperscript{3}$\Pi$ $\rightarrow$ X\textsuperscript{1}$\Sigma^+$; note that there is a spin change) rate coefficient, which are: (i) every carbon monoxide molecule in the ground state (X\textsuperscript{1}$\Sigma^+$) populates to a\textsuperscript{3}$\Pi$ after UV photoexcitation into the spin and dipole allowed A\textsuperscript{1}$\Sigma$ state; (ii) phosphorescence of CO(a\textsuperscript{3}$\Pi$) down to the ground state CO(X\textsuperscript{1}$\Sigma^+$) follows an exponential decay pattern determined by the radiative lifetime of the a\textsuperscript{3}$\Pi$ state; and (iii) there are no significant changes in the radiative rate coefficients for the molecules at different temperatures. In order to implement the assumption (i), we assumed the quantum yield ($\Phi_2$) of the CO(a\textsuperscript{3}$\Pi$) after photoexcitation from X\textsuperscript{1}$\Sigma^+$ to A\textsuperscript{1}$\Sigma^+$(i.e. X\textsuperscript{1}$\Sigma^+$ $\rightarrow$ a\textsuperscript{3}$\Pi$) to be unity. With regard to the assumptions (ii) and (iii), the mean lifetime of CO(a\textsuperscript{3}$\Pi$) was taken from \cite{Lee_2020} to be 3 milliseconds.

All these calculated photochemistry-related rate coefficients were later added to the CHEMIKIN format located at the bottom of the CHEMKIN input files indicated as 'Newly Added Photochemistry' (See Supplementary Materials).

\begin{deluxetable}{ccccc}
\tabletypesize{\footnotesize}
\tablewidth{1pt} 
\tablenum{1}
\tablecaption{Rate coefficients of various photochemistry included in the reaction kinetic model (\textit{j}=1 corresponds to phosphorescence, \textit{j}=2 corresponds to photoexcitation, and \textit{j}=3--8 correspond to photodissociation) \label{tab:photochem_rxns}}
\tablehead{
\colhead{\textit{j}} & \colhead{Reactions} & \colhead{\textit{T} [K]}& \colhead{\textit{k}\textsubscript{\textit{j}} [s\textsuperscript{-1}]  \textsuperscript{\textit{a}}} & \colhead{Note}
} 
\startdata
1&{CO(a\textsuperscript{3}$\Pi$) $\rightarrow$ CO(X\textsuperscript{1}$\Sigma^+$)}&-&3.33$\times$10\textsuperscript{2}& mean lifetime of 3 ms taken from \cite{Lee_2020}\\
\hline
{}&{}&573 &4.74$\times$10\textsuperscript{-5}&\\
2&{CO(X\textsuperscript{1}$\Sigma^+$) $\rightarrow$ CO(a\textsuperscript{3}$\Pi$)}&873&5.18$\times$10\textsuperscript{-5}&calculated using the Equation (\ref{eq:photodissrate}),\\
{}&{}&1173 &5.45(5.19)$\times$10\textsuperscript{-5}& $\Phi_2$ of 1 assumed to be unity\\
{}&{}&1473 &5.64(5.41)$\times$10\textsuperscript{-5}&\\
\hline
{}&{}&573 &1.10$\times$10\textsuperscript{-3}&\\
3&{\ce{H2O} $\rightarrow$ \ce{H} + \ce{OH}}&873&1.20$\times$10\textsuperscript{-3}&calculated using the Equation (\ref{eq:photodissrate}),\\
{}&{}&1173 &1.30(1.20)$\times$10\textsuperscript{-3}&$\Phi_3$ of 1 assumed to be unity\\
{}&{}&1473 &1.40(1.30)$\times$10\textsuperscript{-3}&\\
\hline
{}&{}&573 &5.46$\times$10\textsuperscript{-4}&\\
4&{\ce{CH4} $\rightarrow$ \ce{CH3} + \ce{H}}&873&6.72$\times$10\textsuperscript{-4}&calculated using the Equation (\ref{eq:photodissrate}),\\
{}&{}&1173&7.14(6.72)$\times$10\textsuperscript{-4}&$\Phi_4$ of 0.42 taken from \cite{Gans_2011}\\
{}&{}&1473&7.98(7.14)$\times$10\textsuperscript{-4}&\\
\hline
{}&{}&573 &6.24$\times$10\textsuperscript{-4}&\\
5&{\ce{CH4 } $\rightarrow$ \ce{CH2} + \ce{H2}}&873&7.68$\times$10\textsuperscript{-4}&calculated using the Equation (\ref{eq:photodissrate}),\\
{}&{}&1173&8.16(7.68)$\times$10\textsuperscript{-4}&$\Phi_5$ of 0.48 taken from \cite{Gans_2011}\\
{}&{}&1473&9.12(8.16)$\times$10\textsuperscript{-4}&\\
\hline
{}&{}&573 &9.10$\times$10\textsuperscript{-5}&\\
6&{\ce{CH4 } $\rightarrow$ \ce{CH} + \ce{H2} + \ce{H}}&873&1.12$\times$10\textsuperscript{-4}&calculated using the Equation (\ref{eq:photodissrate}),\\
{}&{}&1173&1.19(1.12)$\times$10\textsuperscript{-4}&$\Phi_6$ of 0.07 taken from \cite{Gans_2011}\\
{}&{}&1473&1.33(1.19)$\times$10\textsuperscript{-4}&\\
\hline
{}&{}&573 &3.90$\times$10\textsuperscript{-5}&\\
7&{\ce{CH4} $\rightarrow$ \ce{CH2} + \ce{H} + \ce{H}}&873&4.80$\times$10\textsuperscript{-5}&calculated using the Equation (\ref{eq:photodissrate}),\\
{}&{}&1173&5.10(4.80)$\times$10\textsuperscript{-5}&$\Phi_7$ of 0.03 taken from \cite{Gans_2011}\\
{}&{}&1473&5.70(5.10)$\times$10\textsuperscript{-5}&\\
\hline
{}&{}&573 &1.11$\times$10\textsuperscript{-4}&\\
8&{\ce{CO2} $\rightarrow$ \ce{CO} + \ce{O}}&873&1.23$\times$10\textsuperscript{-4}&calculated using the Equation (\ref{eq:photodissrate}),\\
{}&{}&1173&1.31(1.25)$\times$10\textsuperscript{-4}&$\Phi_8$ of 1 taken from \cite{Venot_2018}\\
{}&{}&1473&1.37(1.32)$\times$10\textsuperscript{-4}&\\
\hline
\enddata
\tablecomments{\footnotesize\textsuperscript{\textit{a}} Numbers in parentheses refer to calculated photochemical reaction rate coefficients under the experimental condition of \cite{Fleury_2020} (i.e. 99.26 \% of \ce{H2}, 0.48 \% of \ce{H2O}, and 0.26 \% of CO) whose optical depth is slightly larger compare to the experimental condition of \cite{Fleury_2019} (i.e. 99.7 \% of \ce{H2} and 0.3 \% of CO).}
\end{deluxetable}

\subsection{Temperature- and pressure-dependent rate coefficients of \ce{H2} + CO(a\textsuperscript{3}$\Pi$) reactions} \label{photochem_rate_calc}
\begin{figure}
    \centering
    \includegraphics[width=1.0\textwidth]{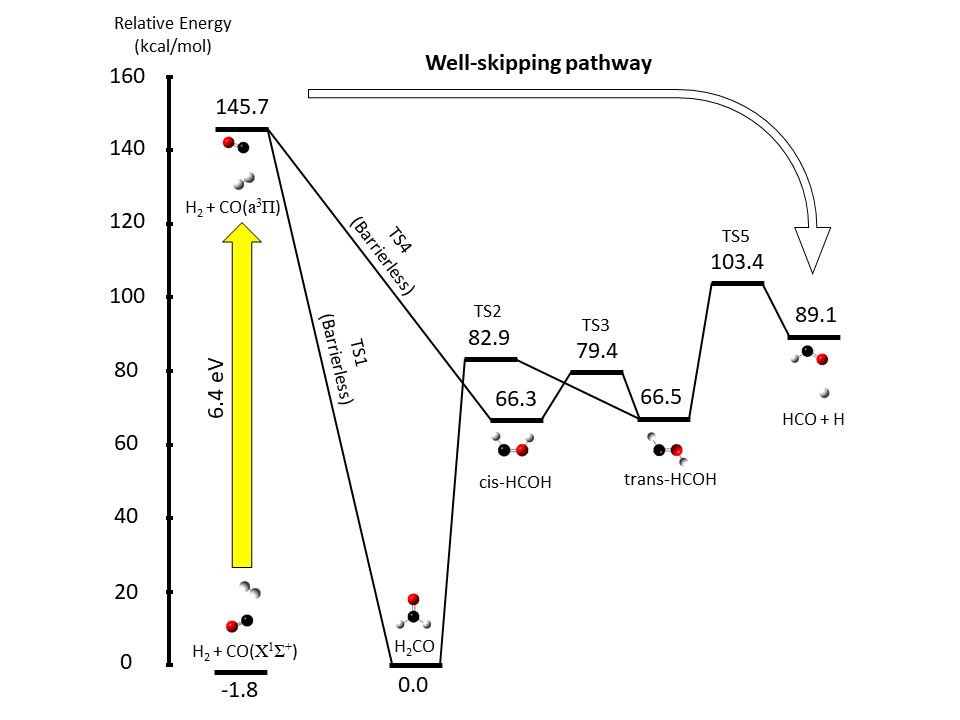}
    \caption{\footnotesize The \ce{CH2O} potential energy surface calculated at the CBS-QB3 level of theory. The calculated energy of Transition State 1 (TS1) and TS4 were originally 81.8 and 104.9 kcal/mol at the CBS-QB3 level of theory, but here assumed to be barrier-less reactions with respect to \ce{H2} + CO(a\textsuperscript{3}$\Pi$). For the definition of "Well-skipping pathway", please refer to section \ref{subsec:H2_CO_rxn}. }
    \label{fig:H2CO_PES}
\end{figure}

Since the goal of this study is to gain a better insight into simultaneous thermally and photochemically driven reaction pathways involving electronically excited CO in its metastable state (a\textsuperscript{3}$\Pi$), we need to assess to what extent these metastable and reactive carbon monoxide molecules (i.e. CO in a\textsuperscript{3}$\Pi$ state) affect the whole chemistry observed in the experiments by \cite{Fleury_2019,Fleury_2020}. In order to achieve this, it is critical to estimate and include reactions between the most dominant gas species (i.e. \ce{H2}) and the excited carbon monoxide in the reaction kinetic modeling. For this reason, we first carried out the potential energy surface (PES) calculations of \ce{H2} + \ce{CO} previously explored by \cite{Euclides_2019}. The CBS-QB3 method was performed on the stationary points and transition states reported by \cite{Euclides_2019} using Gaussian 09 (\cite{g09}) and all these molecular parameter outputs are available in the Supplementary Materials. Then as it is shown in Figure \ref{fig:H2CO_PES}, instead of connecting \ce{H2} + CO(X\textsuperscript{1}$\Sigma^+$) to the entire PES, we connected \ce{H2} + CO(a\textsuperscript{3}$\Pi$) to the entire PES via Transition State 1 (TS1) and TS4 with assumptions that the entrance barriers of the TS1 and TS4 are zero (i.e. barrier-less reactions).

As for next step after finishing the PES calculation, Arkane (\cite{Allen_2012}) was used to calculate temperature- and pressure-dependent rate coefficients \emph{k}(\emph{T,P}) of \ce{H2} + CO(a\textsuperscript{3}$\Pi$) channels based on the \ce{CH2O} PES mentioned above. Briefly describing, Arkane is a tool that can calculate pressure-dependent phenomenological rate coefficients \emph{k}(\emph{T,P}) for unimolecular reaction networks based on the results of quantum chemistry calculations (i.e. PES) and the Rice–Ramsperger–Kassel–Marcus (RRKM) theory (\cite{Marcus_1952}). Arkane first generates a detailed model of the reaction network using the one-dimensional master equation methods (\cite{Miller_2006}) and then applies one of several available model reduction methods (e.g. the modified strong collision approximation) to simplify the detailed model into a set of phenomenological rate coefficients \emph{k}(\emph{T,P}) that are suitable for use in chemical reaction mechanisms. 

The reservoir state method was used in calculating \emph{k(T,P)} from the \ce{CH2O} pressure-dependent networks. Lennard-Jones parameters of \ce{H2} and \ce{CH2O} isomers (approximated to be same as those for \ce{C2H6}) in helium bath gas were taken from \cite{Jasper_2015}. The collisional energy-transfer parameters, 
$\langle\Delta\emph{E}\textsubscript{down}\rangle=\alpha\textsubscript{300}\left(\frac{T}{T_{0}}\right)\textsuperscript{n}$ cm\textsuperscript{-1},  were also taken from \cite{Jasper_2015}. There are four reaction channels from \ce{H2} + CO(a\textsuperscript{3}$\Pi$) and their rate coefficients under the experimental conditions of \cite{Fleury_2019, Fleury_2020} are listed in Table \ref{tab:additional_rxns} as well as the rate coefficients of the reaction of \ce{H2} + CO(X\textsuperscript{1}$\Sigma^+$) for a comparison. These reaction coefficients also can be found in CHEMKIN format in the Supplementary Materials. Since RMG-generated thermochemical reaction model didn't distinguish HCOH isomers (\textit{cis-} and \textit{trans-}HCOH), the rate coefficients of \ce{H2} + CO(a\textsuperscript{3}$\Pi$) $\rightarrow$ \textit{cis}-HCOH and \ce{H2} + CO(a\textsuperscript{3}$\Pi$) $\rightarrow$ \textit{trans}-HCOH were summed up together and appended as the rate coefficient of \ce{H2} + CO(a\textsuperscript{3}$\Pi$) $\rightarrow$ HCOH into the CHEMKIN format file (see the Supplementary Materials).

\begin{figure}
    \centering
    \includegraphics[width=1\textwidth]{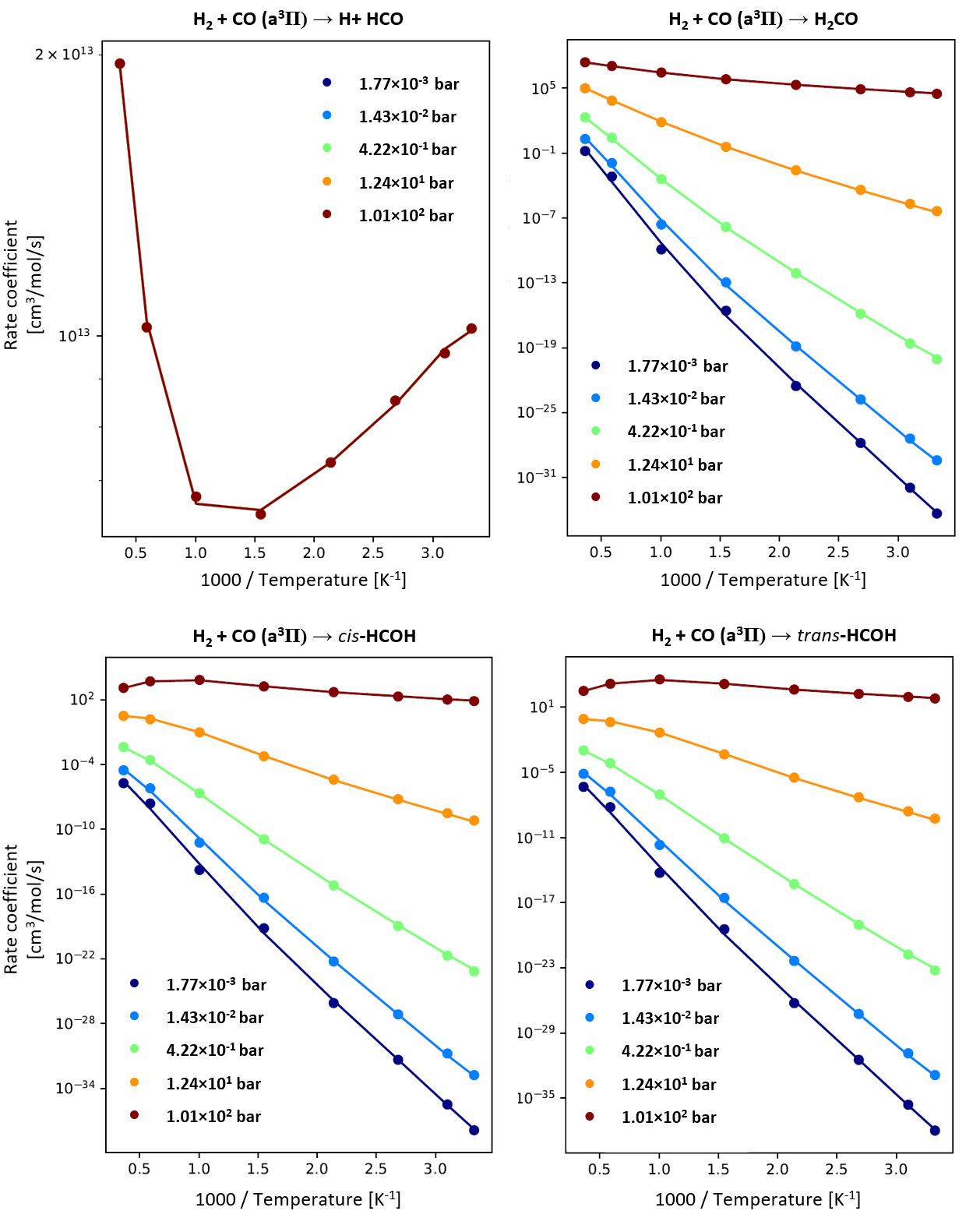}
    \caption{\footnotesize Calculated temperature- and pressure-dependent rate coefficients \emph{k}(\emph{T,P}) of \ce{H2} + CO(a\textsuperscript{3}$\Pi$) channels based on the \ce{CH2O} PES in Figure \ref{fig:H2CO_PES}. Note that \ce{H2} + CO(a\textsuperscript{3}$\Pi$) $\rightarrow$ H + HCO pressure-dependent rate coefficients overlap with each other, thus showing only one plot. Individual behavior of each reaction shown in these plots is described in detail in the section \ref{subsec:H2_CO_rxn}.}
    \label{fig:rate_coefficients_list}
\end{figure}

\begin{deluxetable}{ccccc}
\tablenum{2}
\tablecaption{Reaction rate coefficients of \ce{H2} + CO(a\textsuperscript{3}$\Pi$) and CO(X\textsuperscript{1}$\Sigma^+$) + CO(a\textsuperscript{3}$\Pi$) at the different temperature conditions of \cite{Fleury_2019,Fleury_2020}\textsuperscript{\textit{a,b,c}}. The method to estimate the rate-coefficients of \ce{H2} + CO(a\textsuperscript{3}$\Pi$) is described in detail in the section \ref{photochem_rate_calc}, while individual behavior of each reaction is described in detail in the section \ref{subsec:H2_CO_rxn}. The rate-coefficient of the  CO(X\textsuperscript{1}$\Sigma^+$) + CO(a\textsuperscript{3}$\Pi$) reaction was estimated following the method described in the section \ref{CO_CO_rate_calc} and its behavior is described in detail in the section \ref{subsec:COt_CO_rxn}. \label{tab:additional_rxns}}
\tablehead{
\colhead{Reactions} & \colhead{573 K}& \colhead{873 K} & \colhead{1173 K} & \colhead{1473 K}
} 
\startdata
\ce{H2} + CO(a\textsuperscript{3}$\Pi$) $\rightarrow$ H + HCO&6.83$\times$10\textsuperscript{12}(1.00$\times$10\textsuperscript{-26})&6.44$\times$10\textsuperscript{12}(4.47$\times$10\textsuperscript{-13})&7.32$\times$10\textsuperscript{12}(2.43$\times$10\textsuperscript{-6})&8.94$\times$10\textsuperscript{12}(2.55$\times$10\textsuperscript{-2})\\
\ce{H2} + CO(a\textsuperscript{3}$\Pi$) $\rightarrow$ \ce{H2CO}&1.82$\times$10\textsuperscript{-15}(1.14$\times$10\textsuperscript{-19}) &2.77$\times$10\textsuperscript{-9}(1.25$\times$10\textsuperscript{-9})&6.94$\times$10\textsuperscript{-6}(7.09$\times$10\textsuperscript{-5})&8.21$\times$10\textsuperscript{-4}($3.60\times$10\textsuperscript{-2})\\
\ce{H2} + CO(a\textsuperscript{3}$\Pi$) $\rightarrow$ \textit{cis}-HCOH&8.42$\times$10\textsuperscript{-19}(4.52$\times$10\textsuperscript{-23}) &8.78$\times$10\textsuperscript{-13}(3.50$\times$10\textsuperscript{-13})&1.15$\times$10\textsuperscript{-9}(1.13
$\times$10\textsuperscript{-8})&7.26$\times$10\textsuperscript{-8}(3.31$\times$10\textsuperscript{-6})\\
\ce{H2} + CO(a\textsuperscript{3}$\Pi$) $\rightarrow$ \textit{trans}-HCOH&5.65$\times$10\textsuperscript{-19}(3.09$\times$10\textsuperscript{-23}) &3.65$\times$10\textsuperscript{-13}(1.49$\times$10\textsuperscript{-13})&3.53$\times$10\textsuperscript{-10}(3.58$\times$10\textsuperscript{-9})&1.93$\times$10\textsuperscript{-8}(9.03$\times$10\textsuperscript{-7})\\
CO(X\textsuperscript{1}$\Sigma^+$) + CO(a\textsuperscript{3}$\Pi$) &2.07$\times$10\textsuperscript{6}(3.44$\times$10\textsuperscript{-50}) &2.60$\times$10\textsuperscript{8}(9.19$\times$10\textsuperscript{-29})&3.42$\times$10\textsuperscript{9}(3.34$\times$10\textsuperscript{-18})&1.79$\times$10\textsuperscript{10}(6.88$\times$10\textsuperscript{-12})\\
 $\rightarrow$ C(\textsuperscript{3}P) + \ce{CO2}&&&&\\
\hline
\enddata
\tablecomments{\footnotesize\textsuperscript{\textit{a}}Pressure condition is 15 mbar; \textsuperscript{\textit{b}}Unit is [cm\textsuperscript{3}/mol/s]; \textsuperscript{\textit{c}}Numbers in parentheses refer to calculated rate coefficients of corresponding reactions when CO is in the ground state (X\textsuperscript{1}$\Sigma^+$) }
\end{deluxetable}

\subsection{Temperature-dependent rate coefficients of CO(X\textsuperscript{1}$\Sigma^+$) + CO(a\textsuperscript{3}$\Pi$) $\rightarrow$ C(\textsuperscript{3}P) + \ce{CO2} reaction} \label{CO_CO_rate_calc}
\begin{figure}
    \centering
    \includegraphics[width=0.43\textwidth]{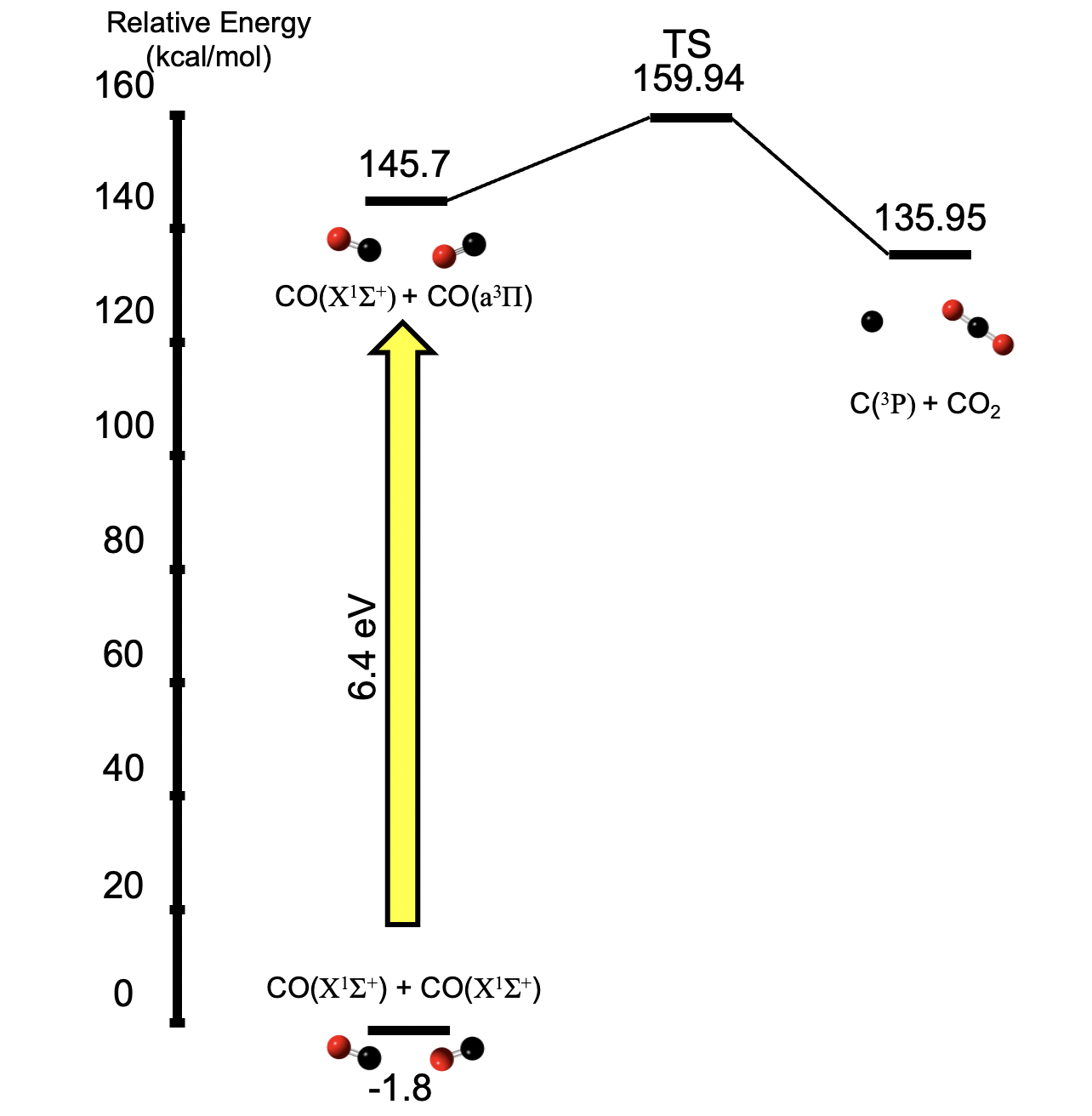}
    \includegraphics[width=0.56\textwidth]{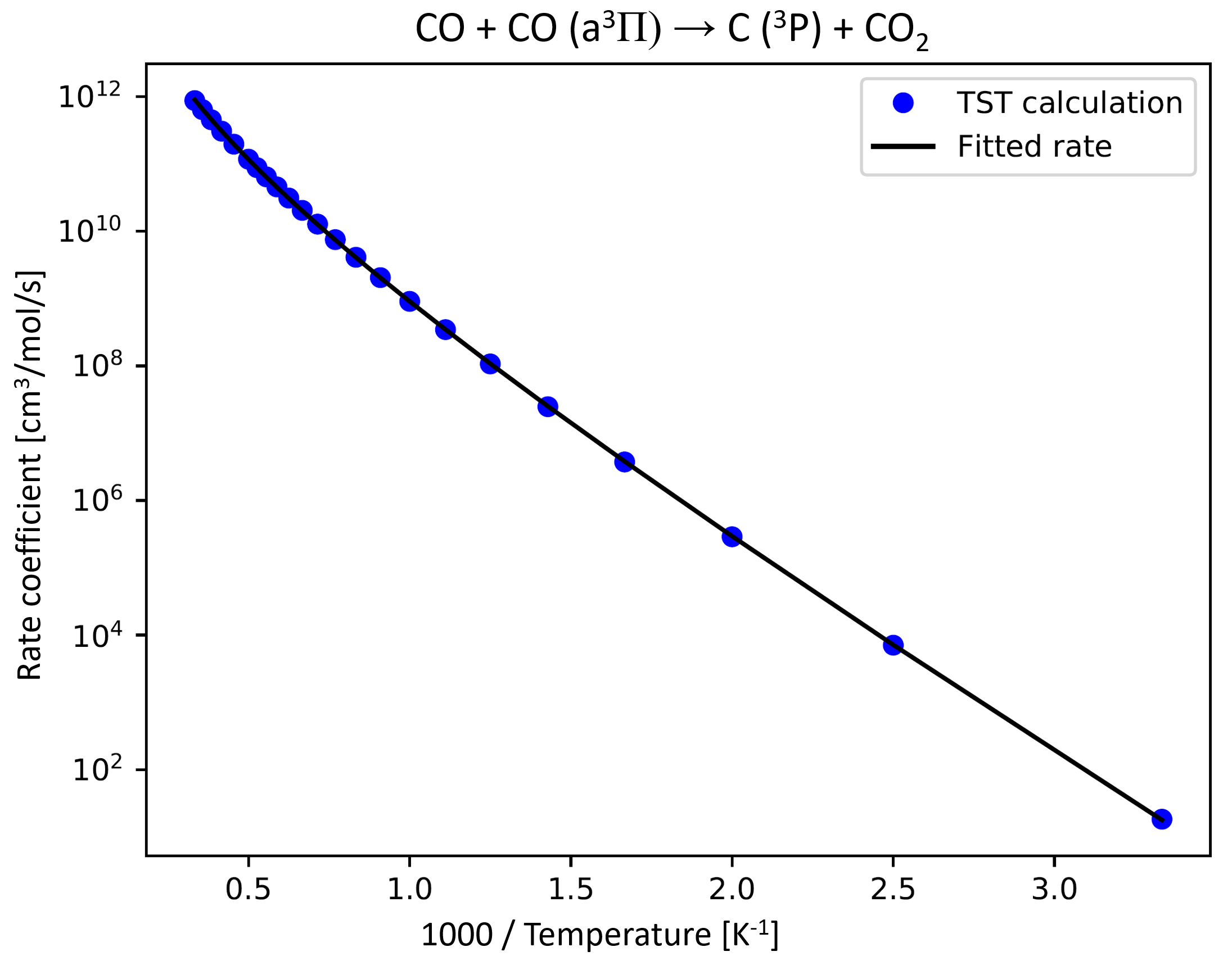}
    \caption{\footnotesize (Left) The triplet \ce{C2O2} potential energy surface calculated at the CCSD(T)/def2-qZVP level of theory with $\omega$B97M-V/6-311+G* zero-point correction by \cite{DeVine-2022}.  (Right) Temperature-dependent rate coefficients of the CO(X\textsuperscript{1}$\Sigma^+$) + CO(a\textsuperscript{3}$\Pi$) $\rightarrow$ C(\textsuperscript{3}P) + \ce{CO2} reaction calculated by Arkane (\cite{Allen_2012}) using conventional transition state theory}
    \label{fig:C2O2_PES}
\end{figure}

In the previous study by \cite{Fleury_2019}, the reaction between the excited CO and the ground-state CO was suggested as the important reaction that might rationalize the formation of \ce{CO2} (along with C(\textsuperscript{3}P)) observed in the experiments. Since carbon monoxide was the second most abundant species in the system, it is important to estimate and include this reaction into the kinetic modeling. Since the potential energy surface of this system was already explored at the CCSD(T)/def2-qZVP level of theory (which is an even higher level of theory compared to CBS-QB3) with $\omega$B97M-V/6-311+G* zero-point correction by \cite{DeVine-2022} (shown in Figure \ref{fig:C2O2_PES}) with all the required parameters (e.g. potential energy differences, rotational constants, vibrational frequencies, symmetry number, etc.) available, all these parameters were manually provided as an Arkane input rather than carrying out ab-initio calculation at the CBS-QB3 level of theory from the beginning.
As long as each potential energy is calculated on the same level of theory, the uncertainty of the estimated reaction rate-coefficients would be good enough for the reaction kinetic modeling purpose. Arkane then generated temperature-dependent rate coefficients using conventional transition state theory \cite{Allen_2012} (Arkane input and output files are available in the Supplementary Materials). Temperature-dependent rate coefficients of the CO(X\textsuperscript{1}$\Sigma^+$) + CO(a\textsuperscript{3}$\Pi$) $\rightarrow$ C(\textsuperscript{3}P) + \ce{CO2} reaction at various temperatures are available in Table \ref{tab:additional_rxns}.

\subsection{Model simulation and analysis}\label{model_simulation}
With the reaction mechanism built following the procedure mentioned above, simulations (solving differential equations) were performed for the reaction time of 18 hours (i.e. 64800 seconds) on eight experimental conditions of \cite{Fleury_2019} (i.e. With UV and without UV at each T = 573, 873, 1173, and 1473 K with the initial composition of 99.7 \% of \ce{H2} and 0.3\% of CO) and four experimental conditions of \cite{Fleury_2020} (i.e. With UV and without UV at each T= 1173 and 1473 K with the initial composition of 99.26 \% of \ce{H2}, 0.26\% of CO, and 0.48 \% of \ce{H2O}) using reaction mechanism simulator (RMS, \cite{rms}), a package included in the RMG (\cite{Gao_2016}) suite package. 
The model output is a set of temperature-dependent molecular mixing ratio profiles of each of the species, summarized in Figures \ref{fig:Fleury_2019_wUV_woUV}, \ref{fig:C2H2_pathway}a, and \ref{fig:Fleury_2020_wUV_woUV}, which add up to 1 (e.g. a molecular mixing ratio of 10\textsuperscript{-6} in Figure \ref{fig:Fleury_2019_wUV_woUV} refers to 1 ppm).

Additional model simulations were performed on the same experimental conditions of \cite{Fleury_2019} except for higher temperatures (i.e. each T = 2000 and 2500 K). The model simulation for longer reaction time (206 hours), higher pressure (81 mbar), and the temperature of 1473 K was also performed using RMS. The reactor was assumed as isobaric, isothermal, and homogeneous. Rate of production analysis (ROP) was done using the RMS and the ROP analysis of the kinetic model describing \cite{Fleury_2019} is available in the Appendix \ref{sec:ROP}. Sensitivity analysis was done using the RMG suite package (\cite{Gao_2016}).

\begin{figure}
    \centering
    \includegraphics[width=1.0\textwidth]{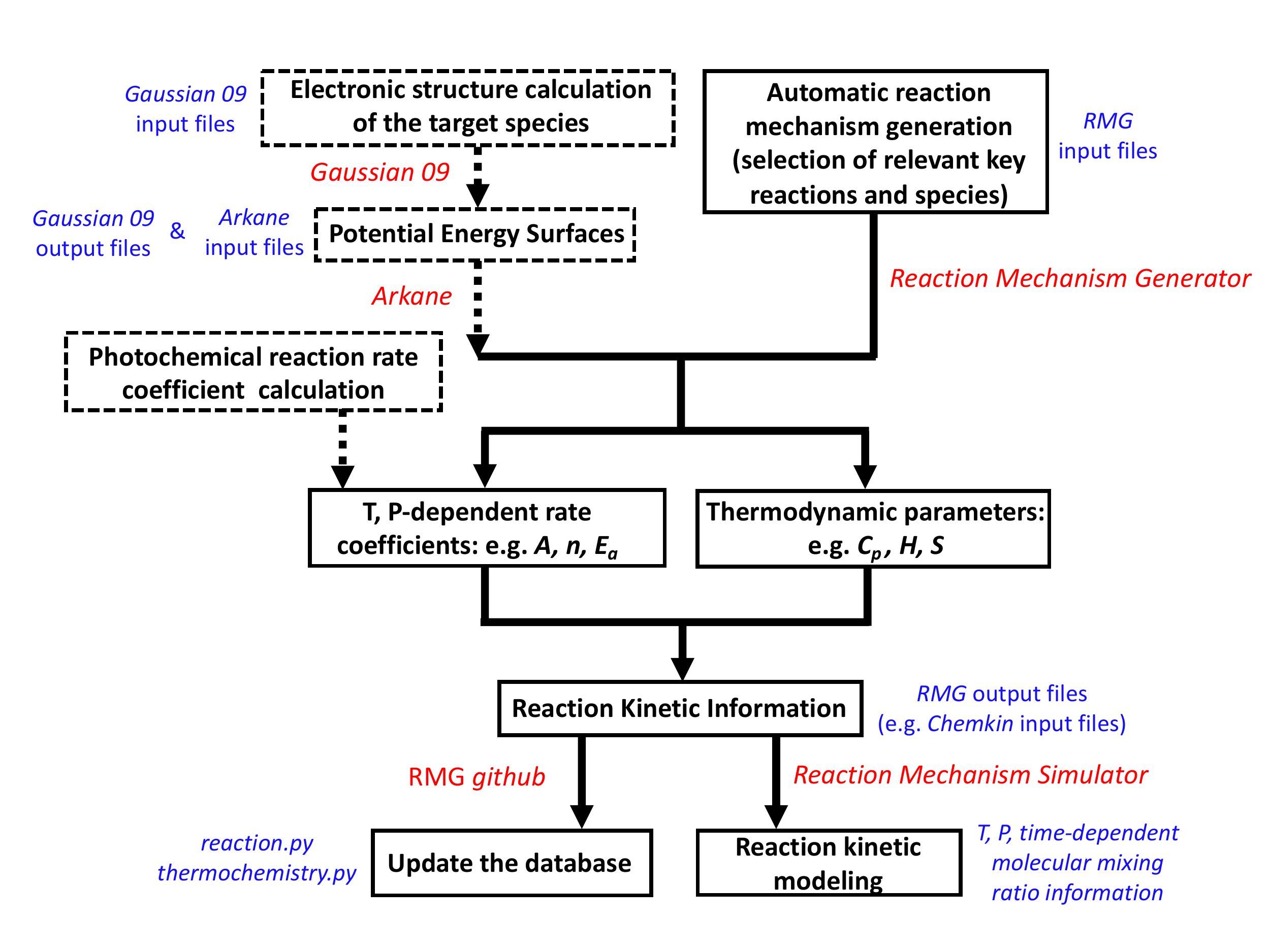}
    \caption{\footnotesize A schematic diagram of the flow of the reaction kinetic modeling carried out in this study. The red-colored texts refer to the software tools used in this study. The blue-colored text refers to the inputs and outputs necessary in this study (some of which are mentioned in the main text). The black-colored text in the boxes refers to the major steps described in the main text. For example, 'Automatic reaction mechanism generation' refers to the section \ref{rmg_library} and \ref{rmg_generation}, while the 'Photochemical reaction rate coefficient calculation' shown in this figure refers to the section \ref{photoexcit_rate_calc}, and the 'Potential energy surfaces' refers to the section \ref{photochem_rate_calc}. The solid line refers to the procedure for generating the thermochemical reaction model. The dashed line refers to the procedure for estimating the photochemical reaction rate-coefficients or any rate-coefficients that need to be estimated from the first-principle methods (i.e. ab-initio calculations) due to various reasons (e.g. There is no any available reaction rate-coefficient in a previous study).}
    \label{fig:modeling_scheme}
\end{figure}

\section{Results and Discussions} \label{sec:resultsanddiscussions}
\subsection{\ce{H2} + CO(a\textsuperscript{3}$\Pi$) reaction rate coefficients}\label{subsec:H2_CO_rxn}
As it is shown in Table \ref{tab:additional_rxns} and Figure \ref{fig:rate_coefficients_list}, the rate-coefficient of the \ce{H2} + CO(a\textsuperscript{3}$\Pi$) $\rightarrow$ H + HCO reaction is at least 10 orders of magnitude faster than the rate coefficients of any other channels, which indicates that the reaction between UV-excited CO and \ce{H2} will dominantly proceed into H and HCO molecules at any temperature and pressure conditions. The \ce{H2} + CO(a\textsuperscript{3}$\Pi$) $\rightarrow$ H + HCO reaction is also called as a "well-skipping reaction" which refers to an elementary reaction that traverses more than one transition-state dividing surface in a single elementary step (see Figure \ref{fig:H2CO_PES}). This well-skipping behavior is mainly attributed to UV-excited CO's excessively high potential energy compared to the potential energy of all the other species and transition states. As shown in Figure \ref{fig:H2CO_PES}, the potential energy of (photo)chemically activated reactants (i.e. \ce{H2} + CO(a\textsuperscript{3}$\Pi$)) is already $\sim$40 kcal/mol above the second highest potential energy (i.e. the potential energy of TS4, 104.9 kcal/mol). For this reason, even at higher pressure condition of 100 bar, increased collisional stabilization into other stable molecules (i.e. HCOH isomers and \ce{H2CO}) is still not efficient enough to beat this well-skipping reaction to form HCO and H. For a similar reason (i.e. excessive energy from UV-photons), the UV-excited well-skipping channel is less sensitive to temperature changes compared to other reaction channels. However, this is not the case when CO is in the ground state. As can be seen from Table \ref{tab:additional_rxns}, in the case of the reaction between \ce{H2} and CO in the ground state (i.e. X\textsuperscript{1}$\Sigma^+$), the reaction channel of \ce{H2} + CO(X\textsuperscript{1}$\Sigma^+$) $\rightarrow$ \ce{H2CO} stays as a main channel up until 1173 K, while the well-skipping reaction (i.e. \ce{H2} + CO(X\textsuperscript{1}$\Sigma^+$) $\rightarrow$ H + HCO) takes over the main channel from the \ce{H2} + CO(X\textsuperscript{1}$\Sigma^+$) $\rightarrow$ \ce{H2CO} channel at 1473 K (see numbers in parentheses in Table \ref{tab:additional_rxns}). This is because the only energy source for the reactants (\ce{H2} + CO(X\textsuperscript{1}$\Sigma^+$) to overcome the reaction barrier of 103.4 kcal/mol (i.e. TS5 in Figure \ref{fig:H2CO_PES}) to form HCO and H is from thermal energy. Up until 873 K, thermal energy is not enough so the intermediates that overcome the reaction barrier of 81.8 kcal/mol (i.e. TS1) will stabilize into \ce{H2CO} rather than skipping all the other wells to form HCO and H. With increasing temperatures above 1173 K, the reactants are more thermally energized so that they are more likely to proceed to the well-skipping channel (i.e. \ce{H2} + CO(X\textsuperscript{1}$\Sigma^+$) $\rightarrow$ H + HCO) and competing with the stabilization channel down to \ce{H2CO}. So the major difference between the \ce{H2} + CO chemistry with and without UV photons would be whether this well-skipping channel (i.e. \ce{H2} + CO $\rightarrow$ H + HCO) is dominant (i.e. with UV) or not (i.e. thermal-only). These results demonstrate the importance of rigorous reaction kinetics modeling, including photochemical reaction pathways. In the previous study by \cite{Fleury_2019}, the CO(X\textsuperscript{1}$\Sigma^+$) + CO(a\textsuperscript{3}$\Pi$) $\rightarrow$ C(\textsuperscript{3}P) + \ce{CO2} reaction was considered as the only important photochemistry regardless of dominant \ce{H2} in the system. It was not immediately evident that the reaction \ce{H2} + CO(a\textsuperscript{3}$\Pi$) $\rightarrow$ H + HCO reaction is the most predominant channel and has several orders of magnitude higher rate coefficients than other reaction channels (e.g. \ce{H2} + CO(a\textsuperscript{3}$\Pi$) $\rightarrow$ \ce{H2CO}). This has the important consequence that the H atoms generated are now very reactive and start radical chain reactions. 

\subsection{CO(X\textsuperscript{1}$\Sigma^+$) + CO(a\textsuperscript{3}$\Pi$) $\rightarrow$ C(\textsuperscript{3}P) + \ce{CO2} reaction reaction rate coefficients}\label{subsec:COt_CO_rxn}
As it is shown in Figure \ref{fig:C2O2_PES}, the CO(X\textsuperscript{1}$\Sigma^+$) + CO(a\textsuperscript{3}$\Pi$) reaction has a non-zero reaction barrier (i.e. 14.24 kcal/mol) even after the photoexcitation of the ground state CO, different from \ce{H2} + CO(a\textsuperscript{3}$\Pi$) reactions (i.e. barrier-less reactions). For this reason, the calculated reaction rate coefficient shows significant positive temperature dependency as shown in Figure \ref{fig:C2O2_PES}. The last row of the Table \ref{tab:additional_rxns} shows $\sim$4 orders of magnitude difference between the rate coefficient at the lowest temperature (i.e. 573 K) and the rate coefficient at the highest temperature (i.e. 1473 K). So we can easily predict that this reaction will play a more significant role at elevated temperatures. It is also clear from the Table \ref{tab:additional_rxns} that this reaction would be insignificant without UV photons (even at 1473 K, the rate-coefficient is still smaller than 10\textsuperscript{-13} cm\textsuperscript{3}/mol/s), which makes a major difference between the chemistry with and without UV photons (i.e. thermal-only).

\subsection{Modeling of the \ce{H2}/CO exoplanet atmosphere analogue of \cite{Fleury_2019}}
\subsubsection{Reaction kinetics of thermal-only chemistry}
\begin{figure}
    \centering
    \includegraphics[width=1\textwidth]{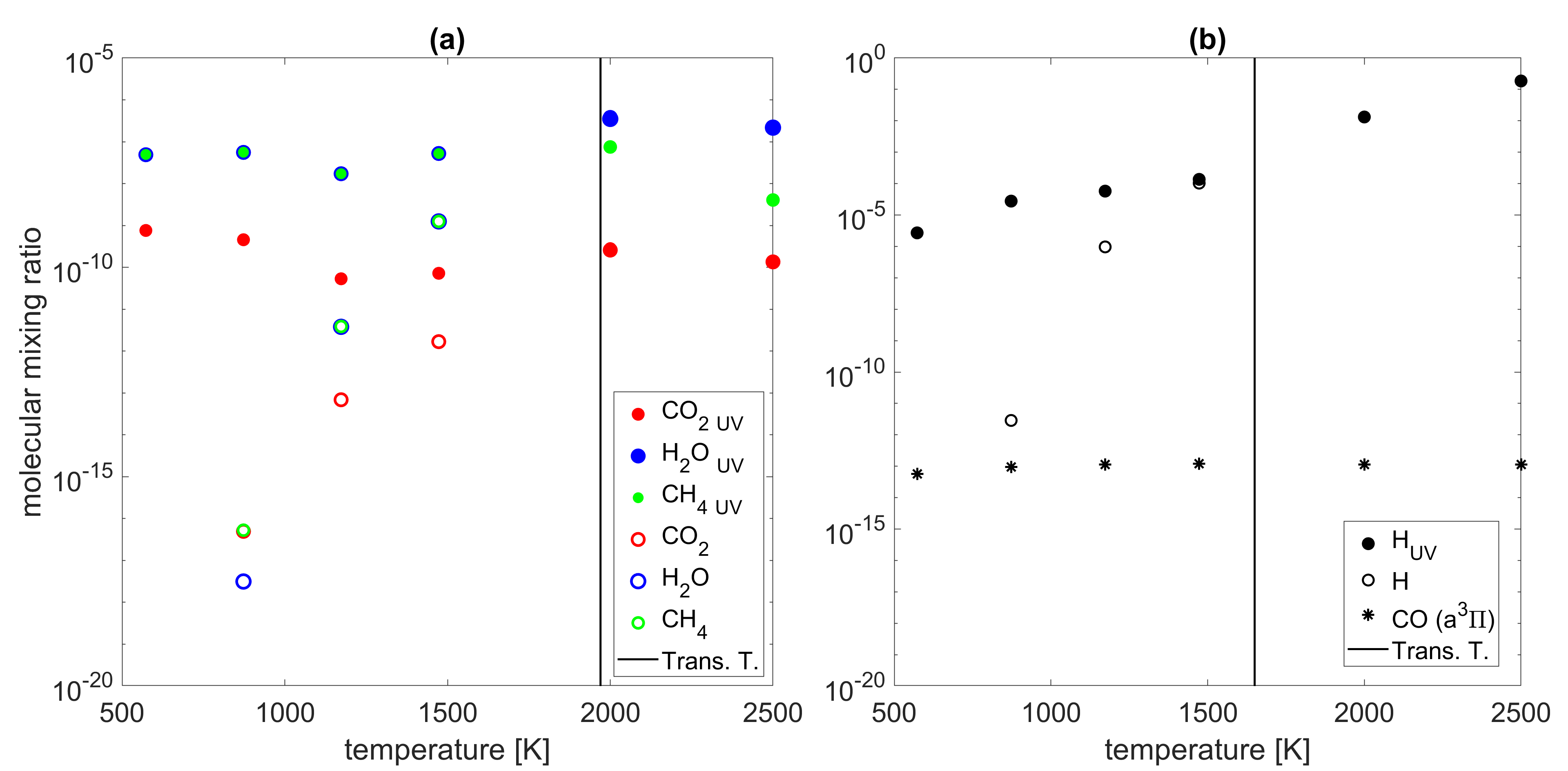}
    \caption{\footnotesize Temperature-dependent molecular mixing ratio profiles predicted by the reaction kinetic modeling for the experimental conditions described in \cite{Fleury_2019} and extra simulation up to 2500 K. Solid symbols indicate predicted mixing ratios of molecular species with both thermal energy and UV photons, while open symbols indicate predicted mixing ratios of molecular species at the thermal-only condition (i.e. without UV photons). Please note that several open symbols (i.e. thermal-only chemistry) overlap with solid symbols. The solid lines refer to the transition temperature at which the molecular mixing ratios of each species formed in different systems (i.e. thermal + UV vs thermal-only) only differ by less than 10 \%): (a) Red circles indicate \ce{CO2}; Blue circles indicate \ce{H2O}; Green circles indicate \ce{CH4}; the transition temperature with regard to \ce{CH4}, \ce{H2O}, and \ce{CO2} is calculated to be 1970 K; \ce{CO2}, \ce{CH4}, and \ce{H2O} mixing ratios with the thermal-only condition at 573 K are predicted to be lower than 10\textsuperscript{-20} (i.e. the absolute tolerance of the differential equation solver) and not shown in the Figure \ref{fig:Fleury_2019_wUV_woUV}a (b) Black circles indicate H-atom (radical species); Asterisks indicate UV-excited CO (i.e. carbon monoxide in the a\textsuperscript{3}$\Pi$ state); the transition temperature with regard to \ce{H2} is calculated to be 1650 K}
    \label{fig:Fleury_2019_wUV_woUV}
\end{figure}

As shown in Figure \ref{fig:Fleury_2019_wUV_woUV}, all major species (i.e. \ce{CH4}, \ce{H2O}, \ce{CO2}, and H radical) formations under thermal-only conditions are predicted to be extremely temperature dependent. This is mainly attributed to elevated reaction rate-coefficients with increasing temperature, leading to the formation of major species. This predicted behavior qualitatively matches well with the previous experimental results of \cite{Fleury_2019} using the quartz cell. For example, the absorption infrared (IR) spectrum of \ce{CO2} and \ce{H2O} show up at 573 K and increase with temperatures. The model predicts that the molecular mixing ratio of these molecules might be too low ($\sim$ 10\textsuperscript{-28} or lower) to be observed experimentally, but it has to be noted that the surface chemistry happening on the wall of the quartz cell might have affected the reaction chemistry. Compared to \ce{H2O} and \ce{CO2}, \ce{CH4} has much lower absorption cross-sections of IR, thus the IR peaks of \ce{CH4} start showing up at 1173 K in the quartz cell used in \cite{Fleury_2019}. The predicted molecular mixing ratio of \ce{CH4} and \ce{H2O} after the reaction time of 18 hours are almost same. This predicted behavior is due to their formation pathways shown in Figure \ref{fig:reactionpathwayswithandwithoutUV}a. According to the ROP analysis (see Appendix \ref{sec:ROP}), the predicted major precursors of \ce{CH4} and \ce{H2O} all over the reaction time (i.e. 18 hours) are \ce{CH3} and \ce{OH}, respectively. These radical species (i.e. \ce{CH3} and \ce{OH}) are simultaneously formed through the reaction HCOH + \ce{H2} $\rightarrow$ \ce{CH3} + \ce{OH} at all temperature conditions (additionally formed through the reaction \ce{CH2OH} + H $\rightarrow$ \ce{CH3} + \ce{OH} at the temperatures above 1173 K), and respectively react with \ce{H2} to form their corresponding stable species along with the H atom. Up until 1173 K, HCOH is directly formed through the reaction \ce{H2} + CO $\rightarrow$ HCOH followed by HCOH + \ce{H2} $\rightarrow$ \ce{CH3} + \ce{OH} or the unimolecular reaction (i.e. isomerization) into \ce{CH2O}. However, at temperatures above 1173 K, \ce{CH2O} starts to isomerize back to HCOH, which eventually leads to increased formation of \ce{CH4} and \ce{H2O}. \ce{CH2O} can also be formed through either \ce{H2} + CO $\rightarrow$ \ce{CH2O} at all temperature conditions or \ce{HCO} + \ce{H2} $\rightarrow$ \ce{CH2O} + H (only at T $\geq$ 873 K). HCO forms through \ce{H2} + CO $\rightarrow$ H + HCO, but readily dissociates back to CO + H. However, as mentioned previously, at T $\geq$ 873, HCO can proceed to \ce{CH2O} by reacting with \ce{H2} and at T $\geq$ 1473 K, HCO can be produced through the additional reaction H + CO $\rightarrow$ HCO.

H radicals are formed in a significant amount at T $\geq$ 873 K through thermal dissociation into H radicals (i.e. \ce{H2} $\rightarrow$ 2H), while the recombination back to \ce{H2} (i.e. H + H $\rightarrow$ \ce{H2}) becomes significant at T $\geq$ 1473 K after H radicals being formed enough in the system.
\ce{CO2} is formed through CO + OH $\rightarrow$ H + \ce{CO2} at T $\leq$ 1173 K. But at T $\geq$ 1473 K, \ce{CO2} is formed through \ce{CH2O}
 + CO $\rightarrow$ \ce{CH2} + \ce{CO2}, while \ce{CO2} reacts with H radical and forms CO and OH (not indicated in Figure \ref{fig:reactionpathwayswithandwithoutUV}a). This reaction (i.e. \ce{CO2} + H $\rightarrow$ CO + OH) is attributed to an increased amount of H radicals due to elevated thermal dissociation of \ce{H2} at elevated temperatures. Overall, as can be seen from Figure \ref{fig:reactionpathwayswithandwithoutUV}a, with increasing temperatures, additional reaction pathways are being added to the system and push the chemistry to the right side of the figure toward the formation of \ce{CH4}, \ce{H2O} and \ce{CO2}. 
 
 It is also interesting to observe that the model predictions of molecular mixing ratios of species at thermal-only conditions becomes similar to those at the condition with UV photons at around 2000 K. At T = 1970 K, the amount of \ce{CH4}, \ce{H2O}, and \ce{CO2} formed in the system with UV photons was only different by less than 10 \% compared to that in the thermal-only system. In this paper, we will call this temperature a "transition temperature" with regard to these species (i.e. \ce{CH4}, \ce{H2O}, and \ce{CO2}). This indicates that at temperatures higher than 2000 K, thermally driven reactions are fast enough so that the system becomes less sensitive to UV photons compared to the system at lower temperatures. This feature has already been suggested from previous theoretical studies (\cite{Moses-2011, Venot_2012, Moses-2014}). However, it has to be noted that there are multiple factors (e.g. other photochemical reactions not considered in this study, temperature-dependent UV-photoabsorption cross-sections of carbon monoxide, etc.) that can change the transition temperature and this will be discussed later. Also, it is predicted from the model that the molecular mixing ratios of all major species (i.e. \ce{CH4}, \ce{H2O}, and \ce{CO2}) decrease at the temperature above this transition temperature, which means that all these major species have reached quasi-equilibrium and are not favored at this temperature condition anymore. In case of \ce{CH4}, the decrease in molecular mixing ratio at temperatures higher than 2000 K is more significant compared to molecular mixing ratios of \ce{H2O} and \ce{CO2}. This behavior is due to the thermal conversion of \ce{CH4} into \ce{C2} species (mainly \ce{C2H2}). This will be discussed in detail in a later section.

\begin{figure*}
 \centering(\textbf{a})
\gridline{\includegraphics[width=1\textwidth]{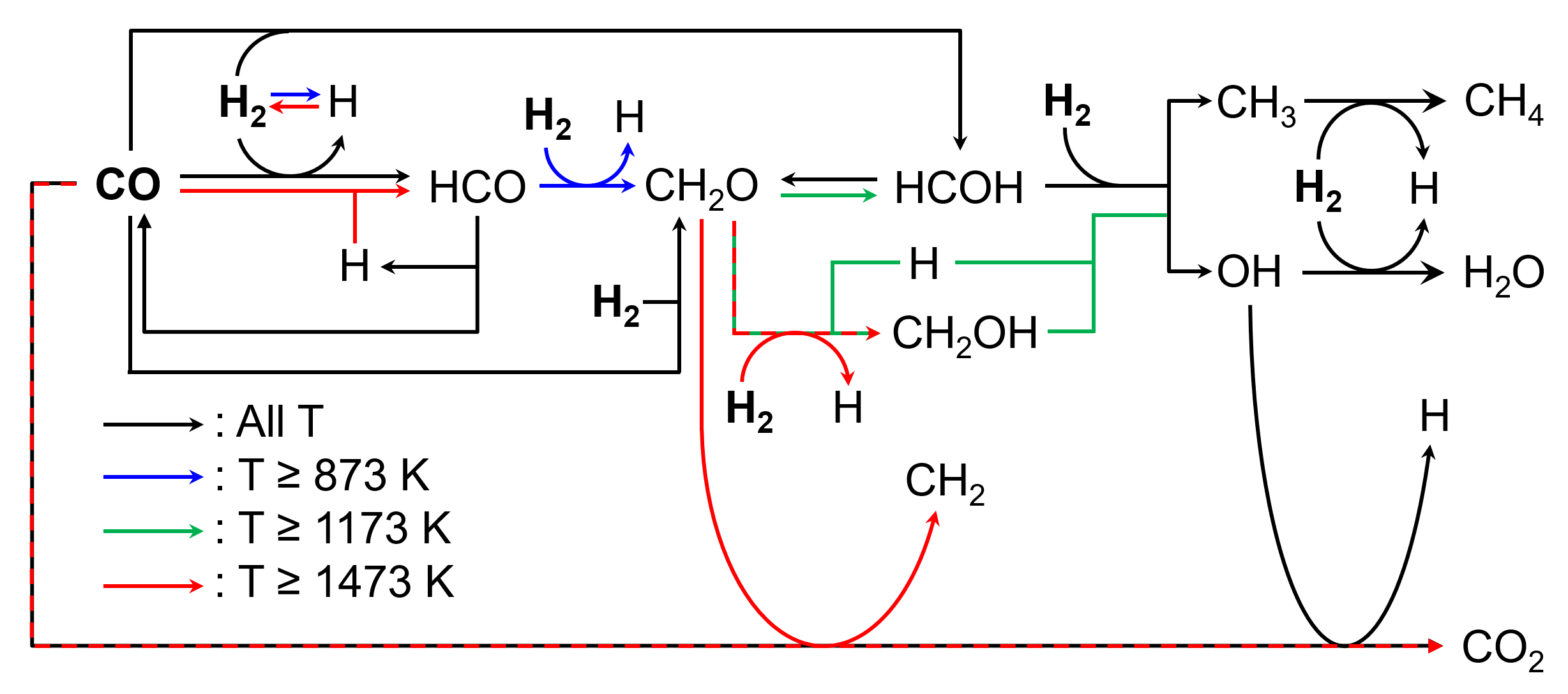}
          }
\centering\textbf{(b)}
\gridline{\includegraphics[width=1\textwidth]{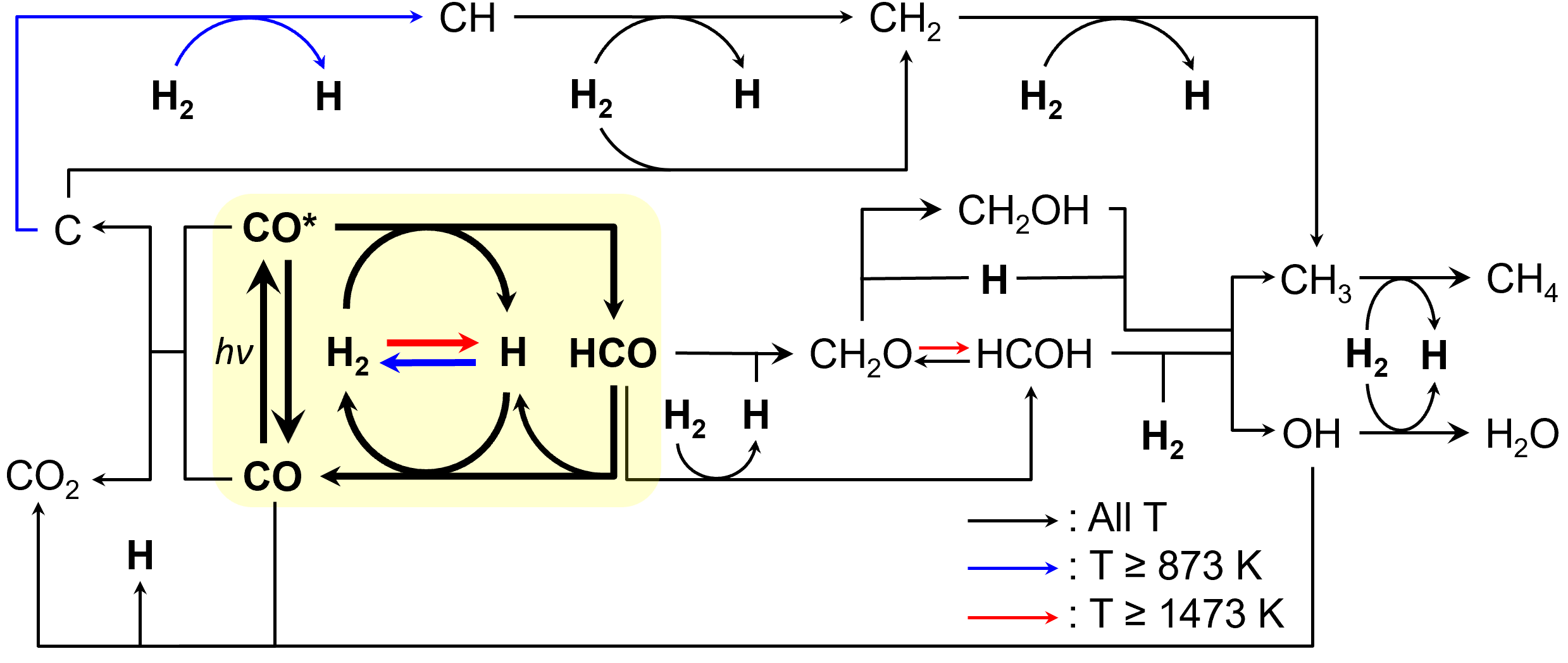}
          }
\caption{Model predicted major reaction pathways based on the ROP analysis describing previous experimental studies of \cite{Fleury_2019,Fleury_2020} at various temperature conditions indicated by the color of lines: (a) without UV photons, and (b) with UV photons. Reaction rates inside the yellow highlighted region in Figure \ref{fig:reactionpathwayswithandwithoutUV}b are at least around three orders of magnitude faster than the rates of any other reactions in the model.
\label{fig:reactionpathwayswithandwithoutUV}}
\end{figure*}

\subsubsection{Reaction kinetics of thermal- and UV-photochemistry}
The most notable feature shown from the reaction kinetic modeling of simultaneous thermal- and UV-photochemistry is the significant increase of the molecular mixing ratios of all the species at all temperatures compared to the case with thermal-only chemistry (see Figure \ref{fig:Fleury_2019_wUV_woUV}). All the species (i.e. \ce{CH4}, \ce{H2O}, \ce{CO2}, and H radicals) are predicted to be produced at least $\sim$20 orders of magnitude, $\sim$7 orders of magnitude, and $\sim$3 orders of magnitude more in the thermal- and UV-photochemistry condition than in the thermal-only condition at temperatures of 573 K, 873 K, and 1173 K, respectively. At temperature of 1473 K, all these species are predicted to be formed $\sim$40 times more in the thermal- and UV-photochemistry condition compared to the predicted productions of the species in the thermal-only condition. Figure \ref{fig:Fleury_2019_wUV_woUV}b shows that the molecular mixing ratio of CO(a\textsuperscript{3}$\Pi$) is not sensitive to temperature difference, which indicates that only the photochemical cycle (mentioned later) determines its abundance in the system. Overall, the model prediction can well explain significantly augmented formations of \ce{CH4}, \ce{H2O}, and \ce{CO2} at all temperature conditions (T $\leq$ 1473 K) under simultaneous thermal- and UV-photochemistry, as shown in Figure 6 of \cite{Fleury_2019}.

If we look at the major reaction pathways based on the ROP analysis as shown in Figure \ref{fig:reactionpathwayswithandwithoutUV}b, we can see that the chemistry of the whole system at all temperature conditions is dominated by one photochemical cycle indicated in the yellow highlighted region: (i) Carbon monoxide in the ground state gets electronically excited to the a\textsuperscript{3}$\Pi$ state by UV-photons, (ii) Excited CO can either go down to the ground state CO within its radiative lifetime of 3 ms (\cite{Lee_2020}) or react with \ce{H2} to form H and HCO, and (iii) HCO can either dissociate into H + CO or react with H radical to disproportionate to \ce{H2} and CO. At elevated temperatures, H radical recombination (i.e. H + H $\rightarrow$ \ce{H2} at T $\geq$ 873 K) and \ce{H2} dissociation (\ce{H2} $\rightarrow$ H + H at T $\geq$ 1473 K) also become dominant as well. Based on the ROP analysis, these major reaction pathways of the highlighted cycle are at least $\sim$3 orders of magnitude faster than any other reactions shown (or not shown if they are minor reactions) in Figure \ref{fig:reactionpathwayswithandwithoutUV}b. For this reason, a significant amount of HCO becomes available with the aid of this photochemical cycle even at very low temperature such as 573 K, which makes the chemistry of the system very different from thermal-only chemistry. As a result, every chemical reaction that is involved with HCO becomes much more efficient with UV-photons than it used to be in thermal-only chemistry, which in turn pushes the whole chemistry to the right side of the Figure \ref{fig:reactionpathwayswithandwithoutUV}b and eventually leads to the augmented formation of \ce{CH4}, \ce{H2O}, and \ce{CO2}, in agreement with the experimental observations of \cite{Fleury_2019}.

The reaction between the photoexcited CO and the ground-state CO that forms a triplet C atom and \ce{CO2} turns out to be very important according to the kinetic model especially at high temperatures above 873 K. As shown in Figure \ref{fig:reactionpathwayswithandwithoutUV}b, this reaction efficiently forms \ce{CO2} and the triplet C atom first. The C atom will then be hydrogenated by reacting with \ce{H2} into CH, \ce{CH2}, and then \ce{CH3} step by step, which will form \ce{CH4}. Along with every hydrogenation step (i.e. CH\textsubscript{n} + \ce{H2} $\rightarrow$ CH\textsubscript{n+1} + H), H radicals are efficiently formed and then attack \ce{CO2} to form OH radicals as well, which will form \ce{H2O}. As mentioned earlier in the section \ref{subsec:COt_CO_rxn}, due to its temperature-dependency of the CO(X\textsuperscript{1}$\Sigma^+$) + CO(a\textsuperscript{3}$\Pi$) $\rightarrow$ C(\textsuperscript{3}P) + \ce{CO2} reaction, this reaction will be more important at elevated temperatures. According to our kinetic model result, significantly augmented formations of major species (\ce{CH4}, \ce{H2O}, and \ce{CO2}) are attributed to the \ce{H2} + CO(a\textsuperscript{3}$\Pi$) $\rightarrow$ H + HCO reaction up until 873 K, while the augmented major species formations are attributed to the CO(X\textsuperscript{1}$\Sigma^+$) + CO(a\textsuperscript{3}$\Pi$) $\rightarrow$ C(\textsuperscript{3}P) + \ce{CO2} reaction at temperatures above 873 K. In short, two different photochemical schemes (i.e. \ce{H2} + CO(a\textsuperscript{3}$\Pi$) $\rightarrow$ H + HCO vs CO(X\textsuperscript{1}$\Sigma^+$) + CO(a\textsuperscript{3}$\Pi$) $\rightarrow$ C(\textsuperscript{3}P) + \ce{CO2}) determines the mixing ratios of \ce{CH4}, \ce{H2O}, and \ce{CO2} at different temperature regions below the transition temperature (i.e. 1970 K for \ce{CH4}, \ce{H2O}, and \ce{CO2}), respectively. With regard to H radicals, however, it has to be noted that \ce{H2} + CO(a\textsuperscript{3}$\Pi$) mainly determines the mixing ratio of H radicals in the system at temperatures below the transition temperature (i.e. 1650 K for H radicals).

In contrast to thermal-only chemistry, the molecular mixing ratio of \ce{CH4}, \ce{H2O}, and \ce{CO2} wiggles as shown in Figure \ref{fig:Fleury_2019_wUV_woUV}a. These molecules have been efficiently formed as a result of UV photochemistry enhanced by metastable a\textsuperscript{3}$\Pi$ state carbon monoxide and remain relatively stable at relatively lower temperatures below 873 K (i.e. UV photochemistry determines their mixing ratio after the reaction time of 18 hours). However, at 1173 K, thermal chemistry starts to partially determine their mixing ratio and this is well-shown at 2000 K in Figure \ref{fig:Fleury_2019_wUV_woUV}a where the molecular mixing ratios of \ce{CH4}, \ce{H2O}, and \ce{CO2} formed with UV photons are not distinguishable from those formed without UV photons (i.e. thermal only). We further simulated the molecular mixing ratio of \ce{CH4}, \ce{H2O}, and \ce{CO2} at 2000 and 2500 K after 18 hours while maintaining other conditions as same as that at 1473 K, and as expected, there was no significant difference in the molecular mixing ratio of \ce{CH4}, \ce{H2O}, and \ce{CO2} formed no matter whether UV photons were available or not (see Figure \ref{fig:Fleury_2019_wUV_woUV}). So based on the model prediction (and as mentioned earlier in the text), we can say that the transition temperature where thermal chemistry becomes dominant compared to photochemistry starts at around 1970 K and it has to be noted that this transition temperature can be shifted due to multiple factors including (i) errors in the UV photon fluxes, (ii) errors in photoabsorption cross-sections (e.g. temperature dependent) of species, (iii) errors in quantum yields of any photochemical reactions, (iv) any missing reaction species or reactions in the system, (v) any errors in included thermal reactions, and (vi) any errors in included thermodynamic parameters of related species. With regard to this, the sensitivity of the model will be discussed later.

\subsubsection{The potential importance of the acetylene (\ce{C2H2}) formation with regard to the organic refractory aerosol formation in hot-Jupiter exoplanet atmospheres}
Although a detectable amount of solid-phase product was not observed from any of the experiments at the temperature below 1473 K conducted by \cite{Fleury_2019} (i.e. 15 mbar; 573--1473 K; with and without UV; reaction time of 18 hours), an observable amount of condensed-phase organic aerosol products as a thin film deposited on sapphire windows was reported by \cite{Fleury_2019} after 204 hours of experiments with increased starting total gas pressure of 81 mbar with UV irradiation at 1473 K. Since there was no any direct information about the molecular structure of this aerosol, it is hard to exclusively say whether this observed organic aerosol is composed of aromatic hydrocarbons (if not polycyclic) or not. However, it is logical to say that \ce{C1} species should first go through \ce{C2} species before growing into aerosol particles, which gives us a rationale to look into any \ce{C2} or larger species predicted to be formed during our model simulation. The model simulation of the corresponding experimental condition (i.e. 81 mbar, 1473 K, with UV-photons, and the reaction time of 206 hours) predicted the molecular mixing ratio of each species as following: [\ce{C2H6}] = 6.62$\times$10\textsuperscript{-15}, [\ce{C2H4}] = 7.32$\times$10\textsuperscript{-12}, and [\ce{C2H2}] = 1.95$\times$10\textsuperscript{-10}. Any species larger than \ce{C2} species (e.g. \ce{C3H3} or larger) were predicted to be around or less than  1.00$\times$10\textsuperscript{-20} (the absolute tolerance of the differential equation solver), which indicates that no significant amount of the species larger than \ce{C2} species are predicted to be formed in the simulated systems even at the temperature of 2500 K.

Among these \ce{C2} species, we focused on the acetylene formation in the model for the following reason: (i) Previous studies have suggested benzene (1-ring aromatic hydrocarbon) and naphthalene (2-ring aromatic hydrocarbon) as precursors to refractory hydrocarbon aerosols (\cite{Trainer-2013,Brem-2015}), (ii) one of the prevailing mechanisms that rationalize the formation up to 2-ringed aromatics (including benzene and naphthalene) is the hydrogen-abstraction-acetylene-addition mechanism (HACA), an aromatic radical formed via hydrogen abstraction adds to acetylene (\ce{C2H2}) to form a larger vinylic radical adduct (\cite{Bittner-1981, Chu_2019, Frenklach_1985, Parker_2014, Yang_2016, Smith_2020}), and (iii) a previous photochemical experiment by \cite{Franklin_2018-thesis} using a \ce{D2}-lamp (115--170 nm) at 25$^{\circ}$C in oxygen-free conditions has shown that the VUV irradiation on \ce{C2H2} is the most efficient in the formation of organic particles compared to the VUV irradiation on other \ce{C2} species (i.e. \ce{C2H6} and \ce{C2H4}). Although it is not clear how the acetylene formation is exactly related to the formation of organic aerosols observed from \cite{Fleury_2019}, as mentioned earlier, many previous studies indicate the positive relationship between the acetylene formation and the aerosol formation. 

\begin{figure}
    \centering
    \includegraphics[width=1\textwidth]{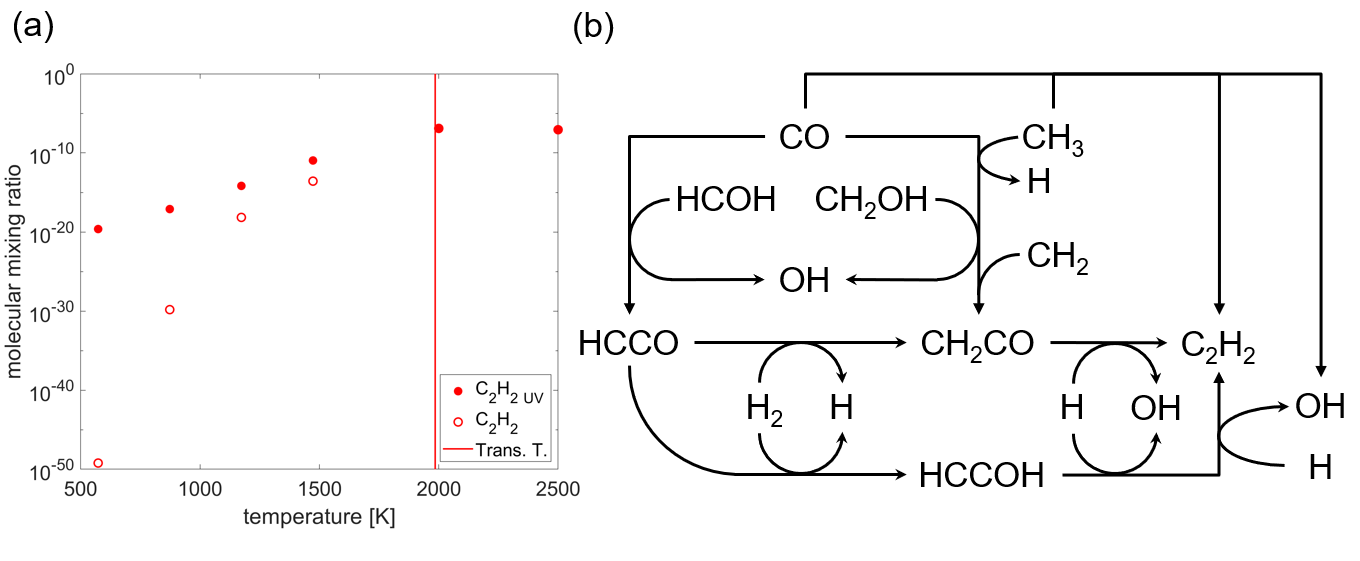}
    \caption{\footnotesize (a) Temperature-dependent mixing ratio profiles of acetylene (\ce{C2H2}) predicted by the reaction kinetic modeling for the experimental conditions described in \cite{Fleury_2019}. Solid symbols indicate the predicted \ce{C2H2} mixing ratio with UV photons, while open symbols indicate the predicted \ce{C2H2} mixing ratio at the thermal-only condition (i.e. without UV photons). The solid line refers to the transition temperature at which the molecular mixing ratio of \ce{C2H2} formed in different systems (i.e. thermal + UV vs thermal-only) only differ by less than 10 \%). The transition temperature with regard to \ce{C2H2} is 1985 K; (b) A schematic diagram of reaction pathways toward \ce{C2H2} formation based on the ROP analysis}
    \label{fig:C2H2_pathway}
\end{figure}

\begin{deluxetable}{lccccc}
\tablenum{3}
\tablecaption{The model predicted \ce{C2H2} mixing ratio comparison among various conditions (initial molecular compositions of \ce{H2} : CO = 99.7 : 0.3) with UV photons available \textsuperscript{\textit{a}} \label{tab:C2H2_comparison}}
\tablehead{
\colhead{} & \colhead{1473 K, 15 mbar}{\textsuperscript{\textit{b}}}& \colhead{1473 K, 81 mbar} & \colhead{1473 K, 81 mbar}\textsuperscript{\textit{b},{\textit{c}}} & \colhead{2000 K, 15 mbar}& \colhead{2500 K, 15 mbar}\\
\colhead{} & \colhead{(18 hr)}& \colhead{(18 hr)} & \colhead{(206 hr)} & \colhead{(18 hr)}& \colhead{(18 hr)}
} 
\startdata
[\ce{C2H2}]&1.10$\times$10\textsuperscript{-11} (2.68$\times$10\textsuperscript{-14})&1.57$\times$10\textsuperscript{-12} (2.18$\times$10\textsuperscript{-13})&1.95$\times$10\textsuperscript{-10} (2.43$\times$10\textsuperscript{-11})&1.37$\times$10\textsuperscript{-7} (1.25$\times$10\textsuperscript{-7})&8.55$\times$10\textsuperscript{-8} (8.55$\times$10\textsuperscript{-8})\\
\hline
\enddata
\tablecomments{\footnotesize\textsuperscript{\textit{a}}Numbers in parentheses refer to calculated molecular mixing ratio of \ce{C2H2} at thermal-only condition; \textsuperscript{\textit{b}}Experimental conditions conducted by \cite{Fleury_2019}; \textsuperscript{\textit{c}}Experimental condition from which a detectable amount of solid-phase product was observed in \cite{Fleury_2019}}
\end{deluxetable}

With this background, if we look at the temperature-dependent mixing ratio profiles of acetylene predicted by the reaction kinetic modeling for the experimental conditions described in \cite{Fleury_2019}, we can see that UV-excited metastable carbon monoxides can significantly enhance the formation of acetylene compared to thermal-only chemistry as shown in Figure \ref{fig:C2H2_pathway}a at the temperature up to 1473 K. Up until 1173 K, acetylene (\ce{C2H2}) is predicted to be produced at least $\sim$4 orders of magnitude more in the thermal- and UV-photochemistry condition than in the thermal-only condition as shown in Figure \ref{fig:C2H2_pathway}a. At the temperature of 1473 K, acetylene is predicted to be formed $\sim$410 times more in the thermal- and UV-photochemistry condition compared to the predicted acetylene production in the thermal-only condition. At the temperature above 2000 K, there was no significant difference in the predicted molecular mixing ratio of \ce{C2H2} formed no matter whether UV photons were available or not, which indicates that the transition temperature is around 2000 K (precisely 1985 K). Figure \ref{fig:C2H2_pathway}b shows major reaction pathways from \ce{C1} species to \ce{C2H2}. As can be seen from Figure \ref{sec:ROP}\ref{fgr:ROP_573K_3}--\ref{sec:ROP}\ref{fgr:ROP_1473K_3} in the Appendix \ref{sec:ROP}, at early timescales (i.e. up to $\sim$10 ms), \ce{CH3} + CO $\rightarrow$ \ce{C2H2} + OH is the major pathway, and then the major \ce{C2H2} formation pathway changes to others such as \ce{HCCOH} + H $\rightarrow$ \ce{C2H2} + OH or \ce{CH2CO} + H $\rightarrow$ \ce{C2H2} + OH. With UV photons available in the system, the UV-elevated amount of HCO results in the elevated formation of HCOH and \ce{CH2OH}, which then results in the elevated formation of \ce{C2H2}. However, at temperatures higher than the transition temperature (i.e. 1985 K), thermal chemistry now determines the \ce{C2H2} mixing ratio like it did with the molecular mixing ratios of \ce{CH4}, \ce{H2O}, and \ce{CO2}. Like the calculated rate-coefficient of the reaction between the excited CO and \ce{H2} forming HCO and H is much faster than the rate-coefficient of the reaction between the ground state CO and \ce{H2} forming HCO and H, the rate-coefficient of the reaction between the excited CO and \ce{CH3} forming \ce{C2H2} and OH might affect the formation of \ce{C2H2} in a significant way. To test this, we recalculated the potential energy surfaces of \ce{C2H2} + OH taken from \cite{Miller_1989} at the CBS-QB3 level of theory using Gaussian 09 (\cite{g09}) and calculated the temperature- and pressure-dependent rate coefficients of \ce{CH3} + CO(a\textsuperscript{3}$\Pi$) reactions following the similar method described in \ref{photochem_rate_calc} (rate-coefficients and the corresponding PES are available in the Supplementary Materials). However, no matter whether \ce{CH3} + CO(a\textsuperscript{3}$\Pi$) chemistry is included in the reaction kinetic model or not, no significant change was observed on the formation of \ce{C2H2}. This might indicate that the molecular mixing ratio of \ce{CH3} is still too low to draw any significant increases in the acetylene formation through \ce{CH3} + CO(a\textsuperscript{3}$\Pi$) chemistry.

Table \ref{tab:C2H2_comparison} shows the model (thermal- and UV-photochemistry incorporated) predicted \ce{C2H2} mixing ratios at different conditions (starting from \ce{H2} : CO = 99.7 : 0.3). You can see from the Table \ref{tab:C2H2_comparison} that the model simulation indicates that the formation of acetylene (\ce{C2H2}) is favored with increasing temperature. The model also shows that \ce{C2H2} formation has not reached quasi-equilibrium state even after 206 hrs (81 mbar) and 18 hrs (15 mbar) at 1473 K, respectively (see Figure \ref{sec:C2H2_mole_fraction_profile}\ref{fig:C2H2_formation_profiles}a--b in the Appendix \ref{sec:C2H2_mole_fraction_profile}). However, at 2000 K, the acetylene formation reaches its quasi-equilibrium state after 18 hrs with its molecular mixing ratio of $\sim$1.37$\times$10\textsuperscript{-7} (see Figure \ref{sec:C2H2_mole_fraction_profile}\ref{fig:C2H2_formation_profiles}c in the Appendix \ref{sec:C2H2_mole_fraction_profile}) and no significant amount of any species larger than \ce{C3} molecules were predicted in the given system. This mainly indicates that (i) the amount of acetylene formed via thermal-only chemistry (up to 2500 K) from the condition of \ce{H2} : CO = 99.7 : 0.3 is not enough to push the chemistry to larger species such as polycyclic aromatic hydrocarbons (PAHs) that are accepted as precursors to organic aerosols in the majority of combustion research (\cite{Frenklach_2020}) and (ii) some major photochemical reactions can bring the chemistry into disequilibrium chemistry to form larger hydrocarbon species (e.g. 1-ring aromatic hydrocarbons or PAHs) or even organic aerosols. With regard to (ii), the results of previous experimental studies (\cite{Franklin_2018-thesis} and \cite{Fleury_2019}) in combination with the current study in this paper might indicate that the reaction kinetic model including the \ce{C2H2}-photochemistry is the key to explain the experimentally observed organic aerosol formations from \cite{Fleury_2019}. Note that our reaction kinetic model including CO(a\textsuperscript{3}$\Pi$)-photochemistry was successful in qualitatively explaining the significant augmented formation of major species (i.e. \ce{CH4}, \ce{H2O}, and \ce{CO2}) observed in \cite{Fleury_2019} at all temperature conditions. As can be seen from Table \ref{tab:C2H2_comparison}, our thermal- and UV-photochemistry model predicts $\sim$3 orders of magnitude more \ce{C2H2} are formed in the system at the 2000 K and 15 mbar condition even with an order of magnitude shorter time scale (i.e. 18 hrs) compared to the 1473 K and 81 mbar condition with the reaction time of 206 hrs. Thus if our hypothesis (i.e. the amount of \ce{C2H2} is closely related to aerosol formations) is correct, we are able to observe aerosol formations even at temperatures lower than 2000 K within the reaction time shorter than 18 hrs from the same experimental device of \cite{Fleury_2019}. Conducting this experiment and including \ce{C2H2}-related photochemistry into our current model would be interesting future studies.

\subsubsection{Sensitivity analysis of the kinetic model}
One of the major features of this study is the assessment of the transition temperature in which thermal chemistry starts to become dominant compared to photochemistry. However, this transition temperature can be affected by multiple factors as mentioned earlier: (i) errors in the UV light sources, (ii) errors in photoabsorption cross-sections, (iii) errors in quantum yields of any photochemical reactions, (iv) any missing reaction species or reactions (even surface chemistry), (v) any errors in thermochemical reactions included in the model, and (vi) any errors in thermodynamic parameters of related species. Although (iv) is important when it comes to model improvements, this was beyond the scope of the current work. With regard to (i)--(iii), these three factors directly affect the calculated carbon monoxide photoexcitation rate-coefficient (i.e. X\textsuperscript{1}$\Sigma^+$ $\rightarrow$ a\textsuperscript{3}$\Pi$) in \ref{photoexcit_rate_calc}. With regard to (v)--(vi), these errors originally came from the embedded errors of rate coefficients imported from the references described in \ref{rmg_library}. We computed the sensitivity of the major species (i.e. \ce{CH4}, \ce{H2O}, and \ce{CO2}) to all of the rate coefficients in the model simulated under the condition of 1473 K, 15 mbar, [\ce{H2}] = 0.997, and [CO] = 0.003, and the most sensitive three parameters that affect the predicted formation of the major species turn out to be photochemical reactions which are
\begin{equation}
    \mathrm{CO(X\textsuperscript{1}\Sigma^+)} \to \mathrm{CO(a\textsuperscript{3}\Pi)}
    \label{eq:rxn1}
\end{equation}
\begin{equation}
    \mathrm{CO(X\textsuperscript{1}\Sigma^+)} + \mathrm{CO(a\textsuperscript{3}\Pi)} \to \ce{CO2} + \mathrm{C(\textsuperscript{3}P)}
    \label{eq:rxn2}
\end{equation} 
\begin{equation}
    \ce{H2} + \mathrm{CO(a\textsuperscript{3}\Pi)} + \mathrm{M} \to \mathrm{H} + \mathrm{HCO} + \mathrm{M}
    \label{eq:rxn3}
\end{equation} 
They each have normalized sensitivities $\frac{d(lnC\textsubscript{n})}{d(\emph{k}\textsubscript{i})}$ with magnitudes of around 1, 0.95, and -0.75, respectively (see Figure \ref{sec:sensitivity_analysis}14). This means that, for example, if one increases the rate-coefficient of reaction \ref{eq:rxn1} by a factor of 10, the predicted molecular mixing ratio of the major species (i.e. \ce{CH4}, \ce{H2O}, and \ce{CO2}) would be increased by 10 times and the transition temperature would be pushed to a much higher temperature. On the contrary, if one reduces this carbon monoxide photoexcitation rate-coefficient (i.e. reaction \ref{eq:rxn1}, \textit{k}\textsubscript{2}) by a factor of 2, the predicted molecular mixing ratio of the major species (i.e. \ce{CH4}, \ce{H2O}, and \ce{CO2}) would be decreased by $\sim$50\% and the transition temperature would be shifted to lower temperature. The former case is possible if the photoabsorption cross-sections of carbon monoxide are underestimated. It has to be noted that the photoabsorption cross-sections of CO used to calculate \textit{k}\textsubscript{2} were taken from \cite{MyerSamson_1970} which experimentally measured the VUV absorption cross-sections at 298 K. Since VUV absorption cross-sections usually increase with increasing temperatures due to thermally increased populations at higher vibrational states, which then might result in increased access to upper electronic states. \cite{Venot_2018} experimentally showed that the photoabsorption cross-sections of \ce{CO2} around Lyman-$\alpha$ can vary by two orders of magnitude between 150 and 800 K. For this reason, it might be the case if the photoabsorption cross-sections of CO around Lyman-$\alpha$ increase by more than an order of magnitude from 298 K to 1473 K, which can potentially increase the carbon monoxide photoexcitation rate-coefficient (i.e. \textit{k}\textsubscript{2}) by a factor of 10, which means the calculated \textit{k}\textsubscript{2} is 10 times underestimated. However, the latter case (i.e. reducing \textit{k}\textsubscript{2} by a factor of 2) is also possible. It has to be noted that the quantum yield (i.e. $\Phi_2$) used in the calculation of \textit{k}\textsubscript{2} in \ref{photoexcit_rate_calc} is assumed to be unity since we couldn't find any theoretically or experimentally determined quantum yield of this photoexcitation reaction. However, it wouldn't be surprising if $\Phi_2$ is 0.5. In this case, the model predicted molecular mixing ratio would be overestimated by $\sim$50\%. These cases suggest interesting future studies (i.e. a measurement of temperature-dependent VUV photoabsorption cross-sections of carbon monoxide or a determination of the $\Phi_2$).

It has to be noted that there are too many model parameters and too few experimental data (e.g. a measurement of the CO absorption cross-section at various temperatures or a measurement of $\Phi_2$) to allow a perfect determination of the model parameters (and keep in mind that there are several other moderately sensitive rate coefficients and thermochemistry in the model, not just these 3 most sensitive reactions). However, even considering these, our kinetic modeling work indicates that electronically excited CO in its metastable state (a\textsuperscript{3}$\Pi$) can push the whole chemistry to the augmented formation of \ce{CH4}, \ce{CO2}, and \ce{H2O} in an obvious way, which qualitatively matches well with the experimental observation from \cite{Fleury_2019}. The quantitative discrepancy doesn't indicate a fundamental issue with the reaction kinetic model, but rather reflects our imperfect knowledge of the values of the model parameters.

\subsection{Modeling of the \ce{H2}/CO/\ce{H2O} exoplanet atmosphere analogue of \cite{Fleury_2020}}
\begin{figure}
    \centering
    \includegraphics[width=0.75\textwidth]{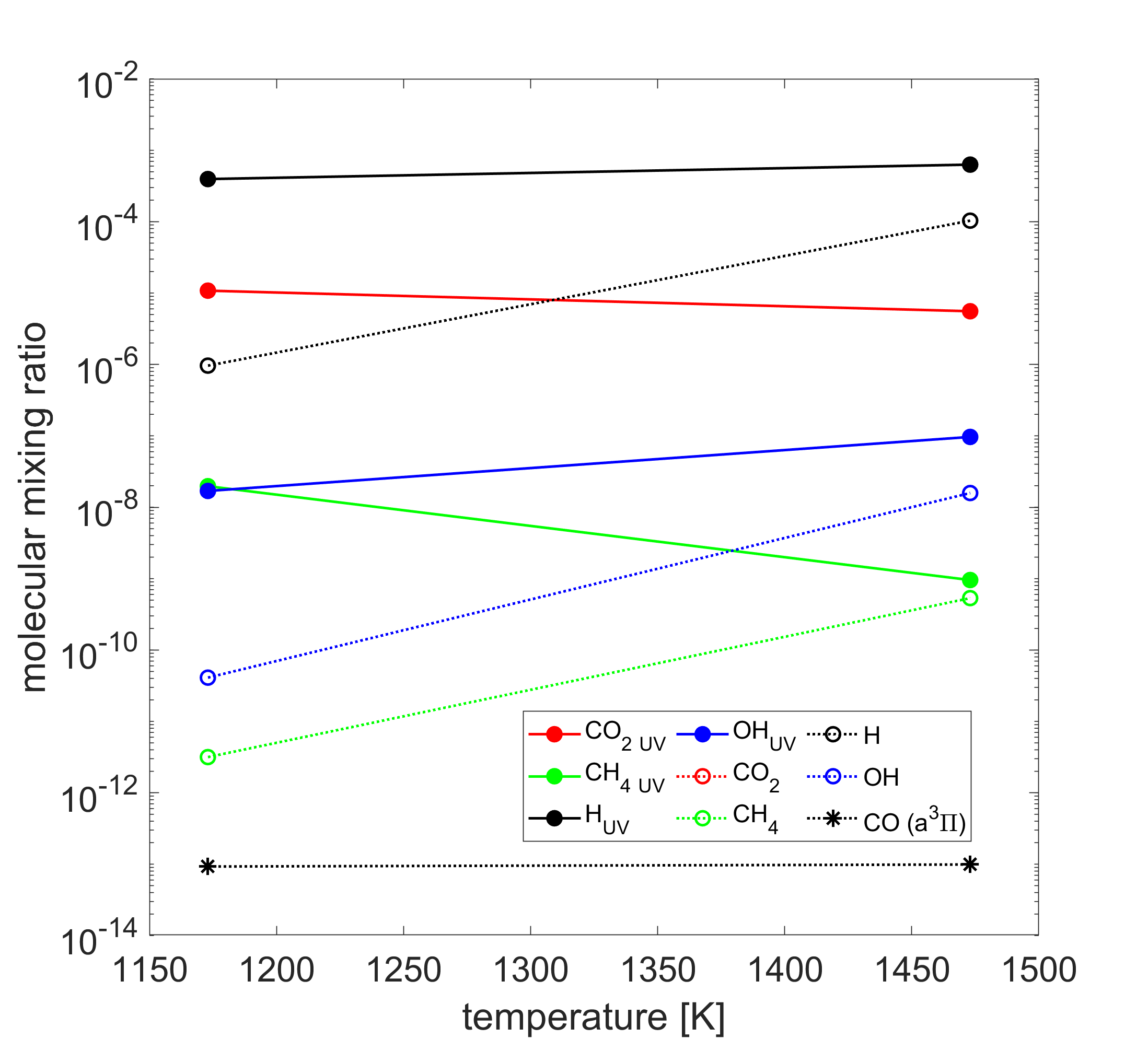}
    \caption{\footnotesize Temperature-dependent molecular mixing ratio profiles predicted by the reaction kinetic model for the experimental conditions described in \cite{Fleury_2020}. Solid symbols indicate predicted mixing ratio of molecular species with both thermal energy and UV photons, while open symbols indicate predicted mixing ratio of molecular species at the thermal-only condition (i.e. without UV photons): (a) Red circles indicate \ce{CO2}. Predicted \ce{CO2} mixing ratio at the thermal-only condition overlaps with the mixing ratio of \ce{CO2} with UV photons, thus not shown in the Figure; Blue circles indicate \ce{OH}; Green circles indicate \ce{CH4}; Black circles indicate H-atom (radical species); Asterisks indicate UV-excited CO (i.e. carbon monoxide in the a\textsuperscript{3}$\Pi$ state).}
    \label{fig:Fleury_2020_wUV_woUV}
\end{figure}

When \ce{H2O} is added to the \ce{H2}/CO mix to change the C/O ratio to more realistic solar elemental abundance, both of the kinetic model-predicted thermal-only and UV-incorporated thermal chemistry of \cite{Fleury_2020} were almost similar to those of \cite{Fleury_2019}. The major reaction pathway forming \ce{CH4} (note that \ce{H2O} was included from the beginning of the experiments at much higher mixing ratio, so not focused as a product) in \cite{Fleury_2020} are predicted to be almost similar to those described in Figure \ref{fig:reactionpathwayswithandwithoutUV} and the previous section of describing the modeling result of \cite{Fleury_2019}. The major reaction pathways forming \ce{CO2} was exclusively CO being oxidized by OH radicals forming \ce{CO2} and H radicals. This behavior is due to a huge amount of \ce{H2O} included in the system from the beginning, which then can either thermally or photochemically dissociate into H and OH radicals in the system. Although the \ce{H2O} formation is predicted from our model simulation of \cite{Fleury_2019}, the amount of the \ce{H2O} predicted in the simulation of \cite{Fleury_2019} is still way smaller than the amount of \ce{H2O} introduced into the system from the beginning of the simulation of \cite{Fleury_2020}. For this reason, the simulated system of \cite{Fleury_2020} would contain much more OH radicals compared to the simulated system of \cite{Fleury_2019}. This leads to the notable difference between the model simulation of \cite{Fleury_2020} and that of \cite{Fleury_2019} in the form of the reversed molecular mixing ratio of \ce{CH4} to \ce{CO2}. As shown in Figure \ref{fig:Fleury_2020_wUV_woUV}, the molecular mixing ratio of \ce{CO2} is $\sim$3 orders and $\sim$4 orders of magnitude larger compared to that of \ce{CH4} at 1173 and 1473 K, respectively. In the case of \cite{Fleury_2019}, the molecular mixing ratio of \ce{CH4} is always at least 2 orders of magnitude higher than that of \ce{CO2} at temperatures higher than 1173 K, no matter UV photons are available or not. As mentioned above, this reversed ratio is attributed to the inclusion of \ce{H2O} from the beginning of the kinetic simulation of \cite{Fleury_2020}. According to the kinetic modeling, \ce{H2O} already thermally dissociates to form both H and OH radicals at 1173 K. These OH radicals reach the quasi-equilibrium state within $\sim$11 hours when the energy source is thermal-only, while they reach quasi-equilibrium much faster within 200 seconds with the aid of UV photons (i.e. photodissociation of \ce{H2O} into H + OH). This tendency gets more severe at higher temperature (i.e. 1473 K) in which thermal dissociation becomes much more efficient (since the thermal dissociation rate-coefficient is positively temperature dependent). This can be observed in the form of increasing OH mixing ratio in Figure \ref{fig:Fleury_2020_wUV_woUV}. Compared to 1173 K, the experimental condition of \cite{Fleury_2020} is much more oxidizing, so that the predicted \ce{CH4} mixing ratio decreases rapidly by 20 times (i.e. from [\ce{CH4}] = 1.97$\times$10\textsuperscript{-8} at 1173 K to [\ce{CH4}] = 9.55$\times$10\textsuperscript{-10} at 1473 K). It also has to be noted that the \ce{CO2} mixing ratio doesn't show any difference whether UV irradiation is available or not (see red symbols and lines in Figure \ref{fig:Fleury_2020_wUV_woUV}). This feature indicates that the \ce{CO2} formation has already reached its thermal quasi-equilibrium after 18 hours. If you look at Figure \ref{sec:CO2_mole_fraction_profile}\ref{fig:CO2_t_profile} in Appendix \ref{sec:CO2_mole_fraction_profile}, \ce{CO2} mixing ratio at 1173 K reaches equilibrium within 7 hours in thermal-only condition, and 50 seconds with UV-irradiation, while \ce{CO2} mixing ratio at 1473 K reaches equilibrium within 200 seconds in thermal-only condition, and 30 seconds with UV-irradiation. According to our model prediction, the predicted methane molecular mixing ratio at 1173 K is enough to be detected (i.e. [\ce{CH4}] = 1.97$\times$10\textsuperscript{-8} which is around the same to the predicted mixing ratio of \ce{CH4} plotted on Figure \ref{fig:Fleury_2019_wUV_woUV}a at 1173 K with UV irradiation). However, it was surprising that no methane formation was observed in any of experiments in \cite{Fleury_2020}. Two plausible explanations can be (i) errors in the current model overestimated the amount of \ce{CH4} formation or (ii) significant amount of \ce{H2O} included in the system from the beginning has disturbed the detection of any IR or mass-spectrometry peaks of \ce{CH4} at 1173 K. With regard to (i), we can simply assess this by sensitivity analysis. According to sensitivity analysis, the most sensitive reaction and the most sensitive thermochemistry that affect the predicted formation of \ce{CH4} are the reaction \ce{CH3} + OH $\rightarrow$ \ce{CH2OH} + H and the thermochemistery of \ce{H2} with their normalized sensitivities $\frac{d(lnC\textsubscript{n})}{d(\emph{k}\textsubscript{i})}$ of 0.5 and 1.25, respectively (see Figure \ref{sec:sensitivity_analysis}\ref{fig:SA_1173K}c and d in Appendix \ref{sec:sensitivity_analysis}). This means that we have to reduce the rate-coefficient of reaction \ref{eq:rxn1} by 20 times or we have to reduce the Gibbs free energy of \ce{H2} by 72 \% in order to decrease the predicted mixing ratio of \ce{CH4} by 5 times. Although it is obvious that those parameters embed a certain amount of errors inside, it is less likely that these errors can solely explain the non-detection of \ce{CH4} in \cite{Fleury_2020}. Rather, it would be more reasonable to lean to the explanation (ii) (i.e. an interference on the IR peaks and mass-spec peaks of \ce{CH4} due to significant amount of pre-existing \ce{H2O} in their experimental condition of \cite{Fleury_2020}). Other than that, the current model's reaction kinetic interpretation qualitatively matches well with the experimental observation of \cite{Fleury_2020}.

\subsection{Quantitative comparison on the mixing ratio of \ce{CO2} between the model simulation results and the experimental results of \cite{Fleury_2019} and \cite{Fleury_2020}}
Since \cite{Fleury_2019, Fleury_2020} quantified the mixing ratios of \ce{CO2} at various experimental conditions using the Beer-Lambert law on the IR spectroscopy results, we tried to derive some quantitative comparisons of \ce{CO2} between the model simulation results and the experimental results at corresponding experimental conditions. Although the discrepancy between the model simulation results and the experimental results of the C/O=0.35 condition is within 2 orders of magnitude as it is shown in Table \ref{tab:mixing_ratio_comparison}, the discrepancy between the model simulation results and the experimental results of the C/O=1 condition is way larger compared to the discrepancy in the C/O=0.35 condition. There are multiple factors that can derive this discrepancy. First, as mentioned in the main text, the products that are observed in gas-phase static cell experiments conducted by \cite{Fleury_2019,Fleury_2020} are formed far beyond tertiary or even further reaction chemistry including surface chemistry on the wall since these experiments were conducted for more than 18 hrs. Surface chemistry is considered to be more efficient and faster compared to gas-phase chemistry and very difficult to simulate in detail. Although surface chemistry is indeed very important when it comes to precisely describe experimental study, it was beyond the scope of the current work to gain a better insight into simultaneous thermally and photochemically driven reaction pathways involving electronically excited CO in its metastable state (a\textsuperscript{3}$\Pi$). Natural isotopic contamination when using enriched isotopic gases was mentioned in \cite{Fleury_2019} and affected the chemistry that might have led to the discrepancy as well. The calculation of the absorption cross-section of \ce{CO2} described in the section 2.5 in \cite{Fleury_2019} can also affect the discrepancy as well. We conclude it is likely that discrepancy is mainly due to combined effects of these multiple factors mentioned above as well as experimental uncertainties. As mentioned earlier, the quantitative discrepancy doesn't indicate a fundamental issue with the reaction kinetic model, but rather reflects our imperfect knowledge of the values of the model parameters.

Keeping the quantitative comparison aside that can be influenced by the experimental conditions discussed above, our modeling work clearly predicts the experimental observations: (a) for C/O=1, where only \ce{H2} and CO molecules were present in the beginning, the thermochemical formation of \ce{CO2} is significantly less efficient than thermally augmented photochemical production of \ce{CO2}, as observed by \cite{Fleury_2019}; (b) for C/O=0.35, which was achieved through the addition of \ce{H2O} to the \ce{H2} + CO starting composition, chemistry is mainly driven by the \ce{H2O} dissociation (both thermally and photochemically) and no significant difference is observed between thermal and thermally augmented photochemical reaction pathways as a result. Our model again confirms the experimental observations of \cite{Fleury_2020}. 

This analysis also highlights the importance of both experimental work and rigorous theoretical reaction kinetics modeling. Further, this work confirms that though experimental conditions have limitations that are unavoidable (such as reaction cell boundaries), relative equilibrium mixing ratios obtained from the experiments are similar to the predicted ones.

\begin{deluxetable}{ccccccccc}
\tablenum{4}
\tablecaption{The molecular mixing ratio comparison of \ce{CO2} at various conditions between the model simulation results and the experimental results of \cite{Fleury_2019} and \cite{Fleury_2020}. \label{tab:mixing_ratio_comparison}}
\tablehead{
\colhead{} & \multicolumn{4}{c}{C/O=1}&\multicolumn{4}{c}{C/O=0.35}\\
\colhead{} & \multicolumn{2}{c}{Thermal only}&\multicolumn{2}{c}{Thermal + UV}& \multicolumn{2}{c}{Thermal only}&\multicolumn{2}{c}{Thermal + UV}\\
\colhead{} & \colhead{This study}&\colhead{Exp. 2019\textsuperscript{\textit{a}}}& \colhead{This study}&\colhead{Exp. 2019\textsuperscript{\textit{b}}}&\colhead{This study}&\colhead{Exp. 2020\textsuperscript{\textit{c}}}&\colhead{This study}&\colhead{Exp. 2020\textsuperscript{\textit{c}}}} 
\startdata
573 K&1.83$\times$10\textsuperscript{-31}&1.50$\times$10\textsuperscript{-5}&7.60$\times$10\textsuperscript{-10}&1.00$\times$10\textsuperscript{-3}& \multicolumn{4}{c}{N/A}\\
\hline
873 K&4.96$\times$10\textsuperscript{-17}&6.40$\times$10\textsuperscript{-5}&4.56$\times$10\textsuperscript{-10}&8.50$\times$10\textsuperscript{-4}& \multicolumn{4}{c}{N/A}\\
\hline
1173 K&6.88$\times$10\textsuperscript{-14}&3.80$\times$10\textsuperscript{-5}&5.33$\times$10\textsuperscript{-11}&1.20$\times$10\textsuperscript{-3}&1.08$\times$10\textsuperscript{-5}&3.50$\times$10\textsuperscript{-4}&1.08$\times$10\textsuperscript{-5}&8.80$\times$10\textsuperscript{-4}\\
\hline
1473 K&1.68$\times$10\textsuperscript{-12}&3.40$\times$10\textsuperscript{-5}&7.21$\times$10\textsuperscript{-11}&4.30$\times$10\textsuperscript{-4}&5.56$\times$10\textsuperscript{-6}&3.60$\times$10\textsuperscript{-4}&5.56$\times$10\textsuperscript{-6}&5.60$\times$10\textsuperscript{-4}\\
\hline
\enddata
\tablecomments{\footnotesize\textsuperscript{\textit{a}}Mixing ratios of \textsuperscript{13}\ce{CO2} taken from Table 2 in \cite{Fleury_2019}; \textsuperscript{\textit{b}}Mixing ratios of \textsuperscript{13}\ce{CO2} taken from Table 3 in \cite{Fleury_2019}; \textsuperscript{\textit{c}}Mixing ratios of \textsuperscript{13}\ce{CO2} taken from Table 3 in \cite{Fleury_2020}}
\end{deluxetable}

\subsection{Impact of the current study on the field of astrochemistry}
In this work, we have implemented the state-of-the-art computer-aided automatic construction of an astrochemical reaction network and successfully demonstrated how this computer-aided reaction kinetic model can help us precisely interpret the previous photochemical experiment in detail. Not only we have figured out the photochemically important role of the UV-excited state of carbon monoxide, but also we suggest future studies such as the measurement of temperature-dependency of the cross-sections of carbon monoxide or the experiment and modeling combined kinetic study of \ce{C2H2}-UV-photochemistry. Our work clearly demonstrates the importance of coupling photochemical reaction pathways to thermochemical models to better understand exoplanet atmospheres. Only then will we be able to realistically describe atmospheric chemistry of exoplanets that receive significant amount of UV photons from their parent stars. The JWST transmission spectroscopy data will keep providing us with the deluge of exoplanetary data that need to be efficiently and precisely interpreted by atmospheric chemical models. As we can see from the case of the recent study by \cite{Tsai_2022}, it is important to properly implement photochemical reactions in chemical models to properly interpret observational data and extract valuable insight on exoplanet atmospheres. We hope our study would be the benchmark for future exoplanet atmospheric photochemical modeling study that will consequently lead to a rapid innovation in the field of astrochemistry.

\section{Conclusions} \label{sec:conclusions}
In this work, we utilized an automatic chemical reaction mechanism generator to build a large and complex thermo- and photochemical network that can qualitatively rationalize the augmented chemistry observed from previous experimental works by \cite{Fleury_2019, Fleury_2020}. Our model has demonstrated that Lyman-$\alpha$-aided electronically excited carbon monoxide in its metastable state (a\textsuperscript{3}$\Pi$) can significantly enhance the chemistry in the exoplanet atmosphere-like conditions through two different reactions which are (i) \ce{H2} + CO(a\textsuperscript{3}$\Pi$) $\rightarrow$ H + HCO and (ii) CO(X\textsuperscript{1}$\Sigma^+$) + CO(a\textsuperscript{3}$\Pi$) $\rightarrow$ \ce{CO2} + C(\textsuperscript{3}P). The first reaction (i.e. \ce{H2} + CO(a\textsuperscript{3}$\Pi$) $\rightarrow$ H + HCO) leads to significantly augmented formation of HCO radicals in \ce{H2}-dominated system, which can push the whole chemistry to further \ce{CH4}, \ce{H2O}, and \ce{CO2} formation even at a very low temperature of 573 K. At temperatures above 873 K, the second reaction (i.e. CO(X\textsuperscript{1}$\Sigma^+$) + CO(a\textsuperscript{3}$\Pi$) $\rightarrow$ \ce{CO2} + C(\textsuperscript{3}P)) forms \ce{CO2} and the triplet C radicals which rapidly hydrogenate into \ce{CH4}, which mainly contributes to significantly augmented formation of major species. However, at temperatures above 2000 K, thermal chemistry then becomes efficient enough to dominate the whole chemistry. Given the experimental conditions of \cite{Fleury_2019, Fleury_2020}, the transition temperature in which thermal chemistry becomes dominant compared to photochemistry starts at $\sim$1970 K, and this transition temperature can be shifted due to multiple factors. Finally, under the experimental conditions of \cite{Fleury_2019, Fleury_2020}, our model might suggest that thermal-only chemistry up to 2500 K cannot push the chemistry to larger species such as PAHs or even organic aerosols, and instead, \ce{C2H2} photochemistry and the photochemistry of higher carbon species that are produced from \ce{C2H2} photochemistry might be the key to explain the experimentally observed hydrocarbon aerosol formations observed from the previous experiments by \cite{Fleury_2019}. The model results have demonstrated the importance of electronically excited metastable carbon monoxide in exoplanet atmospheres and that adding more photochemical reactions and species to current 1D photochemical models would provide a more comprehensive understanding of exoplanet atmospheres.

\vspace{5mm}
This research work was carried out at the Jet Propulsion Laboratory, California Institute of Technology, under a contract with the National Aeronautics and Space Administration. This research work was funded by the NASA Exoplanet Research Program grant to MSG. BF thanks the Universit\'e Paris-Est Cr\'eteil (UPEC) for funding support (postdoctoral grant).

\bibliography{references}{}
\bibliographystyle{aasjournal}

\appendix
\section{Rate of Production (ROP) Analysis}
\label{sec:ROP}

In this section, the rate of production (ROP) analysis data are provided. Every table contains the rate of production of two species at two different conditions (i.e. the thermal-only condition and the thermal- and UV-combined condition). Each row represents the corresponding time and each column represents the corresponding condition. The unit is mol/m\textsuperscript{3}/s. For example, if you take a look at the second row of the left side of Figure \ref{sec:ROP}\ref{fgr:ROP_573K_3}f, you can see the rate of formation of \ce{CO2} at the reaction time of 100 $\mu$s (10\textsuperscript{-4} s) under two different conditions (i.e. the thermal-only condition and the thermal- and UV-combined condition). You can see that at 100 $\mu$s under the thermal-only condition at 573 K, \ce{CO2} is mainly formed through \ce{OH} + \ce{CO} $\leftrightarrow$ \ce{H} + \ce{CO2}. However, it has to be noted that the rate of the production through this reaction is almost insignificant (i.e.$\sim$10\textsuperscript{-44} mol/m\textsuperscript{3}/s) at 100 $\mu$s. In contrast to this, with UV at 573 K, \ce{CO2} is efficiently formed through \ce{CO} + CO(a\textsuperscript{3}$\Pi$) $\leftrightarrow$ \ce{C} + \ce{CO2} with the production rate of 10\textsuperscript{-16} mol/m\textsuperscript{3}/s at 100 $\mu$s, which tells us the importance of the CO photochemistry in the system as mentioned in the main text.

\begin{figure}[hb!]
   \centering(\textbf{a})
\gridline{\includegraphics[width=1\textwidth]{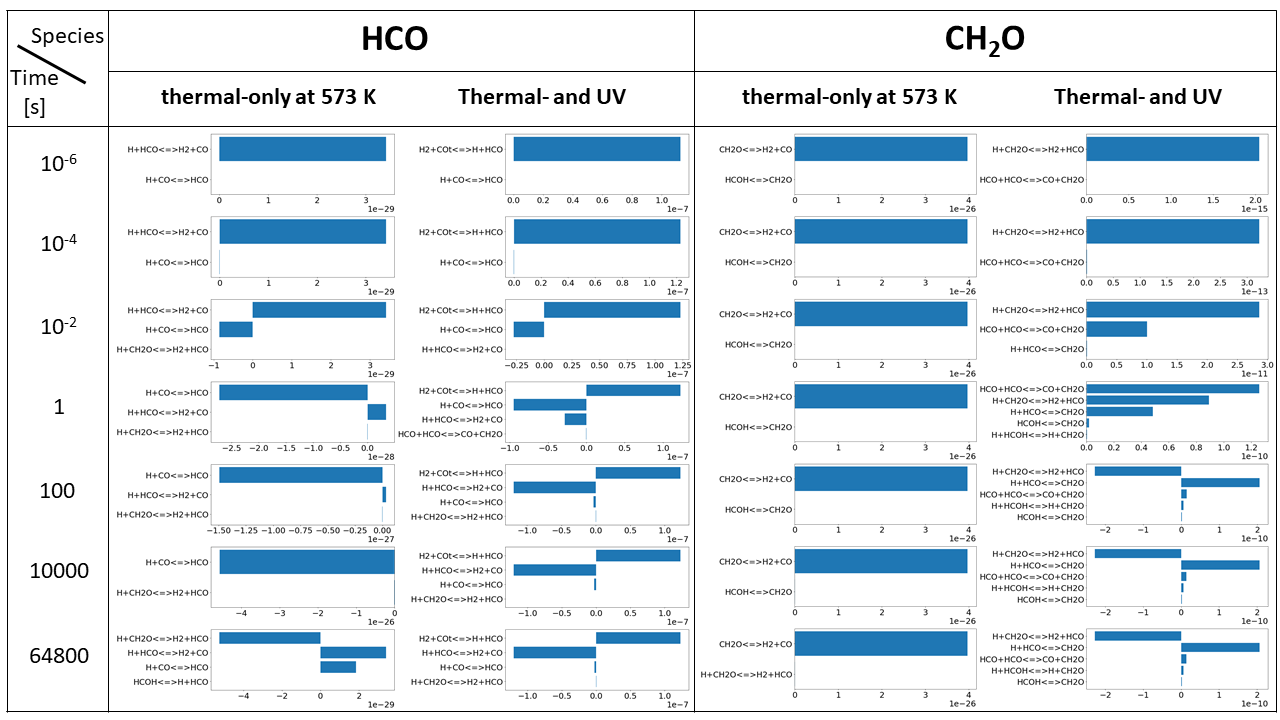}
          }
   \centering\textbf{(b)}
\gridline{\includegraphics[width=1\textwidth]{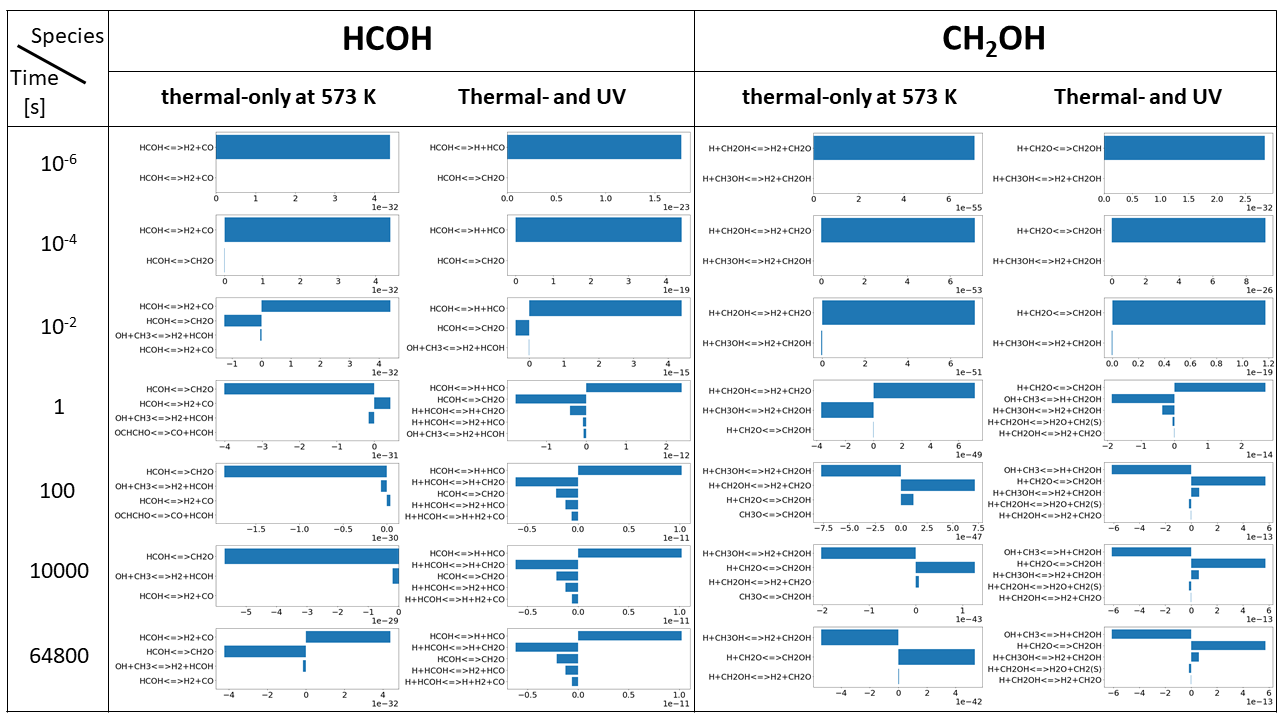}
          }
  \label{fgr:ROP_573K_1}
\end{figure}          

\begin{figure}[hb!]
   \centering\textbf{(c)}
\gridline{\includegraphics[width=1\textwidth]{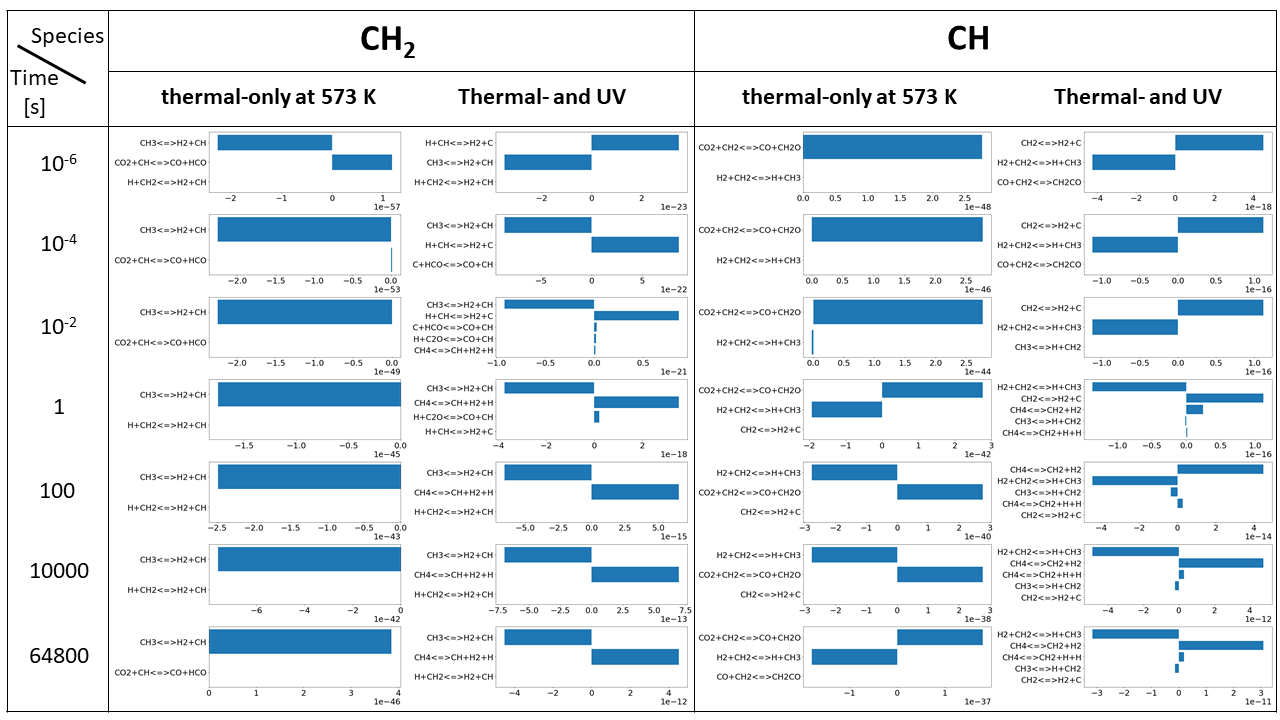}
          }

   \centering\textbf{(d)}
\gridline{\includegraphics[width=1\textwidth]{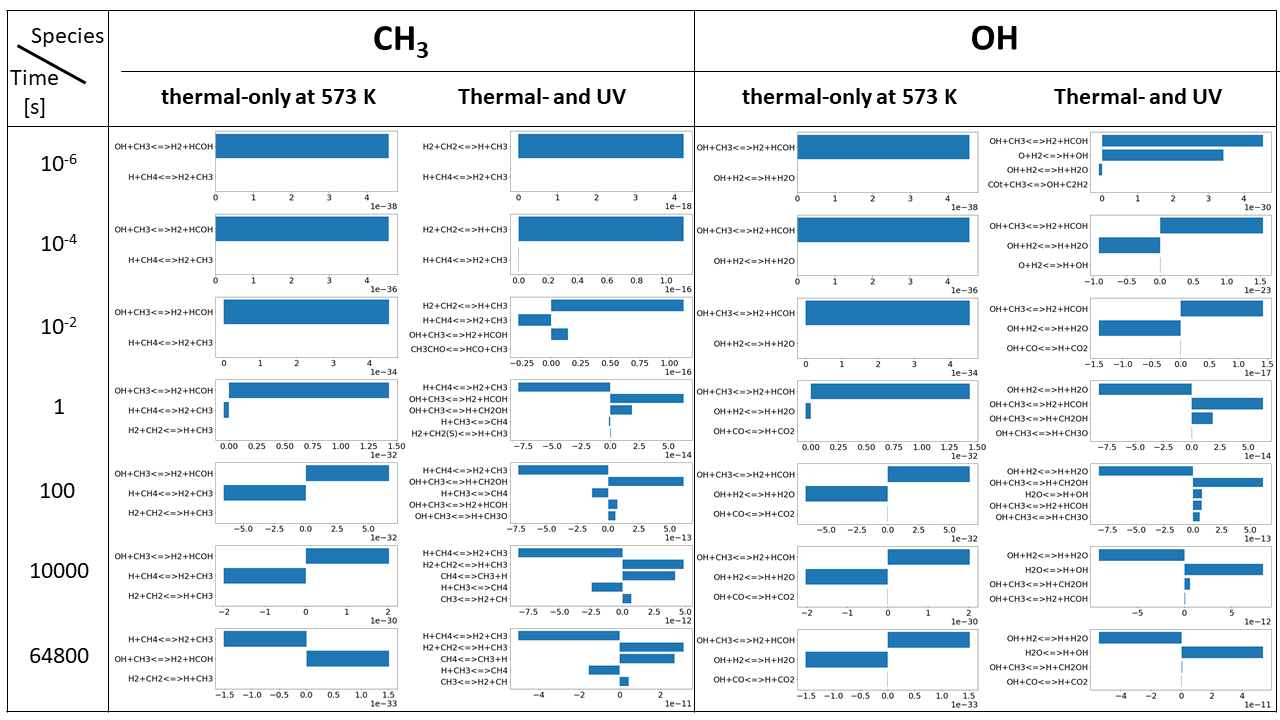}
          }
  \label{fgr:ROP_573K_2}
\end{figure}
\begin{figure}[hb!]
  \renewcommand{\figurename}{Figure A}            
   \centering\textbf{(e)}
   \centering
\gridline{\includegraphics[width=0.95\textwidth]{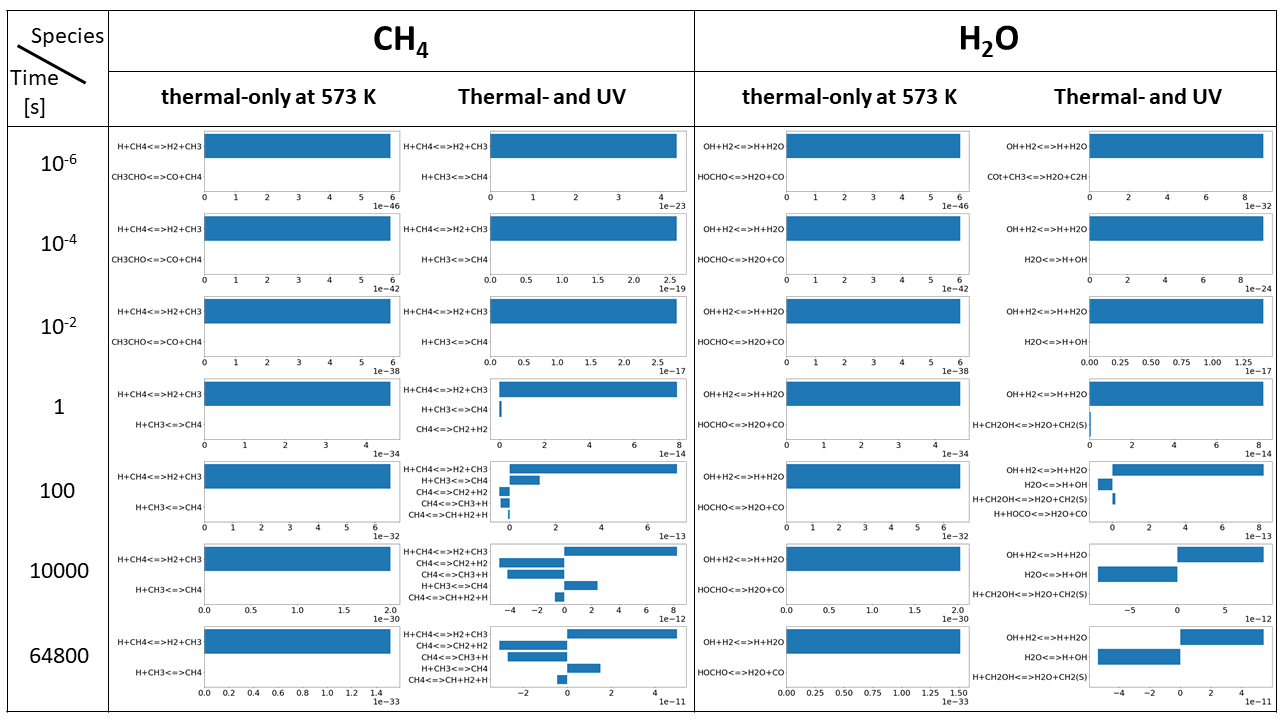}
          } 
          
   \centering\textbf{(f)}
\gridline{\includegraphics[width=0.95\textwidth]{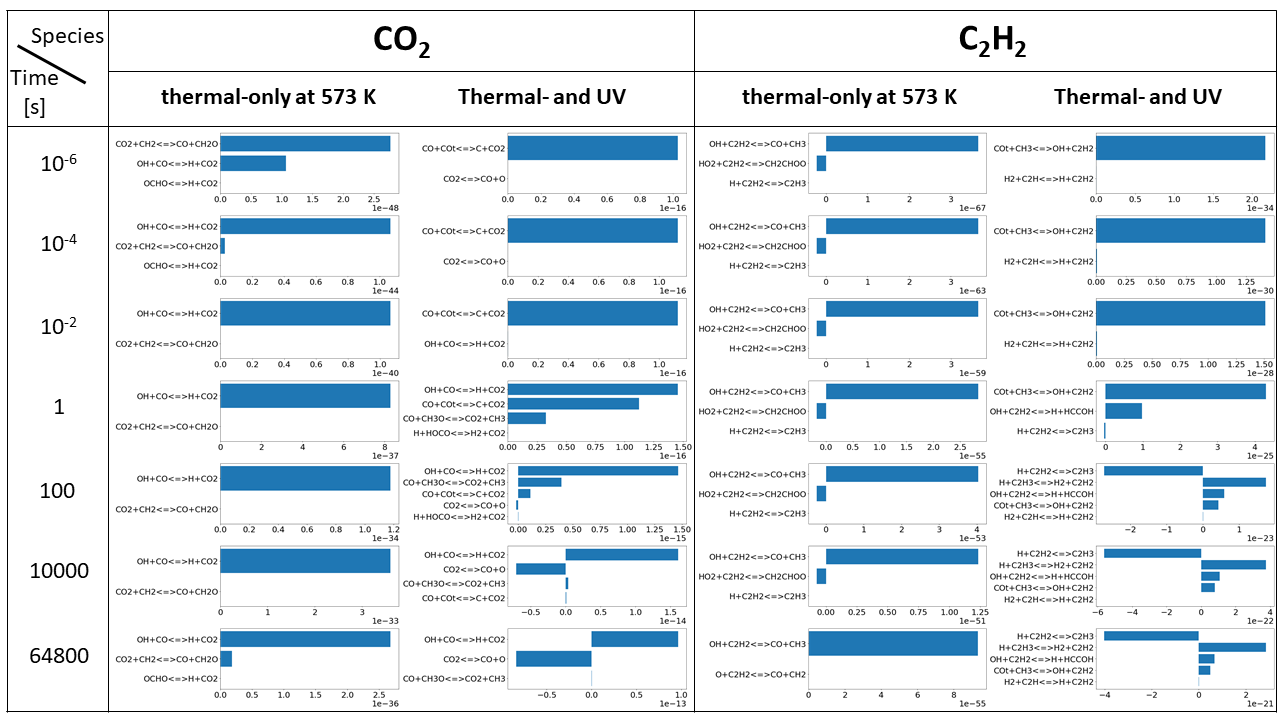}
          }
  \caption{Rate of production analysis on each species: (a) HCO and \ce{CH2O}; (b) HCOH and \ce{CH2OH}; (c) \ce{CH2} and CH; (d) \ce{CH3} and OH; (e) \ce{CH4} and \ce{H2O}; and (f) \ce{CO2} and \ce{C2H2} at temperatures of 573 K of the system of \cite{Fleury_2019}. Each row represents corresponding time and each column represents corresponding condition (i.e. thermal only or thermal- and UV photochemistry). The unit of numbers in the figure is mol/m\textsuperscript{3}/s.}
  \label{fgr:ROP_573K_3}
\end{figure}

\begin{figure}[hb!]
   \centering(\textbf{a})
\gridline{\includegraphics[width=1\textwidth]{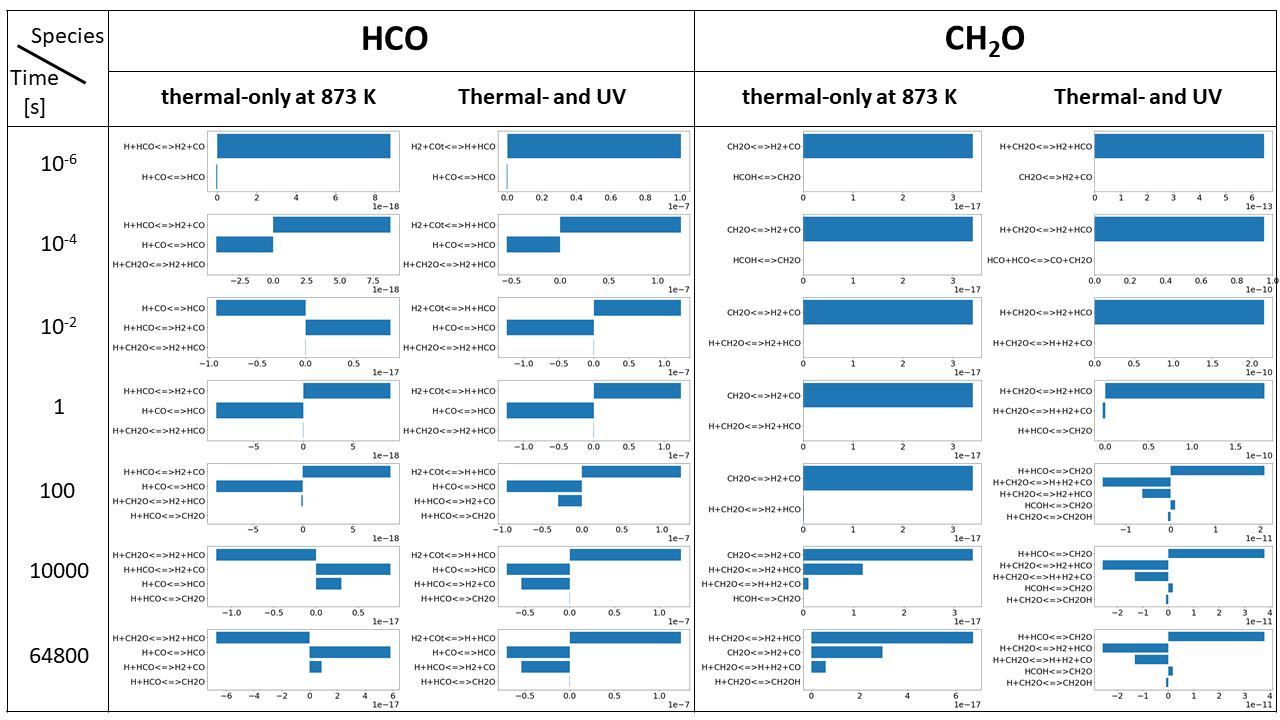}
          }
   \centering\textbf{(b)}
\gridline{\includegraphics[width=1\textwidth]{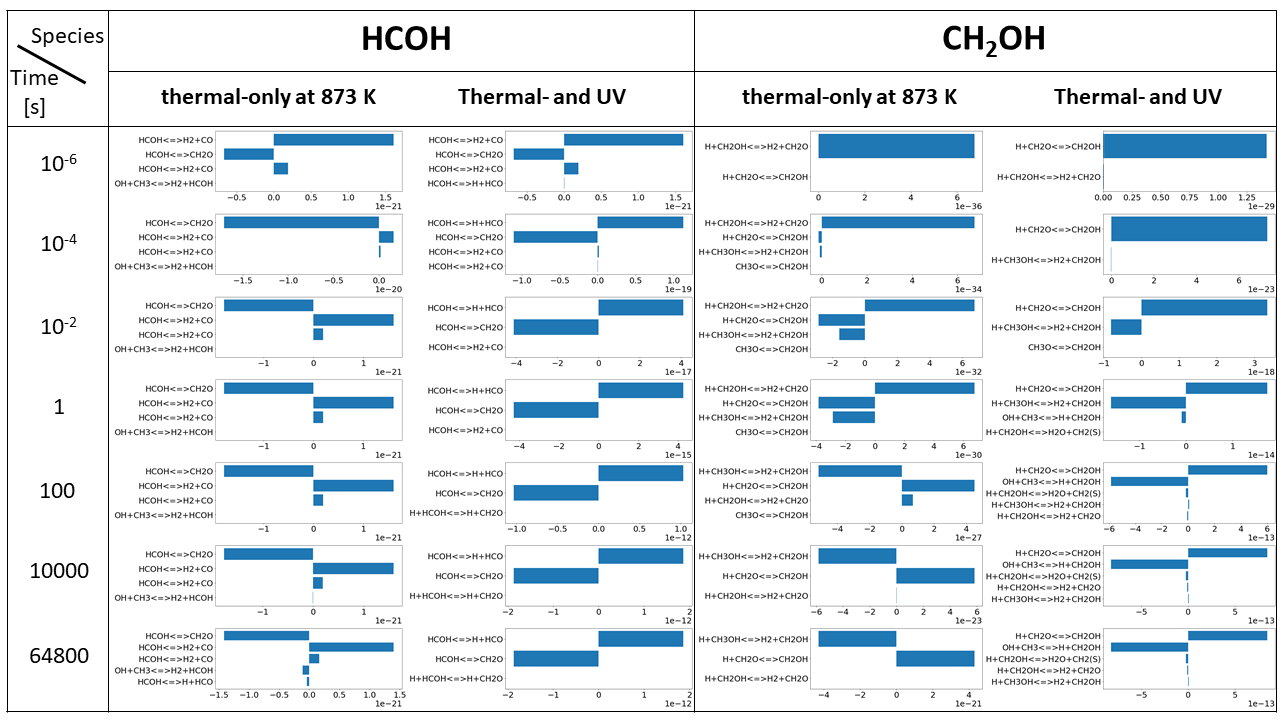}
          }
  \label{fgr:ROP_873K_1}
\end{figure}          

\begin{figure}[hb!]
   \centering\textbf{(c)}
\gridline{\includegraphics[width=1\textwidth]{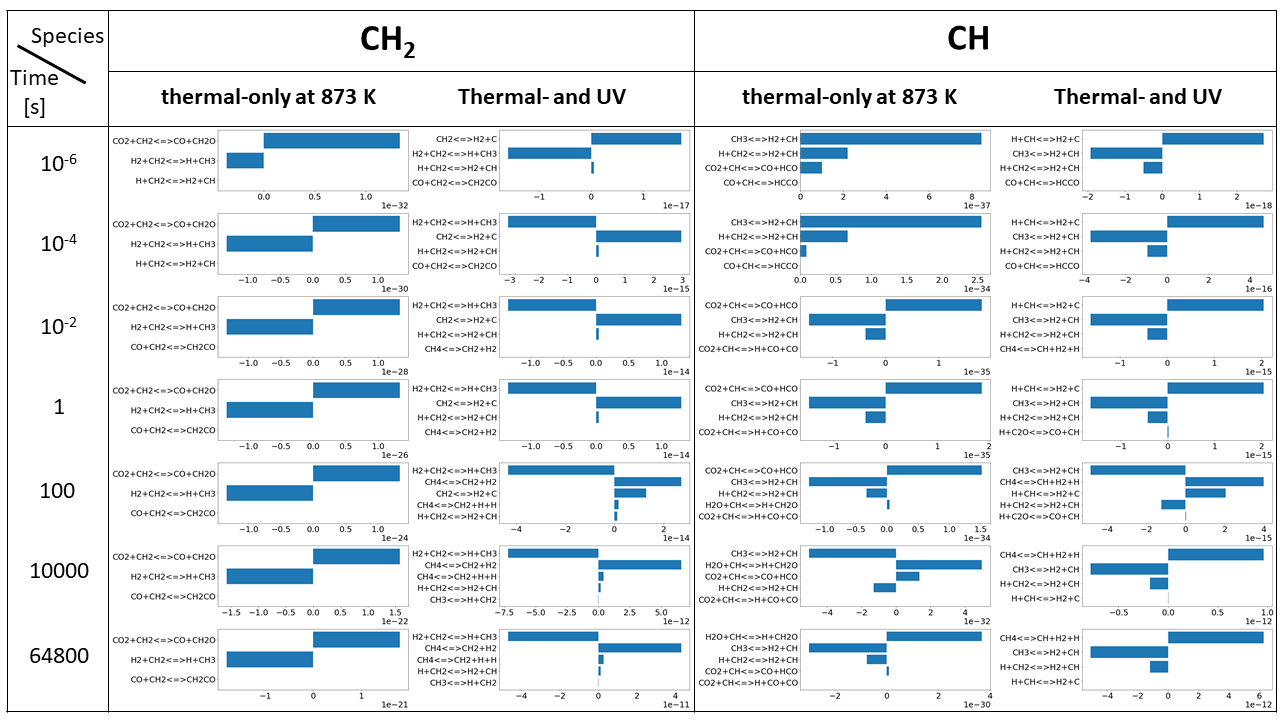}
          }

   \centering\textbf{(d)}
\gridline{\includegraphics[width=1\textwidth]{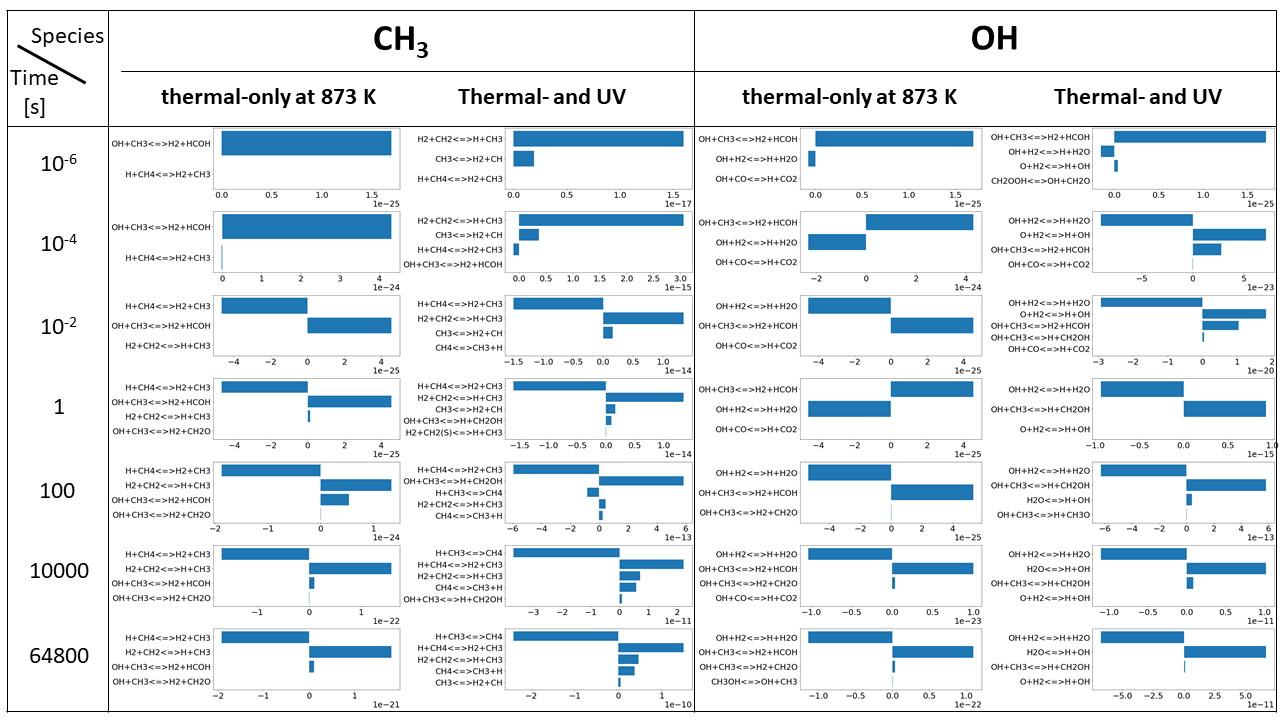}
          }
  \label{fgr:ROP_873K_2}
\end{figure}
\begin{figure}[hb!]
  \renewcommand{\figurename}{Figure A}            
   \centering\textbf{(e)}
\gridline{\includegraphics[width=0.95\textwidth]{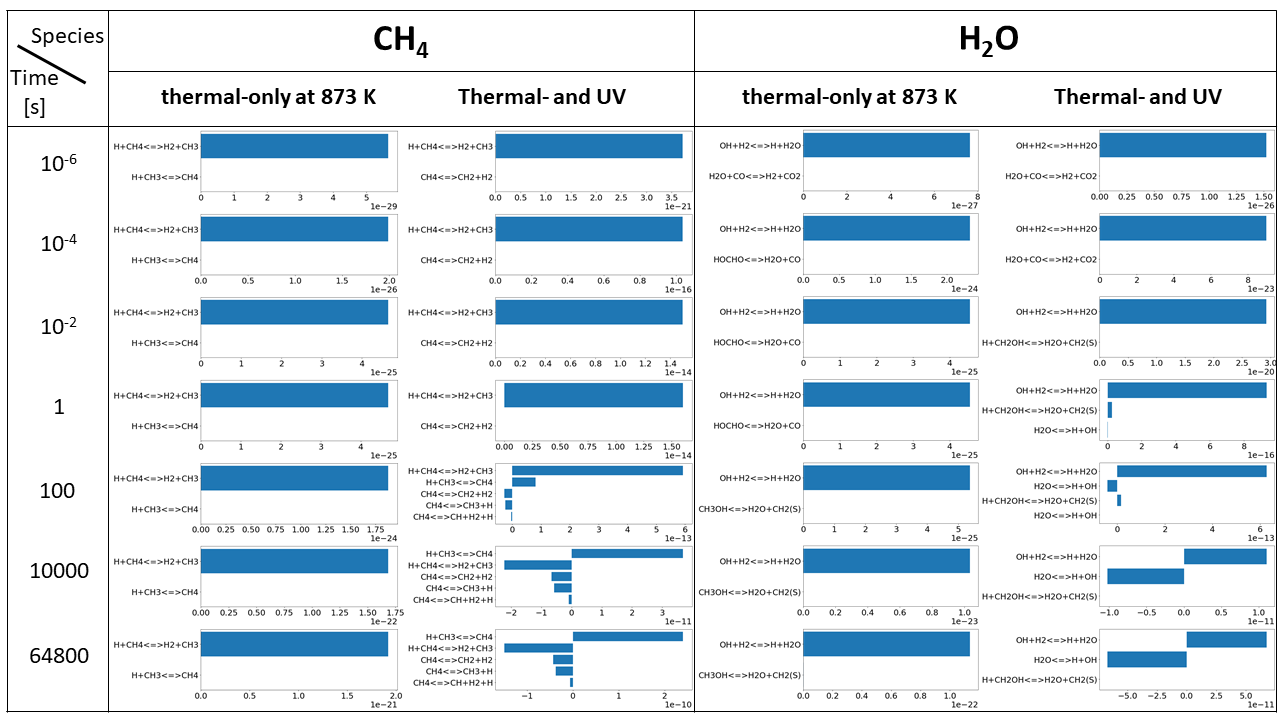}
          } 
          
   \centering\textbf{(f)}
\gridline{\includegraphics[width=0.95\textwidth]{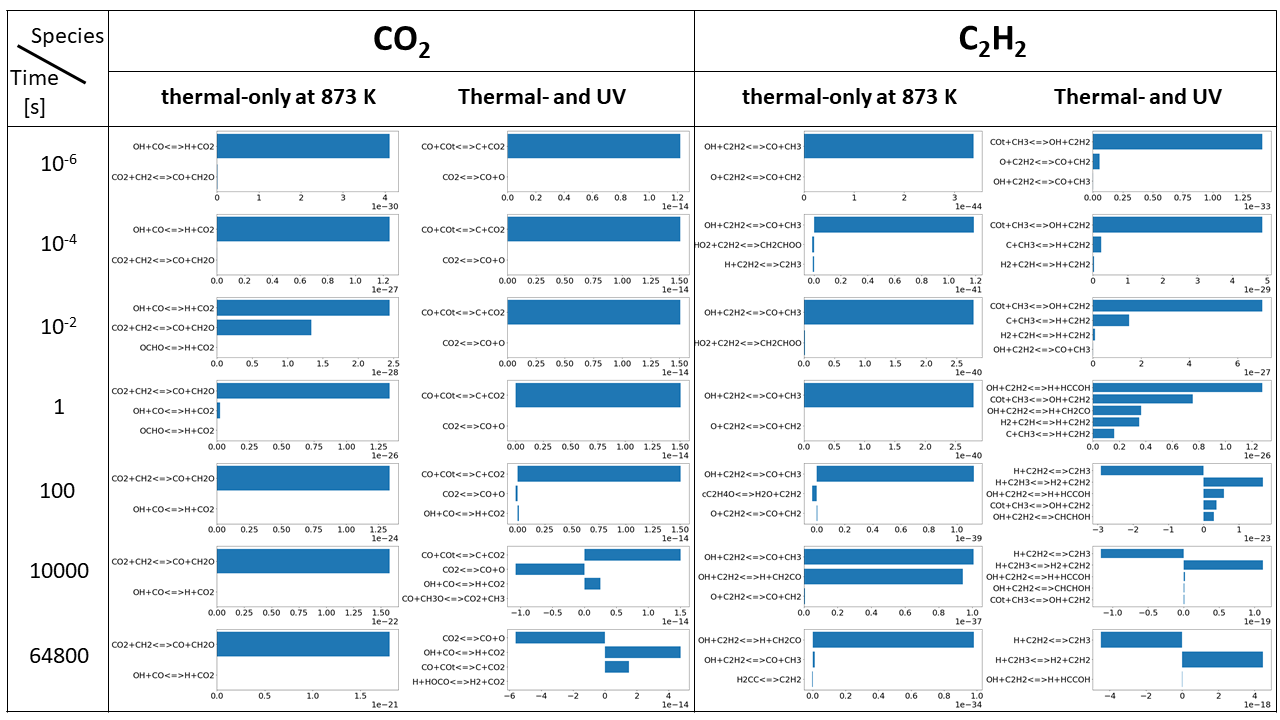}
          }
  \caption{Rate of production analysis on each species: (a) HCO and \ce{CH2O}; (b) HCOH and \ce{CH2OH}; (c) \ce{CH2} and CH; (d) \ce{CH3} and OH; (e) \ce{CH4} and \ce{H2O}; and (f) \ce{CO2} and \ce{C2H2} at temperatures of 873 K of the system of \cite{Fleury_2019}. Each row represents corresponding time and each column represents corresponding condition (i.e. thermal only or thermal- and UV photochemistry). The unit of numbers in the figure is mol/m\textsuperscript{3}/s.}
  \label{fgr:ROP_873K_3}
\end{figure}

\begin{figure}[hb!]
   \centering(\textbf{a})
\gridline{\includegraphics[width=1\textwidth]{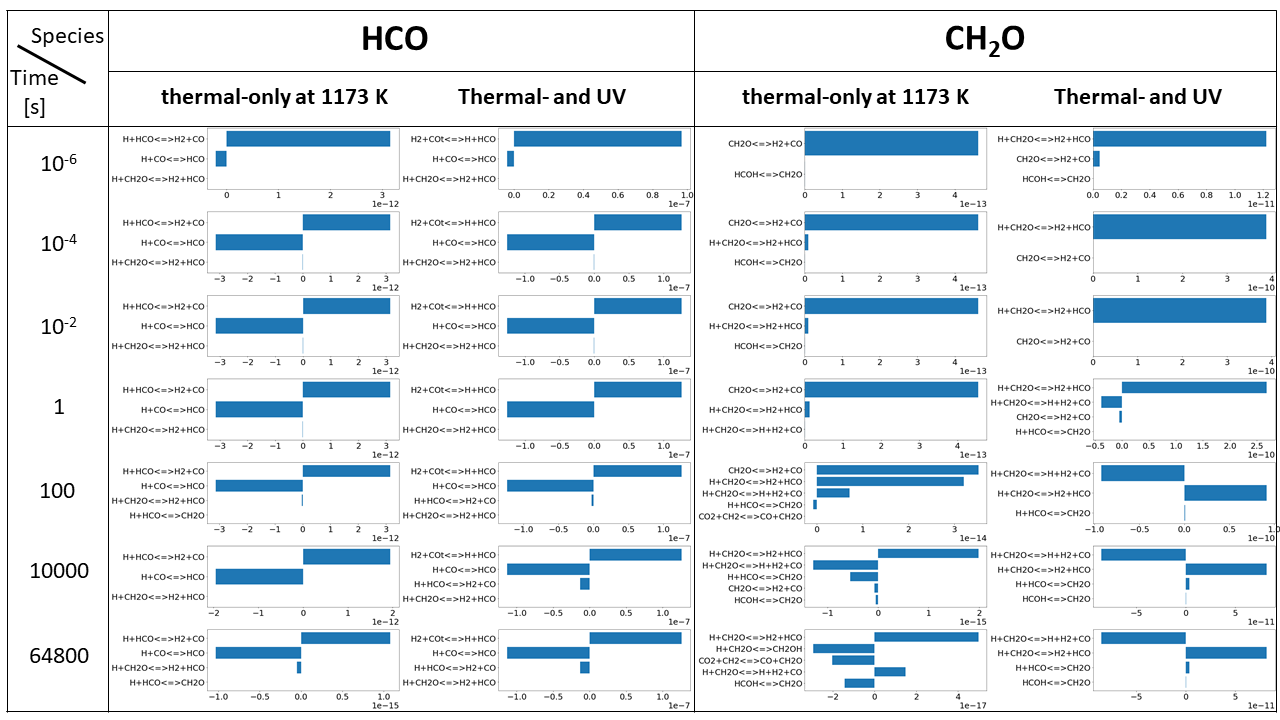}
          }
   \centering\textbf{(b)}
\gridline{\includegraphics[width=1\textwidth]{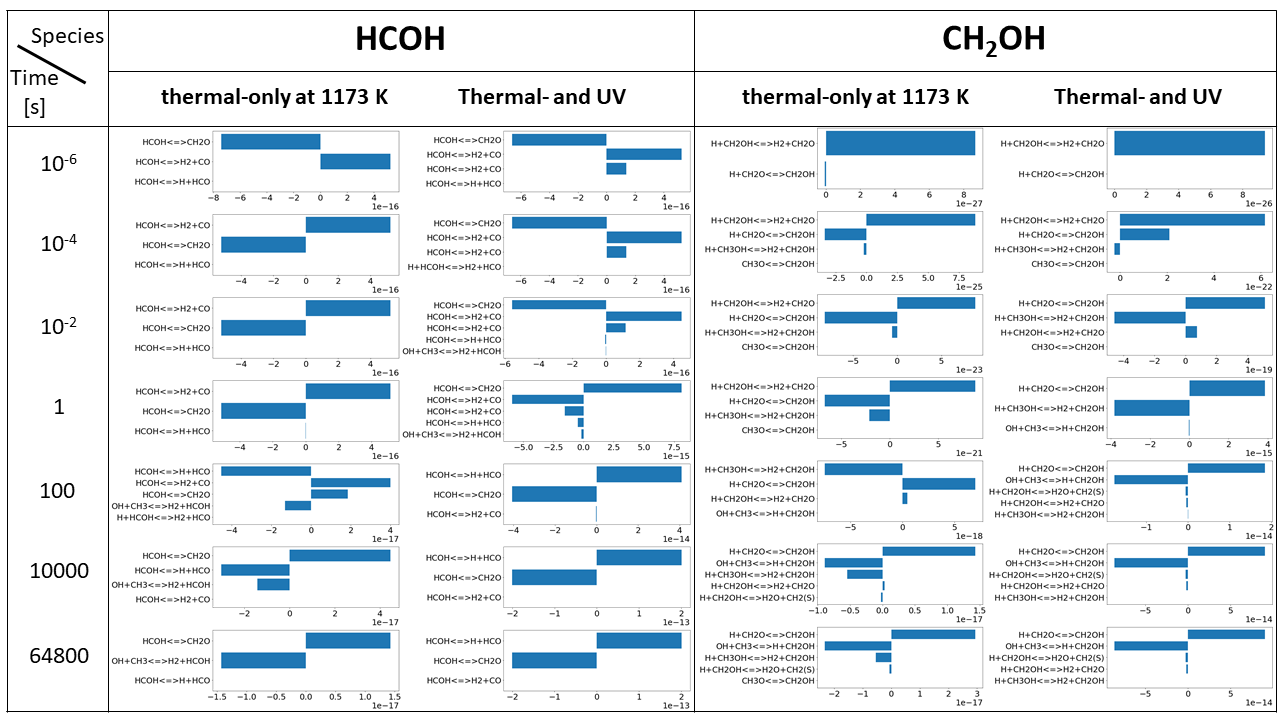}
          }
  \label{fgr:ROP_1173K_1}
\end{figure}          

\begin{figure}[hb!]
   \centering\textbf{(c)}
\gridline{\includegraphics[width=1\textwidth]{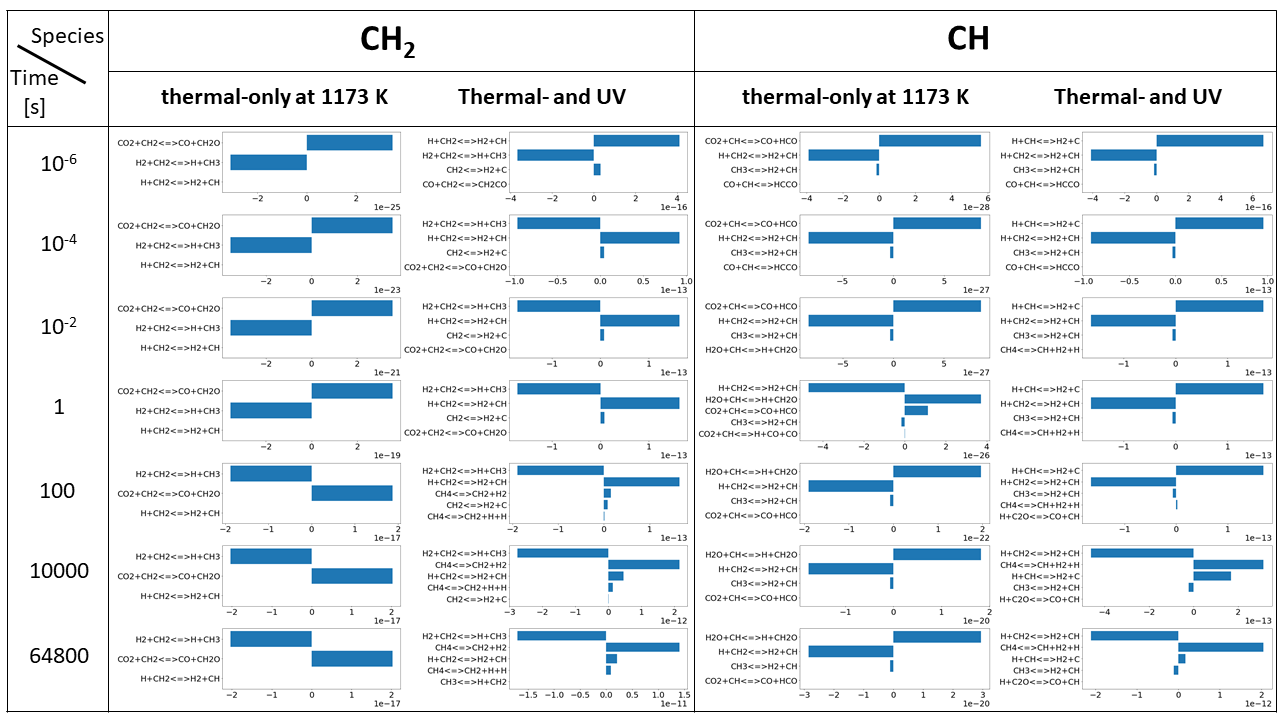}
          }

   \centering\textbf{(d)}
\gridline{\includegraphics[width=1\textwidth]{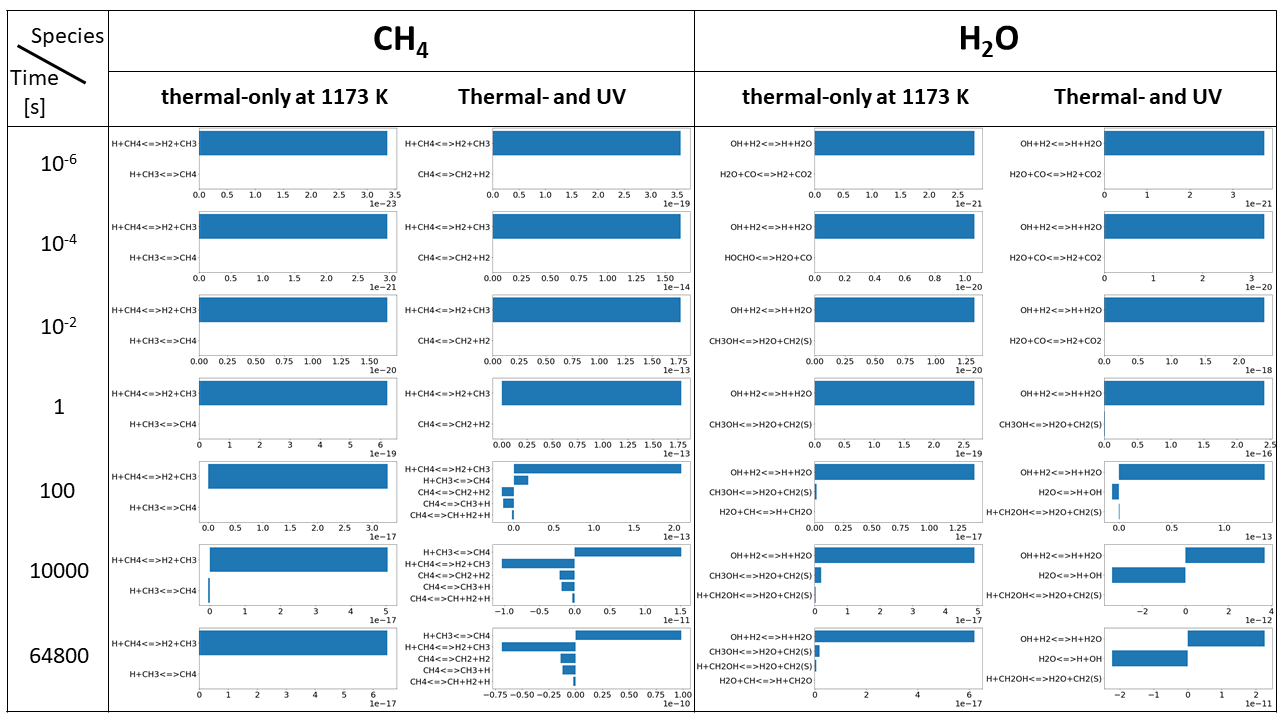}
          }
  \label{fgr:ROP_1173K_2}
\end{figure}
\begin{figure}[hb!]
  \renewcommand{\figurename}{Figure A}            
   \centering\textbf{(e)}
\gridline{\includegraphics[width=0.95\textwidth]{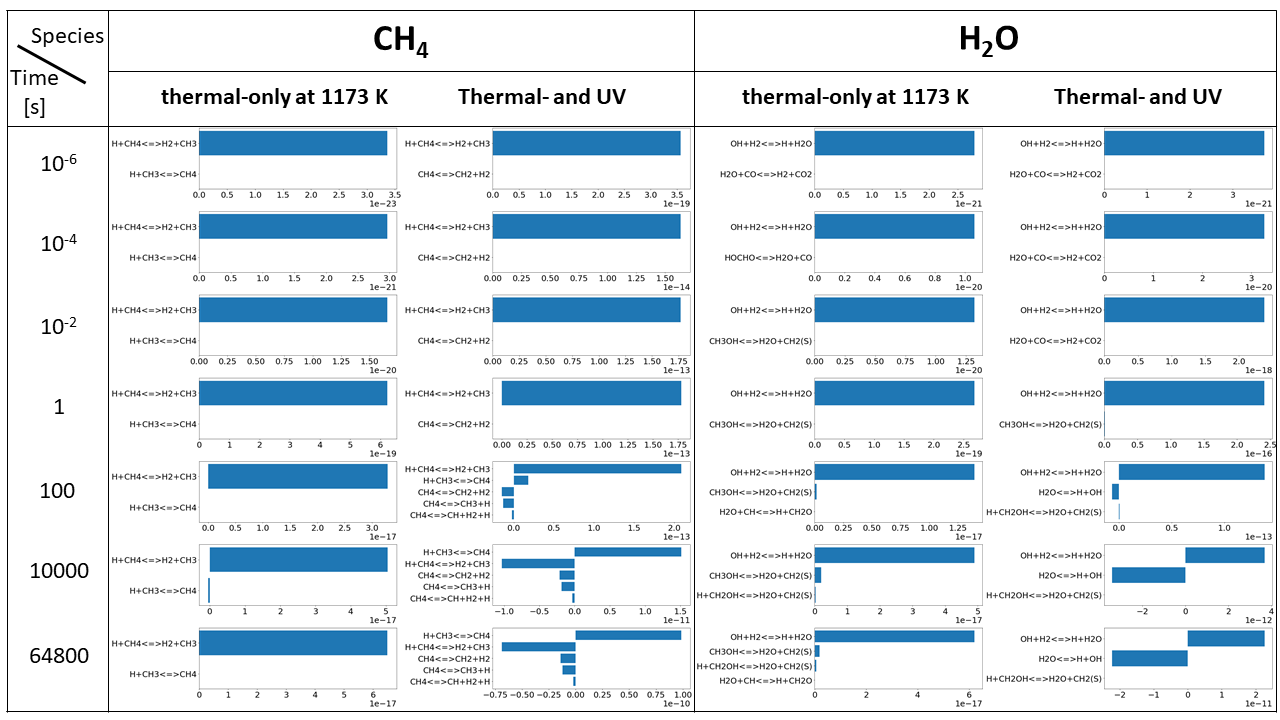}
          } 
          
   \centering\textbf{(f)}
\gridline{\includegraphics[width=0.95\textwidth]{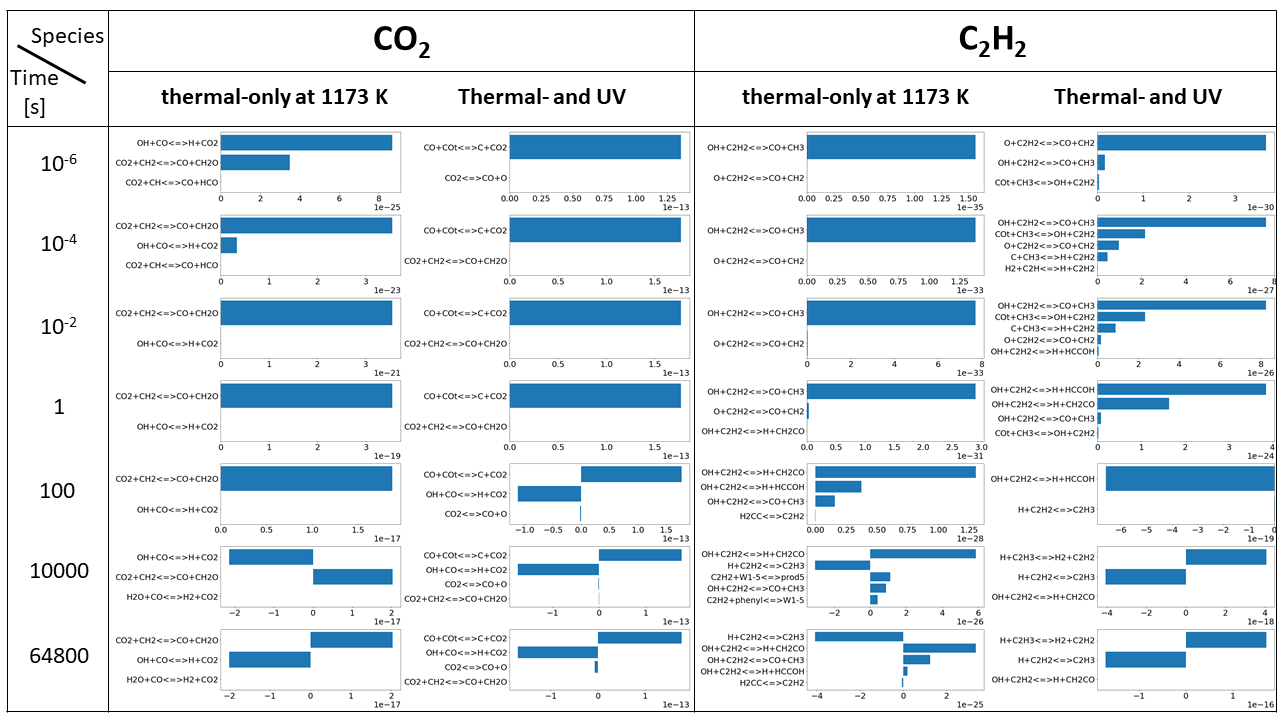}
          }
  \caption{Rate of production analysis on each species: (a) HCO and \ce{CH2O}; (b) HCOH and \ce{CH2OH}; (c) \ce{CH2} and CH; (d) \ce{CH3} and OH; (e) \ce{CH4} and \ce{H2O}; and (f) \ce{CO2} and \ce{C2H2} at temperatures of 1173 K of the system of \cite{Fleury_2019}. Each row represents corresponding time and each column represents corresponding condition (i.e. thermal only or thermal- and UV photochemistry). The unit of numbers in the figure is mol/m\textsuperscript{3}/s.}
  \label{fgr:ROP_1173K_3}
\end{figure}

\begin{figure}[hb!]
   \centering(\textbf{a})
\gridline{\includegraphics[width=1\textwidth]{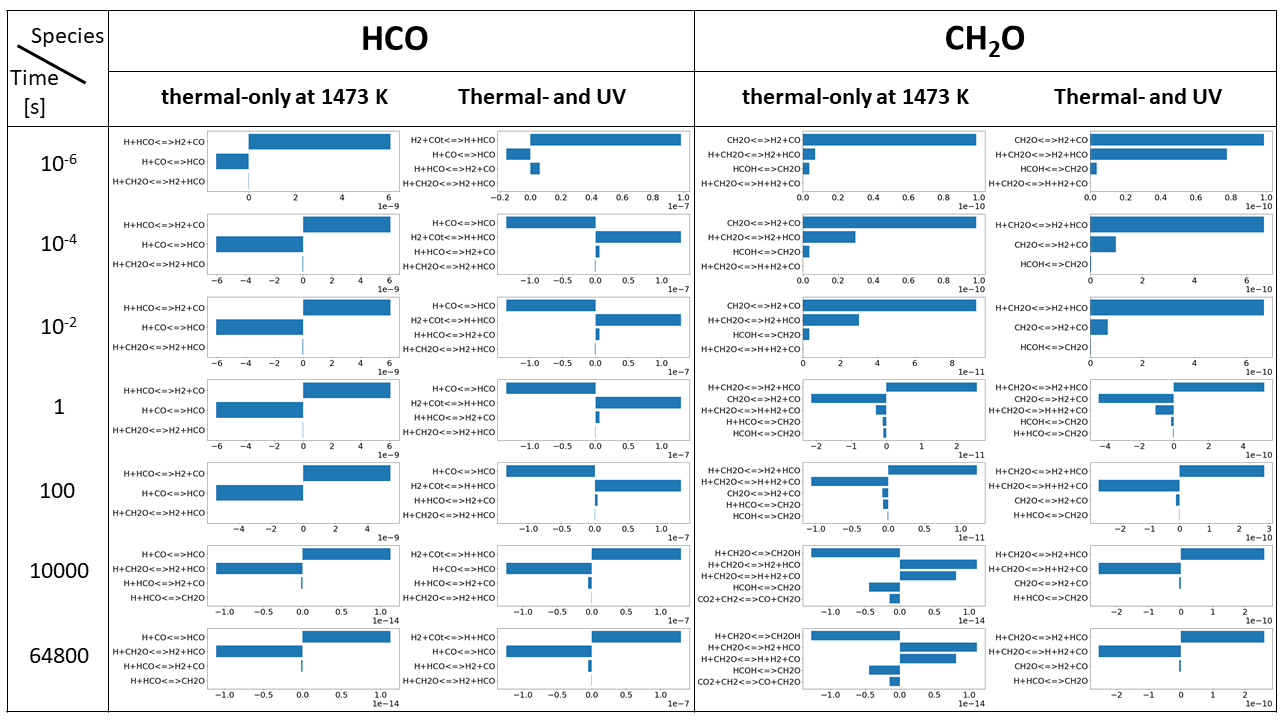}
          }
   \centering\textbf{(b)}
\gridline{\includegraphics[width=1\textwidth]{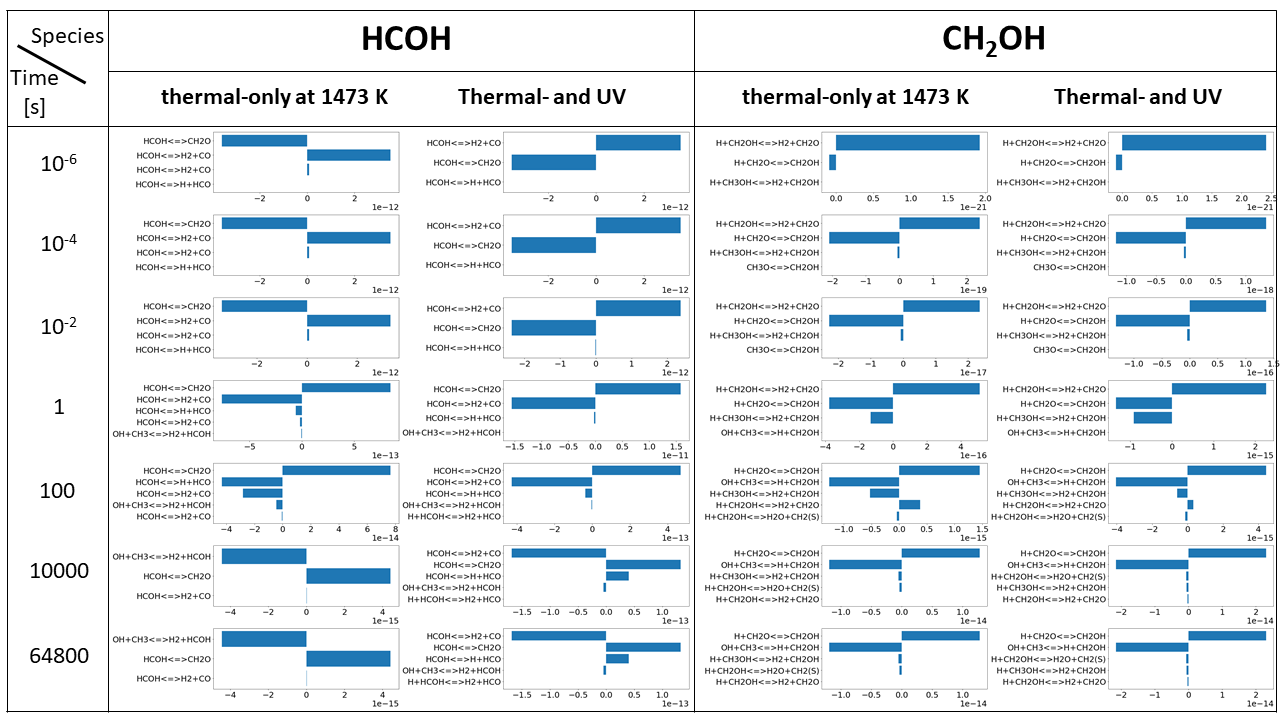}
          }
  \label{fgr:ROP_1473K_1}
\end{figure}          

\begin{figure}[hb!]
   \centering\textbf{(c)}
\gridline{\includegraphics[width=1\textwidth]{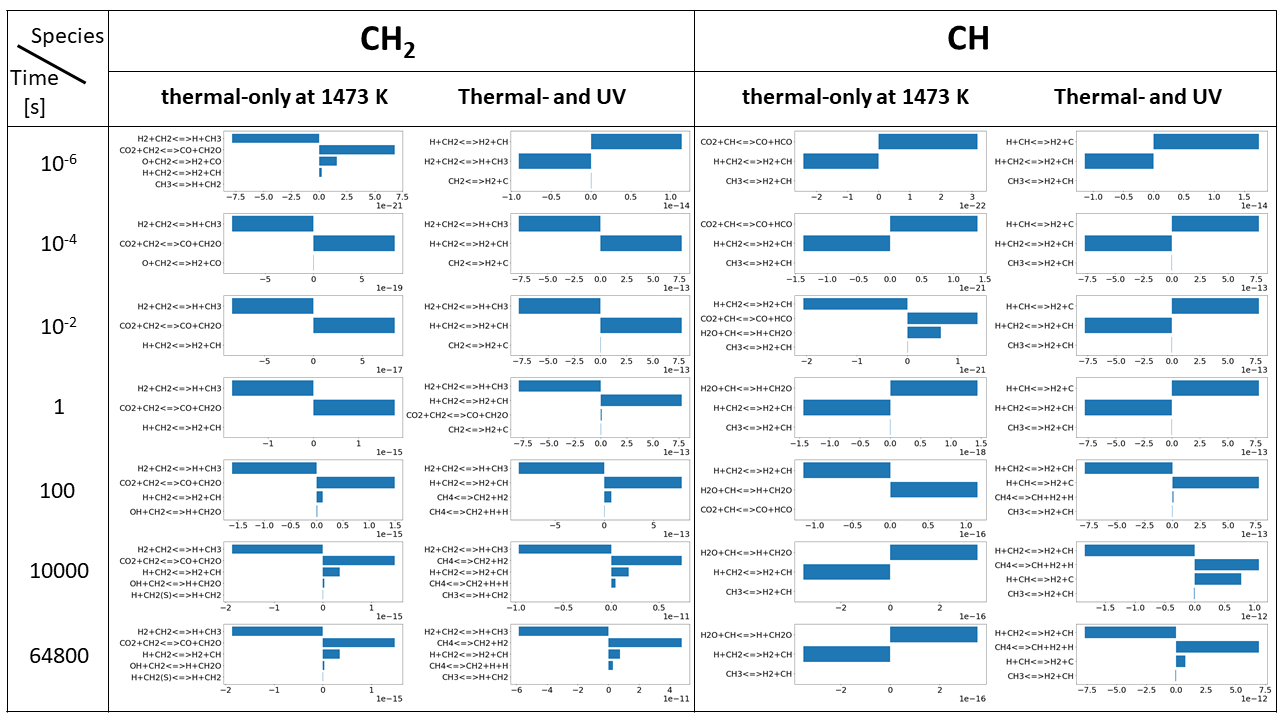}
          }

   \centering\textbf{(d)}
\gridline{\includegraphics[width=1\textwidth]{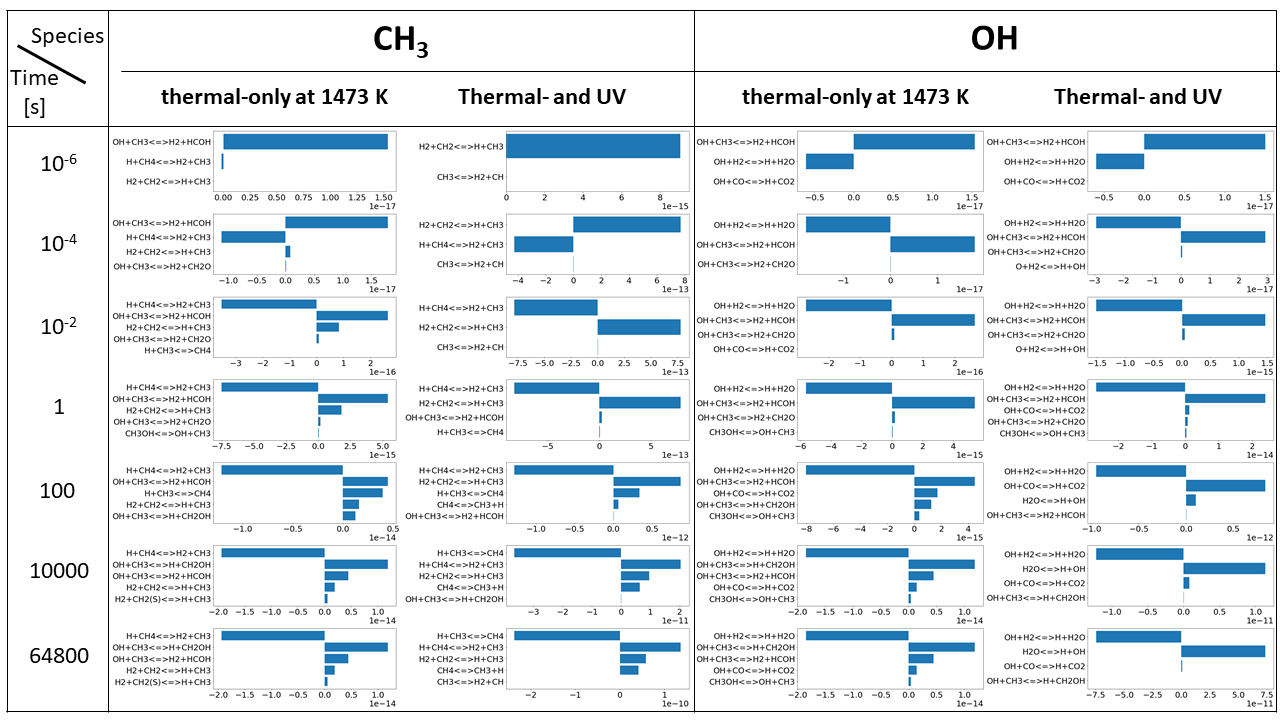}
          }
  \label{fgr:ROP_1473K_2}
\end{figure}
\begin{figure}[hb!]
  \renewcommand{\figurename}{Figure A}            
   \centering\textbf{(e)}
\gridline{\includegraphics[width=0.95\textwidth]{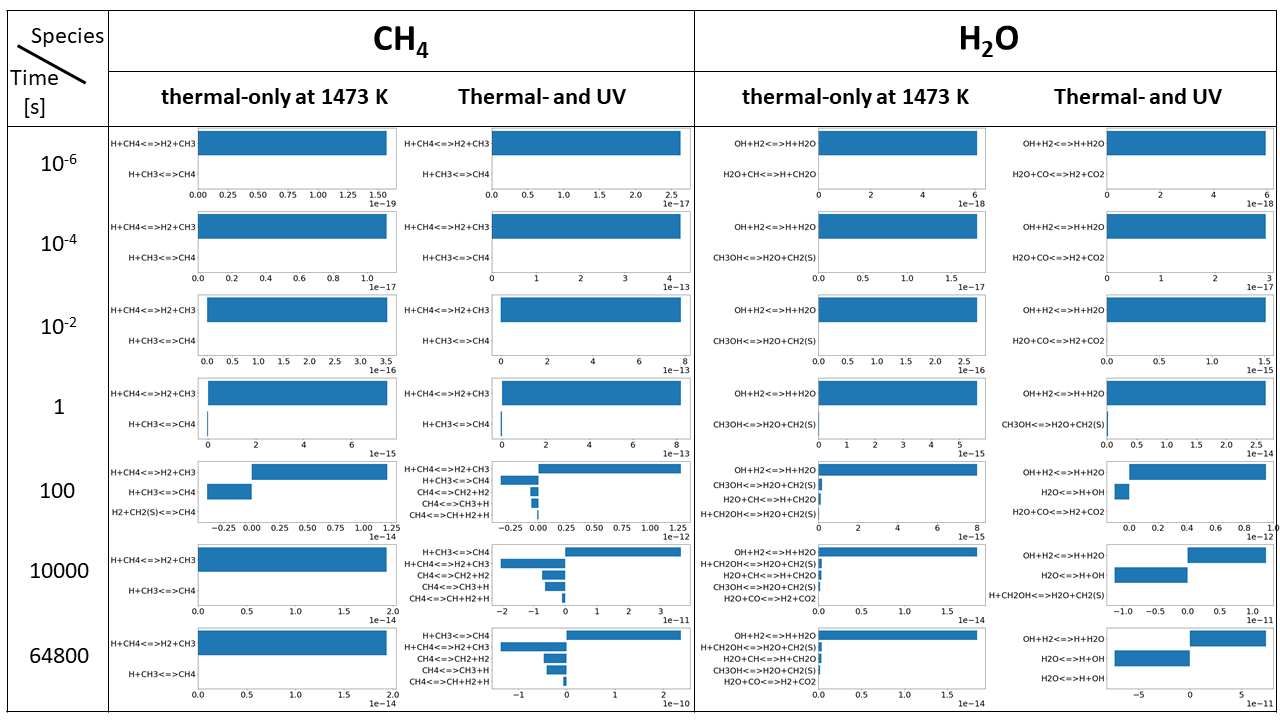}
          } 
          
   \centering\textbf{(f)}
\gridline{\includegraphics[width=0.95\textwidth]{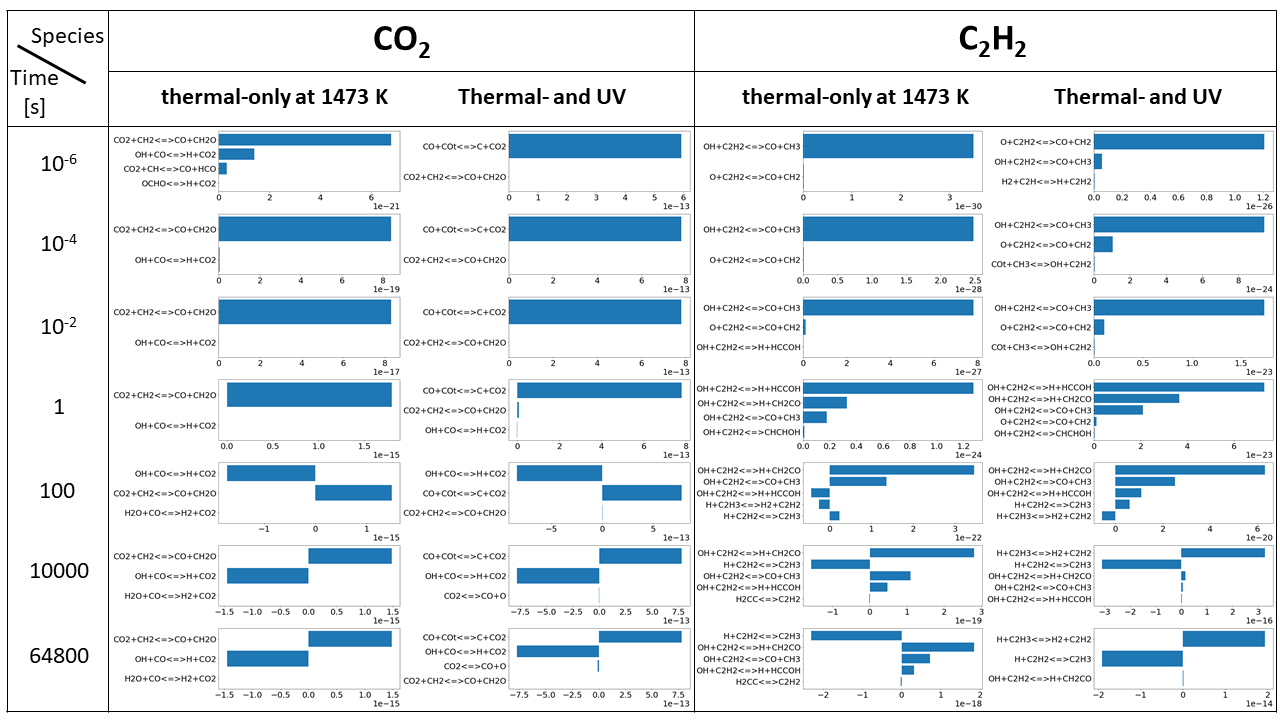}
          }
  \caption{Rate of production analysis on each species: (a) HCO and \ce{CH2O}; (b) HCOH and \ce{CH2OH}; (c) \ce{CH2} and CH; (d) \ce{CH3} and OH; (e) \ce{CH4} and \ce{H2O}; and (f) \ce{CO2} and \ce{C2H2} at temperatures of 1473 K of the system of \cite{Fleury_2019}. Each row represents corresponding time and each column represents corresponding condition (i.e. thermal only or thermal- and UV photochemistry). The unit of numbers in the figure is mol/m\textsuperscript{3}/s.}
  \label{fgr:ROP_1473K_3}
\end{figure}
\clearpage
\section{Time-dependent mole-fraction profiles of \ce{C2H2} under various experimental conditions} \label{sec:C2H2_mole_fraction_profile}
\begin{figure}[hb!]
  \renewcommand{\figurename}{Figure B}   
    \centering
    \includegraphics[width=1\textwidth]{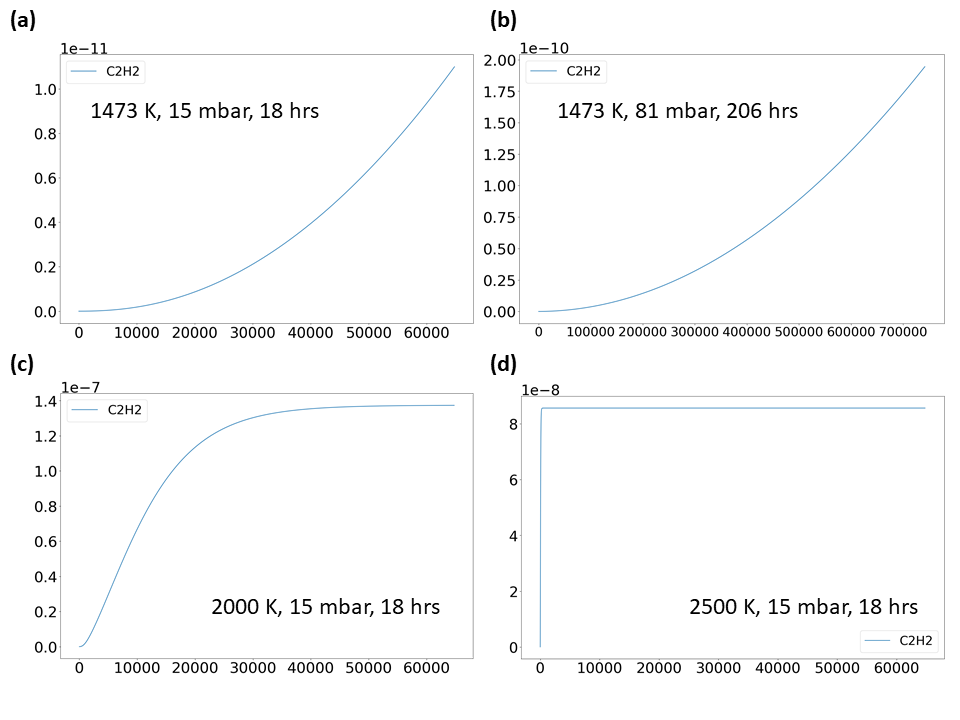}
    \caption{\footnotesize Time-dependent (the unit of the x-axis is [s]) molecular mixing ratio profiles of acetylene predicted by the kinetic model under various experimental conditions with UV photons avaialable (initial mole-fractions of \ce{H2} : CO = 99.7 : 0.3): (a) 1473 K, 15 mbar, 18hrs; (b) 1473 K, 81 mbar, 206 hrs; (c) 2000 K, 15 mbar, 18 hrs; (d) 2500 K, 15 mbar, 18 hrs}
    \label{fig:C2H2_formation_profiles}
\end{figure}
\clearpage
\section{Sensitivity analysis} \label{sec:sensitivity_analysis}
\begin{figure}[hb!]
  \renewcommand{\figurename}{Figure C}   
    \centering
    \includegraphics[width=1\textwidth]{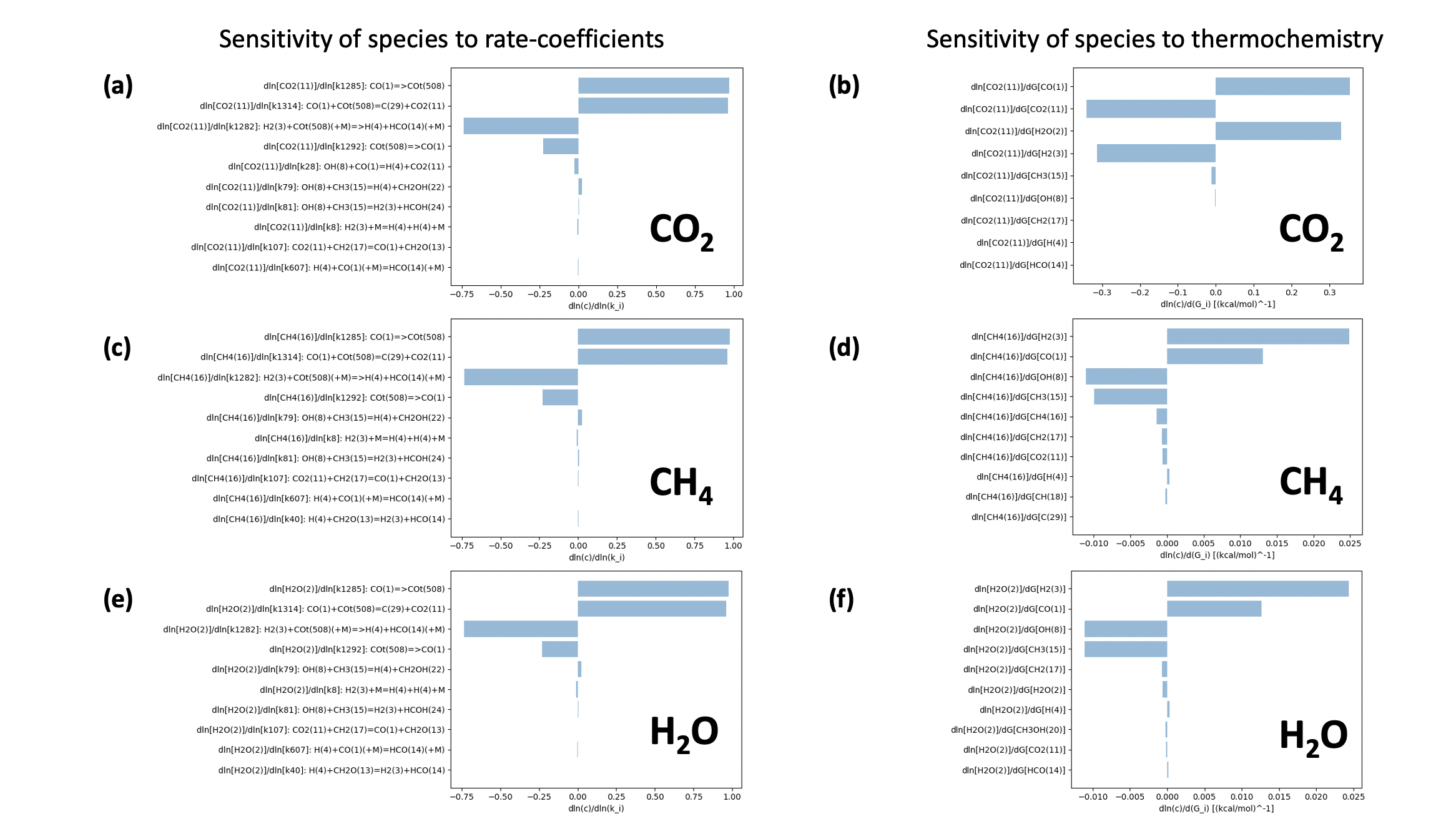}
    \caption{\footnotesize Sensitivity analysis on (a) \ce{CO2} to rate-coefficients; (b) \ce{CO2} to thermochemistry; (c) \ce{CH4} to rate-coefficients; (d) \ce{CH4} to thermochemistry; (e) \ce{H2O} to rate-coefficients; (f) \ce{H2O} to thermochemistry; Simulated under the condition of 1473 K, 15 mbar, [\ce{H2}] = 0.997, [CO] = 0.003.}
    \label{fig:SA_1473K}
\end{figure}

\begin{figure}
  \renewcommand{\figurename}{Figure C}   
    \centering
    \includegraphics[width=1\textwidth]{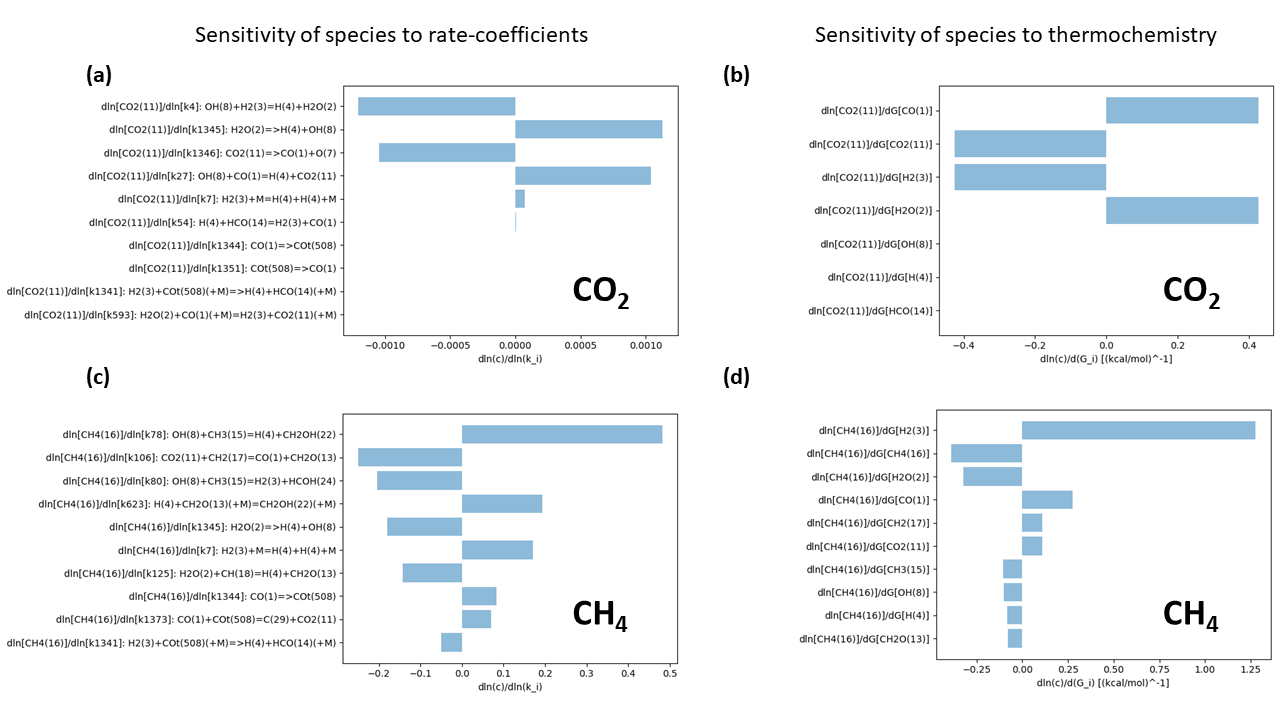}
    \caption{\footnotesize Sensitivity analysis on (a) \ce{CO2} to rate-coefficients; (b) \ce{CO2} to thermochemistry; (c) \ce{CH4} to rate-coefficients; (d) \ce{CH4} to thermochemistry; Simulated under the condition of 1473 K, 15 mbar, [\ce{H2}] = 0.9926, [CO] = 0.0026, [H2O] = 0.0034.}
    \label{fig:SA_1173K}
\end{figure}
\clearpage
\section{Time-dependent mole-fraction profiles of \ce{CO2} under various experimental conditions} \label{sec:CO2_mole_fraction_profile}

\begin{figure}[hb!]
  \renewcommand{\figurename}{Figure D}   
    \centering
    \includegraphics[width=1\textwidth]{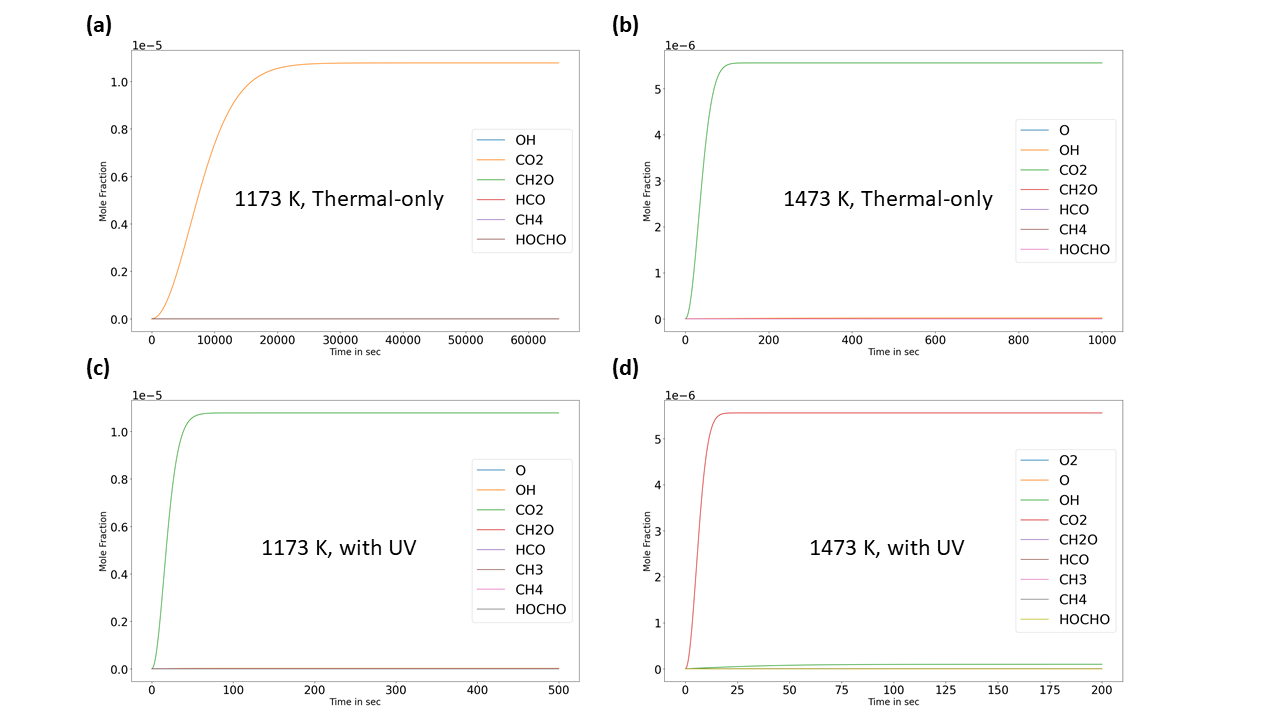}
    \caption{\footnotesize Simulated time-dependent molecular mixing ratio profile of \ce{CO2} under various conditions of \cite{Fleury_2020} : (a) 1173 K, thermal-only; (b) 1473 K, thermal-only; (c) 1173 K, with UV irradiation; (d) 1473 K, with UV irradiation; All simulated under the condition of 15 mbar, [\ce{H2}] = 0.9926, [CO] = 0.0026, [\ce{H2O}] = 0.0034.}
    \label{fig:CO2_t_profile}
\end{figure}


\end{document}